\documentclass[12pt]{article}
\usepackage[]{geometry} %margin=1in
\usepackage{bm}
\usepackage[tableposition=top, font = footnotesize]{caption} 
\usepackage{amsmath}
\usepackage{amsfonts}
\usepackage{amssymb}
\usepackage{multirow}
\usepackage{verbatim}
\usepackage{graphicx}
\usepackage{natbib}
\usepackage{algorithm}
\usepackage{tabularx}
\usepackage[noend]{algpseudocode}
\usepackage[table]{xcolor}
\usepackage{setspace}
\usepackage{subcaption} 

\usepackage[normalem]{ulem}

\graphicspath{{all_images/}}

\makeatletter
\newcommand{\multiline}[1]{%
  \begin{tabularx}{\dimexpr\linewidth-\ALG@thistlm}[t]{@{}X@{}}
    #1
  \end{tabularx}
}
\makeatother

\newcommand{\blind}{1}

\begin{document}

\def\spacingset#1{\renewcommand{\baselinestretch}%
{#1}\small\normalsize} \spacingset{1}

%%%%%%%%%%%%%%%%%%%%%%%%%%%%%%%%%%%%%%%%%%%%%%%%%%%%%%%%%%%%%%%%%%%%%%%%%%%%%%

\if1\blind
{
  \title{\bf Multiple imputation in functional regression with applications to EEG data from a depression study}
  \author{Adam Ciarleglio\thanks{
    Corresponding author.  This work is based upon work supported by the National Institutes of Health under grants NIMH K01 MH113850 and NIMH R01 MH099003.}\hspace{.2cm}\\
    Department of Biostatistics and Bioinformatics, \\ Milken Institute School of Public Health, \\ George Washington University, Washington, DC\\
    \\
    Eva Petkova \\
    Department of Population Health and \\ Department of Child and Adolescent Psychiatry, \\ New York University, New York, NY \\
    \\
    Ofer Harel \\
    Department of Statistics, University of Connecticut, Storrs, CT}
  \maketitle
} \fi

\if0\blind
{
  \bigskip
  \bigskip
  \bigskip
  \begin{center}
    {\LARGE\bf Multiple imputation in functional regression with applications to EEG data from a depression study}
\end{center}
  \medskip
} \fi

\bigskip
\begin{abstract}
Current source density (CSD) power asymmetry, a measure derived from electroencephalography (EEG), is a potential biomarker for major depressive disorder (MDD).  Though this measure is functional in nature (defined on the frequency domain), it is typically reduced to a scalar value prior to analysis, possibly obscuring the relationship between brain function and MDD.  To overcome this issue, we sought to fit a functional regression model to estimate the association between CSD power asymmetry and MDD diagnostic status, adjusting for age, sex, cognitive ability, and handedness using data from a large clinical study.  Unfortunately, nearly 40\% of the observations were missing either their functional EEG data, their cognitive ability score, or both.  In order to take advantage of all of the available data, we propose an extension to multiple imputation by chained equations that handles both scalar and functional data.  We also propose an extension to Rubin's Rules for pooling estimates from the multiply imputed data sets in order to conduct valid inference.  We investigate the performance of the proposed extensions in a simulation study and apply them to our clinical study data.  Our analysis reveals that the association between CSD power asymmetry and diagnostic status depends on both age and sex.
\end{abstract}

\noindent%
\vfill

\newpage
\spacingset{1} % DON'T change the spacing!

\section{Introduction} \label{s_intro}
Functional data analysis (FDA) \citep{fda2} has become an important tool for understanding a host of complex data types generated in medical \citep{harezlak_2008, sorensen_2013}, economic \citep{ramsay_2002_econ, jank_2006}, environmental \citep{henderson_2006, ikeda_2008}, and other application areas \citep{hutchinson_2004, torres_2011}.  In particular, regression methods for functional data \citep{cardot_1999, yao_2005, morris_2006, reiss_2007, james_2009, GoldsmithPenalized, gertheiss_2013, ivanescu2015, scheipl_2015} have been widely developed and applied due to their ability to reveal complex patterns of association.  It is reasonable to assume that FDA methods will become increasingly important as both the collection and storage of large amounts of functional data become simpler and cheaper.  Though many useful and powerful functional regression methods have been developed, they all assume that the data to which they are being applied consist of complete observations (i.e., no missing outcomes or predictors).  There has been limited work on how to handle incomplete data in functional regression.      

Our interest in this problem is motivated by an investigation of how characteristics of electroencephalography (EEG) data differ between those diagnosed with major depressive disorder (MDD) and healthy controls (HCs) in the EMBARC clinical trial (NCT01407094). The EMBARC trial was conducted to search for biomarkers of antidepressant treatment response, but the rich data generated from the study can be probed to address other research questions related to MDD.  Like many psychiatric conditions, MDD is a disease for which no clear biological markers currently exist.  The state-of-the-art for diagnosis of MDD typically relies on self-reported symptom check-lists like those available in the DSM-5 \citep{dsm5}.  It has been argued that the identification of reliable and specific biomarkers of MDD may provide better understanding of the disease, which in turn may lead to development of improved treatment strategies.  To this end, investigators have focused their attention on trying to use neuroimaging to extract information about brain structure and function that may be useful in understanding the disease.  
%\citep{EMBARCDesc}

EEG is a neuroimaging modality that is particularly attractive since it is relatively low-cost, can be administered in resource limited settings, and is non-invasive.  EEG measures changes in voltage across the scalp assumed to be related to gross neuronal activity.  In the EMBARC trial, measurements were taken from multiple electrodes placed at various locations on the scalp using a standard headcap while subjects had their eyes closed.  The time-series data collected at each electrode underwent a sequence of processing steps and were transformed to the frequency domain in order to obtain current source density (CSD) power curves. These curves provide information on the intensity of different frequencies (rhythms) of neuronal activity for a subject \citep{Tenke1}.  Frequency values are often divided into frequency bands: from about 4 to 8 Hz is the theta band, about 8 to 16 Hz is the alpha band, and about 16 to 32 Hz is the beta band.  Values outside of these bands may be of interest, but we restrict our attention to theta, alpha, and beta rhythms.  A full description of the collection and processing of these data can be found in \cite{tenke2017}.      

\begin{figure}[t]
\center
\begin{minipage}{0.20\textwidth}
\vspace{-0.6cm}
\center
\resizebox{1.5\linewidth}{!}{\includegraphics{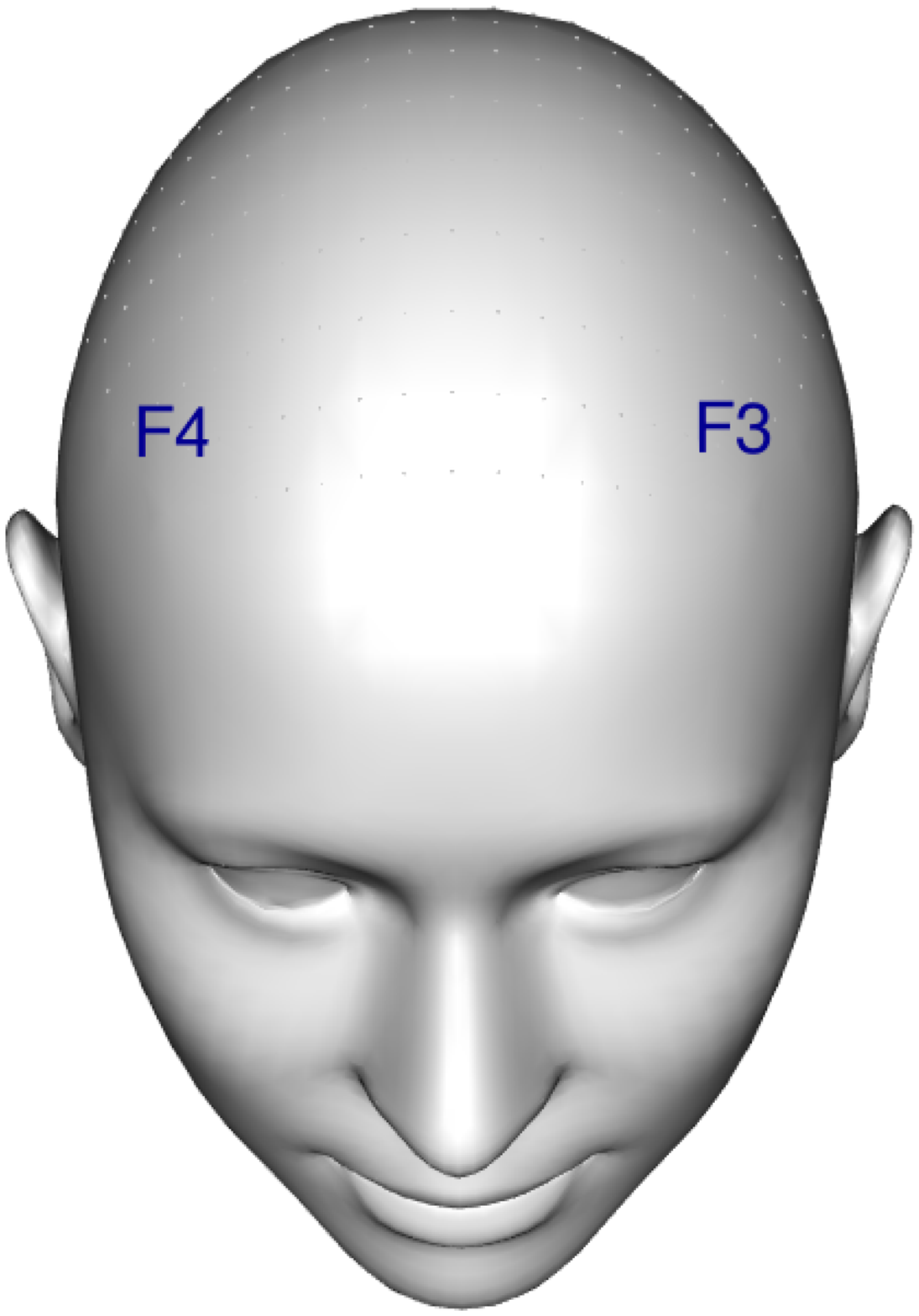}}
\end{minipage}
\begin{minipage}{0.20\textwidth}
\end{minipage}
\begin{minipage}{0.60\textwidth}
\center
\includegraphics[width=0.8\linewidth]{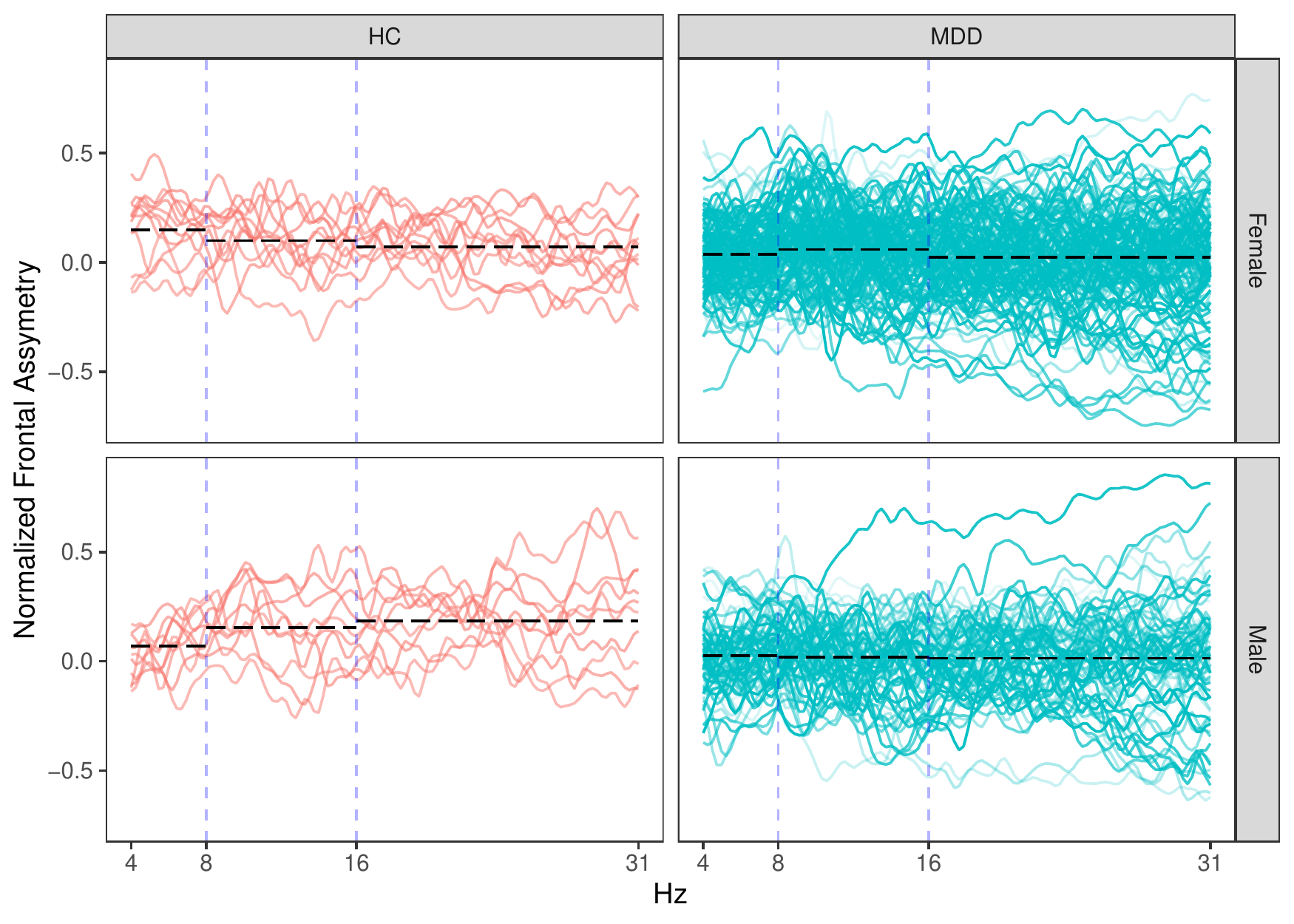} 
\end{minipage}
\caption{\label{fig:raw} \textbf{Left:} Locations of F3 and F4 electrodes. \textbf{Right:} Normalized frontal asymmetry curves. HC = healthy control, MDD = major depressive disorder.  Blue vertical dashed lines separate the theta (4 - 8 Hz), alpha (8 - 16 Hz), and beta (16 - 31 Hz) frequency bands.  Black dashed lines show mean values within a frequency band.}
\end{figure}

There is a large literature on the relationship between various summary measures derived from resting state EEG and depression.  However, taken collectively, the results are generally inconclusive.  One measure that has been studied repeatedly in relation to MDD is frontal alpha asymmetry (F$\alpha$A): the difference in alpha power ($\mu V^2$) between right and left electrodes that are symmetrically located on the frontal region of the scalp.  \cite{vanderVinne17} conducted a meta-analysis of the association between F$\alpha$A, using the F3 and F4 electrodes (shown in Figure \ref{fig:raw}), and depression status (MDD vs. HCs).  They argue that gender, as well as age, may modify the association between depression status and F$\alpha$A, but state that many previous studies have failed to account for these effects.  Another limitation of previous analyses is that F$\alpha$A was analyzed as a single scalar value equal to the difference between the average power in the alpha band from the F4 electrode and that from the F3 electrode (divided by the sum of the values to normalize inter-individual differences).  We argue that this approach potentially discards relevant information by aggregating the CSD power curve to a single scalar summary measure.  Since the CSD power curves are functional data, observed in the frequency domain, it may be advantageous to assess the association between frontal asymmetry (FA) and depression status using a functional data analytic approach.    

The right panels of Figure \ref{fig:raw} show the FA curves over the theta, alpha, and beta frequency bands for MDD and HC subjects in the EMBARC study stratified by gender.  The FA curves are computed as $\frac{F4(t) - F3(t)}{F4(t) + F3(t)}$ for $t \in [4, 31.75]$ Hz where $F4(t)$ and $F3(t)$ are the CSD power values at frequency $t$ for the F4 and F3 electrodes, respectively.  FA values are available at 112 frequencies ranging from 4 to 31.75 Hz in 0.25 Hz increments.  An analysis on the scalar summary F$\alpha$A values similar to that in \cite{vanderVinne17} is equivalent to comparing the group-mean values corresponding to the dashed black lines over the 8 to 16 Hz band.  We propose to fit a functional regression model with the FA curve as the response and diagnostic status as the primary predictor, adjusting for gender, age, and other relevant factors.       

Unfortunately, some of the EEG data collected in the EMBARC study were flagged during quality control assessment as being ``Unacceptable'' or ``Marginal'' and therefore should not be used in our analysis.  Still others were wholly missing.  In fact, Figure \ref{fig:raw} displays only those FA curves for subjects with ``Acceptable'' or ``Good'' quality control designations.  In addition, some of the values for the covariates that we wish to adjust for are missing for some subjects.  Instead of throwing out observations from subjects with incomplete data and conducting a complete case analysis, we have developed a multiple imputation (MI) procedure \citep{rubin_1987, schafer_1997} for imputing the missing scalar and functional data and propose an approach for pooling estimates derived from the multiply imputed data sets.  

MI methods for scalar data have been broadly developed and research in this area remains active, particularly in employing machine learning methods \citep{doove_2014, xu_2016, zhao_2016}.  Furthermore, there are many user-friendly software packages that perform MI when the data consist only of scalar quantities, including software developed for \verb+R+ \citep{R_software} such as \emph{mice} \citep{vanburren_2011} and \emph{Amelia} \citep{honaker_2011} and procedures developed for \verb+SAS+ \citep{SAS_9.3} such as \emph{PROC MI} and \emph{PROC MIANALYZE}.  Similar methods and software should be available for handling functional data.  To our knowledge, approaches for MI of functional data have not been investigated in the literature nor has software been developed to perform MI with functional data.    

The rest of the paper is structured as follows.  In Section \ref{s_formal} we provide an overview of our target analysis models: functional regression models with either functional or scalar outcomes.  In Section \ref{s_framework} we extend the missing data framework to include functional data.  In Section \ref{s_meth} we present a method for performing MI via chained equations with scalar and functional data.  Section \ref{s_sim} presents the results of a simulation study, showing the performance of the proposed imputation and pooling procedures.  Section \ref{s_app} illustrates the proposed method using data from the EMBARC trial.  We conclude in Section \ref{s_disc} with a discussion and comments on possible directions for future research.  A supplementary appendix includes further detail on fitting functional regression as well as additional simulation and application results.     

Throughout, we use the following notational conventions: upper-case bold letters for matrices or collections of functions (distinction should be obvious by context), lower-case bold variables represent vectors, upper-case non-bold Roman letters for functions, lower-case non-bold Roman letters for scalars. Non-bold Greek letters are used for both scalar and functional parameters but the distinction should be obvious by context.

\section{Review of Functional Regression Models} \label{s_formal}
Functional regression refers to a broad category of models.  If the response is a function and the predictors are functions, scalars, or both, then we refer to these as functional response models (FRMs).  If the response is a scalar and the predictors are functions or both functions and scalars, then we refer to these as scalar response models (SRMs).  Here, we briefly outline broad classes of FRMs and SRMs and methods for fitting them.    

\subsection{Functional Response Models (FRMs)} \label{sec:frms}

Suppose we collect a sample of $n$ independent observations from a population of interest.  For each observation, we have a function designated as the response, denoted by $Y_i$, $p$ scalar variables, denoted by the $p$-dimensional vector $\bm{z}_{i} = (z_{i,1}, \ldots, z_{i,p})^{\intercal}$, and $q$ functional variables, denoted by the $q$-element set of functions $\bm{X}_{i} = \{X_{i,1}, \ldots, X_{i,q} \}$ for $i = 1, \ldots, n$.  Assume that $Y_i$ is a one-dimensional functional random variable that is square integrable on a compact support $I_Y \subset \mathbb{R}$ (i.e, $\int_{I_{Y}}{Y_{i}^2(t)}dt < \infty$).  Similarly, assume that $X_{i,1}, \ldots, X_{i,q}$ are one-dimensional functional random variables that are each square integrable on a compact support $I_{k} \subset \mathbb{R}$ (i.e, $\int_{I_{k}}{X_{i,k}^2(t)}dt < \infty, k = 1, \ldots, q$).  For clarity, we assume that the functional predictors are observed without error.  

We can model the relationship between the functional response and predictors as:  
\begin{equation} \label{fr_mod}
Y_{i}(t) \sim EF(\mu_{i}(t), \bm{\eta}), \ \  \textrm{such that} \ \ 
g\{\mu_{i}(t) \} = \beta_{0}(t) + \sum_{j = 1}^{p} z_{i,j} \beta_{j}(t) + \sum_{k =1}^q \int X_{i,k}(s)\rho_{k}(s,t)ds.
\end{equation}
In (\ref{fr_mod}), $EF(\mu_{i}(t), \bm{\eta})$ corresponds to an exponential family distribution with mean $\mu_{i}(t)$ and a set of nuisance parameters given by the vector $\bm{\eta}$, $g(\cdot)$ is a link function, $\beta_{0}(t)$ is the intercept function, $\beta_{j}(t)$ for $j = 1, \ldots, p$ are the coefficient functions corresponding to the functional effects of the scalar predictors on the functional response, and $\rho_{k}(s,t)$ for $k = 1, \ldots, q$ are the functional effects of the functional predictors on the functional response.     

With complete data, there are several methods available for estimating the coefficients in (\ref{fr_mod}).  In subsequent sections, we employ a fitting approach described in \citep{ivanescu2015} for settings where $Y_{i}(t)$ are normally distributed and more generally in \cite{scheipl2016} where $Y_{i}(t)$ can be from any exponential family distribution.  These fitting methods are referred to as penalized function-on-function regression (PFFR).     

\subsection{Scalar Response Models (SRMs)}
Suppose that we observe  $\bm{z}_{i}$ and $\bm{X}_{i}$ for $i = 1, \ldots, n$ as above, but have a scalar response, $y_i$.  We can assume that the model of interest is the generalized functional linear model \citep{MNglm, cardot_1999}:

\begin{equation} \label{e_mod}
y_{i} \sim EF(\mu_{i}, \eta), \ \  \textrm{such that} \ \
g(\mu_{i}) = \theta_{0} + \bm{z}_{i} \bm{\theta} + \sum_{j = 1}^{q} \int X_{i,j}(t)\beta_{j}(t)dt. 
\end{equation}
In (\ref{e_mod}), $EF(\mu_{i}, \eta)$ corresponds to an exponential family distribution with mean $\mu_{i}$ and dispersion parameter $\eta$, $g(\cdot)$ is a link function, $\theta_{0}$ is the intercept, $\bm{\theta}$ is a $p$-vector of scalar coefficients, and $\beta_{j}(t)$, $j = 1, \ldots, q$, are square integrable on a compact support.   
   
As with FRMs, there are several methods for estimating SRMs.  In the following sections, we consider penalized functional regression (PFR) \citep{GoldsmithPenalized} since it is able to handle both scalar and functional predictors as well as generalized scalar outcomes.  A brief overview of both the PFFR and PFR fitting procedures is provided in Section 1 of the online Appendix.  

\section{Missing Scalar and Functional Data} \label{s_framework}
Missing data mechanisms and models have been discussed extensively elsewhere (e.g., \cite{rubin_1987, little_rubin2002, vanBur_2012}).  Here we provide an overview of these concepts in settings with both scalar and functional data as well as an overview of multiple imputation.

\subsection{Notation and Missingness Mechanisms} \label{ss_notation}
We start by relabeling the scalar and functional variables.  If the response of interest is a scalar, then we let $y_{i} = w_{i,1}$, $z_{i,1} = w_{i,2}, \ldots, z_{i,p} = w_{i,p+1}$, and $X_{i,1} = w_{i, p+2}, \ldots, X_{i,q} = w_{i, p+q+1}$.  If the response of interest is a function, then we let $z_{i,1} = w_{i,1}, \ldots, z_{i,p} = w_{i,p}$, $Y_{i} = w_{i,p+1}$, and $X_{i,1} = w_{i, p+2}, \ldots, X_{i,q} = w_{i, p+q+1}$.  Either way, the $p+q+1$ variables can be gathered into an $n \times (p+q+1)$ matrix of components, which are a mix of scalars and functions, $\bm{W} = (\bm{w}_{1}, \ldots, \bm{w}_{p+q+1})$.  $\bm{w}_{i} = (w_{i,1}, \ldots, w_{i, p+q+1})^{\intercal}$ represents a random draw from a multivariate distribution having a set of unknown parameters denoted by $\bm{\xi}$.  Let $\bm{R} = (\bm{r}_{1}, \ldots, \bm{r}_{p+q+1})$ be an $n \times (p+q+1)$ indicator matrix with entries $r_{i,j}$ such that $r_{i,j} = 1$ if $w_{i,j}$ is observed and $r_{i,j} = 0$ if $w_{i,j}$ is missing.  Let $\bm{W}^{obs}$ and $\bm{W}^{mis}$ denote the observed and missing components of $\bm{W}$, respectively.  The expression for the missing data model is $P(\bm{R} | \bm{W}^{obs}, \bm{W}^{mis}, \bm{\psi})$, where $\bm{\psi}$ is the collection of model parameters.  

The classification of missing data given in \cite{little_rubin2002} can be extended to include functional data.  Data are missing completely at random (MCAR) if $P(\bm{R} = \bm{r} | \bm{W}^{obs}, \bm{W}^{mis}, \bm{\psi}) = P(\bm{R} = \bm{r} | \bm{\psi})$,  missing at random (MAR) if  $P(\bm{R} = \bm{r} | \bm{W}^{obs}, \bm{W}^{mis}, \bm{\psi}) = P(\bm{R} = \bm{r} | \bm{W}^{obs}, \bm{\psi})$, or missing not at random (MNAR) if  $P(\bm{R} = \bm{r} | \bm{W}^{obs}, \bm{W}^{mis}, \bm{\psi})$ does not simplify.  

Under MCAR and some MNAR settings \citep{bartlett_2014}, a complete case analysis (CCA) yields unbiased estimates for the model parameters.  However, these estimates may be inefficient since incomplete cases are being discarded.  When there are many predictors, each prone to missing values, the number of complete cases can be much smaller than the full set of observations.  This can greatly limit one's ability to extract information from data with complex associations.  

When data are MAR, a CCA can yield both inefficient and biased parameter estimates in some settings (e.g., if missingness in covariates depends on the value of the response \citep{white2008}).  Under MAR mechanism, missing data can be imputed using imputation models that provide predictions for the missing values.  The assumption that data are MAR is not testable with available data, but previous work suggests that the MAR assumption is approximately valid if the imputation model includes enough relevant variables \citep{schafer_1997, Collins2001, harel2007, white2009, perkins_2018}.  In the following sections, we assume that the data are MAR.  In addition, we also assume that the parameter spaces for $\bm{\psi}$ and $\bm{\xi}$ are distinct (i.e., the joint parameter space is equivalent to the product of the individual parameter spaces). The combination of MAR and parameter distinctness allows us to ignore the missingness indicators, $\bm{R}$, in likelihood or Bayes-type inferences \citep{little_rubin2002}.  

\subsection{Joint and Imputation Models} \label{ss_jandimods}
\cite{rubin_1987} gives a general framework to conduct imputation of missing data: imputation should follow from the specification of a joint model for $[\bm{W}, \bm{R}]$.  Correct specification of such a model can be a complex task in settings with purely scalar data of mixed types (e.g., continuous, categorical, etc.) and is made even more complex here with functional data.  Fortunately, we propose that values can be imputed without directly specifying this joint distribution.  Instead, one can specify an imputation model, $f(\bm{W}^{mis} | \bm{W}^{obs}, \bm{R})$, that describes how missing values are generated.  In MI, one draws from this distribution multiple times to create multiple complete data sets. 

Conceptually, MI in settings with both scalar and functional data is similar to MI of purely scalar data. The goal is to use the distribution of the observed data to fill in plausible values for the missing data.  As with purely scalar data, here MI can be justified using a Bayesian framework.  
  
Under MAR and ignorability assumptions (and also under the stricter MCAR assumption) the posterior predictive distribution for $\bm{W}^{mis}$ is independent of $\bm{R}$ and given by
\begin{equation} \label{pred_mis}
f(\bm{W}^{mis} | \bm{W}^{obs}) = \int f(\bm{W}^{mis} | \bm{W}^{obs}, \bm{\xi}) f( \bm{\xi} | \bm{W}^{obs})d \bm{\xi},
\end{equation}
where $f(\bm{W}^{mis} | \bm{W}^{obs}, \bm{\xi})$ is the predictive distribution of $\bm{W}^{mis}$ given $\bm{W}^{obs}$ and $\bm{\xi}$, 
\begin{equation} \label{post_param}
f( \bm{\xi} | \bm{W}^{obs}) \propto f(\bm{\xi}) \int f(\bm{W}^{mis}, \bm{W}^{obs}| \bm{\xi})d\bm{W}^{mis},
\end{equation}
is the observed-data posterior distribution for $\bm{\xi}$, and $f(\bm{\xi})$ is the prior distribution.  Together, (\ref{pred_mis}) and (\ref{post_param}) point to a two-step method for MI: in the $m$-th imputation ($m = 1, \ldots, M$), first make a random draw for $\bm{\xi}$ from its posterior distribution, denoted by $\hat{\bm{\xi}}^{(m)}$, then impute the missing values in $\bm{W}_{mis}$ by a random draw from $f(\bm{W}^{mis} | \bm{W}^{obs}, \hat{\bm{\xi}}^{(m)})$ to obtain $\bm{W}^{mis (m)}$.

%\cite{Meng_1994} showed the importance of making the imputation model congenial with the analysis model.  In practice, the goal is to make the imputation model at least as general as the analysis model so as to avoid missing important associations being investigated in the analysis model.     

\section{Multiple Imputation for Scalar and Functional Data} \label{s_meth}
\subsection{Imputation Procedures}
\subsubsection{Simplified Case} \label{ss_simp}  
For the sake of clarity, we begin by considering settings in which all but one variable, $\bm{w}_{\cdot j}$ (here the $\cdot j$ subscript indicates column $j$ of matrix $\bm{W}$), are completely observed.  Without loss of generality, assume that the first $r$ values in $\bm{w}_{\cdot j}$ are observed, denoted by $\bm{w}^{obs}_{\cdot j} = (w_{1,j}, \ldots, w_{r,j})^{\intercal}$, and the last $n - r$ values in $\bm{w}_{\cdot j}$ are missing, denoted by $\bm{w}^{mis}_{\cdot j} = (w_{r+1,j}, \ldots, w_{n,j})^{\intercal}$.  Define the complement data set to $\bm{w}_{\cdot j}$ by $\bm{W}_{-j} = (\bm{w}_{\cdot 1}, \ldots, \bm{w}_{\cdot j-1},\bm{w}_{\cdot j+1}, \ldots, \bm{w}_{\cdot p+q+1}) = [ \bm{W}^{obs \intercal}_{-j}, \bm{W}^{mis \intercal}_{-j}]^{\intercal}$.

The observed data are $\bm{W}^{obs} = \{\bm{w}^{obs}_{\cdot j}, \bm{W}^{obs}_{-j}, \bm{W}^{mis}_{-j}\}$ and the missing data are $\bm{W}^{mis} = \bm{w}^{mis}_{\cdot j}$ so that there are $r$ complete observations and $n - r$ incomplete observations that are missing values for $\bm{w}_{\cdot j}$.  In this setting, the imputation model in (\ref{pred_mis}) can be expressed as $f(\bm{w}^{mis}_{\cdot j} | \bm{w}^{obs}_{\cdot j}, \bm{W}_{-j}) = \int f(\bm{w}^{mis}_{\cdot j} | \bm{W}^{mis}_{-j}, \bm{\xi}_{j})f(\bm{\xi}_{j} | \bm{w}^{obs}_{\cdot j}, \bm{W}^{obs}_{-j})d\bm{\xi}_{j}$. In order to obtain the posterior distribution $f(\bm{\xi}_{j} | \bm{w}^{obs}_{\cdot j}, \bm{W}^{obs}_{-j})$, we can specify and fit a regression model with $\bm{w}^{obs}_{\cdot j}$ as the response and $\bm{W}^{obs}_{-j}$ as the predictors.  

When $\bm{w}_{\cdot j}$ is one of the scalar variables (e.g., $w_{i,j} =  z_{i,j}$ or $y_i$ when the analysis model is a SRM), we can employ a suitable SRM.  For example, we use the model $z_{i,j} \sim EF(\mu_{i,j}, \eta_{j})$ such that,
\begin{align} \label{sc_mod} 
h_{j}(\mu_{i,j}) & = \gamma_{j,0} + \sum_{\ell \neq j} z_{i,\ell} \gamma_{j,\ell} + \alpha_{j} y_{i} + \sum_{k = 1}^{q} \int X_{i,k}(t) \omega_{j,k}(t)dt \ \ \textrm{or} \\ 
h_{j}(\mu_{i,j}) & = \gamma_{j,0} + \sum_{\ell \neq j} z_{i,\ell} \gamma_{j,\ell} + \int \alpha_{j}(t) Y_{i}(t) dt + \sum_{k = 1}^{q} \int X_{i,k}(t) \omega_{j,k}(t)dt, \nonumber
\end{align}
depending on whether the response for the analysis model is a scalar (top) or a function (bottom).  The components of (\ref{sc_mod}) are similar to those in analysis model (\ref{e_mod}) where the collection of parameters is $\bm{\xi}_{j} = \{\gamma_{j,0}, \gamma_{j,\ell (\ell \ne j)}, \alpha_{j}, \omega_{j,1}, \ldots, \omega_{j,q}\}$, which can be estimated using PFR or any other suitable approach.  When the error function corresponds to either the binomial or Poisson distributions, the $i$th subject's missing value for the $j$th variable is imputed with a random draw from either distribution, using the predicted value of $\mu_{i,j}$.  When the error function is normal, the missing value is imputed with the predicted value of $\mu_{i,j}$ with a random error term added.  This error term is drawn from the 
$N(0, \hat{\sigma}_{j}^2)$ distribution, where $\hat{\sigma}_{j}^2$ is estimated from the residuals under model (\ref{sc_mod}).

When $\bm{w}_{\cdot j}$ is a functional variable (e.g., $w_{i,j} = X_{i,j}$ or $Y_i$ when the analysis model is a FRM), we can employ a suitable FRM.  For example, we use the model $X_{i,j}(t) \sim EF(\mu_{i,j}(t), \bm{\eta}_{j})$ such that,
\begin{align} \label{f_mod} 
h_{j}\{\mu_{i,j}(t) \} & = \gamma_{j,0}(t) + \sum_{\ell = 1}^{p} z_{i,\ell} \gamma_{j,\ell}(t) + y_{i}\alpha_{j}(t)  + \sum_{k \neq j} \int X_{i,k}(s)\omega_{j,k}(s,t)ds \ \ \textrm{or}\\ 
h_{j}\{\mu_{i,j}(t) \} & = \gamma_{j,0}(t) + \sum_{\ell = 1}^{p} z_{i,\ell} \gamma_{j,\ell}(t) + \int  Y_{i}(s)\alpha_{j}(s, t)ds  + \sum_{k \neq j} \int X_{i,k}(s)\omega_{j,k}(s,t)ds, \nonumber
\end{align}
depending on whether the response for the analysis model is a scalar (top) or a function (bottom).  The components of (\ref{f_mod}) are similar to those in analysis model (\ref{fr_mod}) and the collection of parameters $\bm{\xi}_{j} = \{\gamma_{j,0}, \gamma_{j,1}, \ldots, \gamma_{j,p}, \alpha_{j}, \omega_{j,k (k \neq j)} \}$ can be estimated using PFFR or any other suitable approach.  When the error function corresponds to either the binomial or Poisson distribution, the $i$th subject's missing value for the $j$th variable at domain value $t$ can be filled in with a random draw from either distribution, using the predicted value of $\mu_{i,j}(t)$.  When the error function is normal then the $i$th subject's missing value for the $j$th covariate at domain value $t$ can be filled in with the predicted value of $\mu_{i,j}(t)$ with a functional error term added to it.  This functional error term is generated as follows. First, we compute estimates of the leading principal component basis functions $\{\hat{\psi}_{1},\ldots,\hat{\psi}_{K}\}$ (accounting for at least 99\% of the variance), the corresponding score variances, $\hat{\bm{\lambda}} = (\hat{\lambda}_{1},\ldots, \hat{\lambda}_{K})^{\intercal}$, and mean function, $\hat{\mu}_{r}$, from a functional principal components decomposition of the collection of residual curves derived from fitting (\ref{f_mod}) on the observations for which the $j$th covariate is observed.  Then we generate subject-specific principal component loadings, $\bm{c}_{i} = (c_{1,i}, \ldots, c_{K, i})^{\intercal}$, from $\bm{c}_{i} \sim N(\bm{0}, \text{diag}(\hat{\bm{\lambda}}))$, and let $r_{i}(t) = \hat{\mu}_{r}(t) + \sum_{k = 1}^{K} c_{k i} \hat{\psi}_{k}(t)$ be the functional error term for the $i$th subject. 

In order to account for uncertainty in the imputation model parameters, we propose to select a bootstrap sample from the complete data and obtain parameter estimates from fitting (\ref{sc_mod}) or (\ref{f_mod}) on the bootstrap sample.  The use of a bootstrap sample is suggested in \cite{vanBur_2012} Section 3.1. The complete procedure is repeated $M$ times to obtain $M$ imputed data sets.  Both (\ref{sc_mod}) and (\ref{f_mod}) can be made more flexible by the addition of various interaction terms or by allowing for less restrictive functional forms for the coefficient functions.  These modifications may increase the computational complexity of the imputation procedure.    

\subsubsection{General Missing Patterns and the fregMICE Algorithm}
When more than one variable are incomplete, we propose to employ an extension of multiple imputation by chained equations (MICE) \citep{vanbuuren_1999} that incorporates functional variables.  MICE is conducted in a variable-by-variable manner via specification of a conditional model for each $\bm{w}_{\cdot j}$ ($j = 1, \ldots, p+q+1$) given by $f(\bm{w}_{\cdot j} | \bm{W}_{-j}, \bm{R}, \bm{\xi}_{j})$.  The proposed functional regression MICE (fregMICE) algorithm is provided in Algorithm \ref{alg_1}.  We assume that the variables are arranged such that those with the least missing data have lower index values ($j$) and those with more missing data have higher index values.  As with the original MICE procedure, any pattern of missingness in the variables can be accommodated.  The fregMICE Algorithm can be run in parallel with $M$ streams yielding $M$ imputed data sets after $V$ iterations. 
      
\begin{algorithm} [ht]
\footnotesize
\caption{fregMICE Procedure for Imputation of Scalar and Functional Variables} \label{alg_1}
\begin{algorithmic}[1]
\State Initialize the imputation procedure by filling in $\bm{w}_{\cdot j}^{mis}$ by a random draw from $\bm{w}_{\cdot j}^{obs}$ for each $j$. Denote each initially complete $\bm{w}_{\cdot j}$ by $\bm{w}_{\cdot j}^{[0]}$.
\For{$v$ in $1, \ldots, V$}
\For{$j$ in $1, \ldots, p + q + 1$}
\If {$\bm{w}_{\cdot j}$ has missing values}
\State \multiline{%
           Set $\bm{D}_{-j}^{[v]} = (\bm{w}_{\cdot 1}^{[v]}, \bm{w}_{\cdot 2}^{[v]}, \ldots, \bm{w}_{\cdot j-1}^{[v]}, \bm{w}_{\cdot j+1}^{[v-1]}, \ldots, \bm{w}_{\cdot p+q+1}^{[v-1]})$. Let $\bm{D}_{-j}^{obs [v]}$ be the components in $\bm{D}_{-j}^{[v]}$
            for which $\bm{w}_{\cdot j}$ are observed (having $n_{j,obs}$ observations) and $\bm{D}_{-j}^{mis [v]}$ be the components for which $\bm{w}_{\cdot j}$ are missing (having $n - n_{j,obs}$ observations).}
\State \multiline{%
           Draw a bootstrap sample of size $n_{j,obs}$, with replacement, from $\{\bm{D}_{-j}^{obs [v]}, \bm{w}_{\cdot j}^{obs}\}$ to obtain 
           $\{\bm{D}_{-j}^{b, obs [v]}, \bm{w}_{\cdot j}^{b, obs}\}$.} 
\State \multiline{% 
           Using $\bm{w}_{\cdot j}^{b,obs}$ as the response and $\bm{D}_{-j}^{b, obs [v]}$ as the predictors fit (\ref{sc_mod}) if $\bm{w}_{\cdot j}$ is a scalar or (\ref{f_mod})
            if $\bm{w}_{\cdot j}$ is a function to obtain parameter estimates denoted by $\hat{\bm{\xi}}_{j}^{[v]}$.}
\State \multiline{% 
            Predict $\bm{w}_{\cdot j}^{mis}$ by randomly drawing from the predictive distribution $f(\bm{w}_{\cdot j}^{mis} | \bm{D}_{-j}^{mis [v]}, \hat{\bm{\xi}}_{j}^{[v]})$ using methods described in Section 
            \ref{ss_simp}.  Fill the missing values of $\bm{w}_{\cdot j}$ with the predicted values. Set this completed vector to $\bm{w}_{\cdot j}^{[v]}$.} 
\Else \ (when $\bm{w}_{\cdot j}$ is completely observed)
\State Set $\bm{w}_{\cdot j}^{[v]} = \bm{w}_{\cdot j}$.
\EndIf
\EndFor
\EndFor
\State When $v = V$, convergence is assumed. (Convergence is discussed in Section \ref{sub_conv})
\State Run this algorithm $M$ times to obtain $M$ complete data sets.
\end{algorithmic}
\end{algorithm}

\subsubsection{Convergence and Diagnostics for the fregMICE Procedure} \label{sub_conv}
In settings with only scalar data, it is common to assess convergence of the MICE procedure via inspection of plots of selected parameters that summarize the imputed data (e.g., mean or standard deviation of the imputed values) vs. iteration number for each of the $M$ parallel sequences.  When the specified values for the parallel sequences are plotted together, the streams should overlap and be free of trend in order to diagnose convergence \citep{vanbuuren_1999}. If a stream trails off, away from the other streams, or shows systematically different variation from the other streams, then this would suggest convergence issues for that stream.  We propose to use the same assessment techniques for imputations of scalar values generated from our fregMICE procedure and similar techniques for the imputations of functional values.  Specifically, for the imputed function values, we propose to plot point-wise summary parameters (e.g., mean or standard deviation) for each iteration of each parallel sequence.  We will diagnose convergence if the plots are free of trend and show adequate overlap across the streams.  As with scalar imputation, if a sequence of curves in a stream trails off (either the entire function or over a restricted domain) relative to the other streams or the sequence of curves shows systematically different variation from the other streams, then this would suggest convergence issues for that stream.      

Aside from assessing convergence, one may also want to assess the fidelity of the imputed values to those observed in the data set.  One way to do this is to create strip-plots.  For imputed scalar data, these plots show the observed and imputed values from each imputed data set in contrasting colors.  This allows one to easily identify whether imputed values are realistic and can help the analyst decide if the imputation model needs to be adjusted.  Strip plots can also be constructed for imputed functional data and can be used to make similar assessments.  We illustrate these strip plots and convergence plots in our application in Section \ref{s_app} and in the supplementary Appendix.    

\subsection{Analysis of Multiply Imputed Datasets} \label{sec:MIanalysis}

The generation of multiple data sets accounts for the inherent uncertainty in the prediction of the missing values.  Once the imputed data sets are constructed, we analyze each one using a method designed for complete data.  Most methods for estimating functional regression models, including the PFFR and PFR approaches that we employ in subsequent sections, represent the coefficient functions using judiciously selected sets of basis functions and then estimate the corresponding basis coefficients.  Specifically, for model (\ref{fr_mod}), we let      
$\beta_j(t) = \bm{B}_j^{\intercal}(t) \bm{b}_j$ for $j = 0, \ldots, p$, where $\bm{B}_j(t) = (B_{j,1}(t), \ldots, B_{j,L_j}(t))^{\intercal}$ is a vector of basis functions for $\beta_j(t)$ and $\bm{b}_j = (b_{j,1}, \ldots, b_{j,L_j})^{\intercal}$ is the corresponding vector of unknown basis coefficients. Similarly, we let $\rho_{k}(s,t) = \bm{U}_{k}^{\intercal}(s,t)\bm{u}_{k}$, for $k = 1, \ldots, q$, where $\bm{U}_{k}(s,t) = (U_{k,1}(s,t), \ldots, U_{k,L_k}(s,t))^{\intercal}$ is a vector of bivariate basis functions (e.g., bivariate thin-plate splines) selected for $\rho_k(s,t)$ and $\bm{u}_k = (u_{k,1}, \ldots, u_{k,L_k})^{\intercal}$ is the corresponding vector of unknown basis coefficients.  The coefficient functions in (\ref{e_mod}) can be represented similarly.

The fitted models from the $M$ imputed data sets can be pooled to provide coefficient and variance estimates that account for both within and between imputation variability.  We will use the variance estimates to construct approximate confidence intervals for the scalar coefficients and point-wise confidence bands for the coefficient functions. 

Rubin's Rules \citep{rubin_1987} provide a method for combining scalar and multivariate estimates after MI and can be extended to settings involving functional data.  For clarity, we illustrate the pooling approach for estimating the univariate functional coefficient parameters in model (\ref{fr_mod}).  

Let $\hat{\beta}_j^{(m)}(t) = \bm{B}_j^{\intercal}(t)\hat{\bm{b}}_j^{(m)}$ be the estimate of $\beta_j(t)$ from the $m^{th}$ imputed data set.  The pooled point estimate for $\beta_j(t)$ is given by $\bar{\beta}_{j,M}(t)  = \frac{1}{M} \sum^{M}_{m = 1} \bm{B}_j^{\intercal}(t) \hat{\bm{b}}_j^{(m)} = \bm{B}_j^{\intercal}(t) \left \{ \frac{1}{M} \sum^{M}_{m = 1} \hat{\bm{b}}_j^{(m)} \right \} = \bm{B}_j^{\intercal}(t) \bar{\bm{b}}_{j,M}$. The total variance of $\bar{\beta}_{j,M}(t)$ should incorporate both within and between-imputation variability.  Let $\hat{V}_{j}^{(m)}(t) = \widehat{var}\{ \hat{\beta}_j^{(m)}(t) \} = \widehat{var}\{ \bm{B}_j^{\intercal}(t) \hat{\bm{b}}_j^{(m)} \} = \bm{B}_j^{\intercal}(t)  \widehat{var}(\hat{\bm{b}}_j^{(m)})    \bm{B}_j(t) = \bm{B}_j^{\intercal}(t) \hat{\Lambda}^{(m)}_{\hat{\bm{b}}_j} \bm{B}_j(t)$, where $\hat{\Lambda}^{(m)}_{\hat{\bm{b}}_j}$ is the estimated covariance matrix of $\hat{\bm{b}}_j^{(m)}$.  With $\hat{V}_{j}^{(m)}(t)$ defined, we have the mean within-imputation variance given by $\bar{V}_{j,M}(t)  = \frac{1}{M} \sum^{M}_{m = 1} \hat{V}_{j}^{(m)}(t) = \frac{1}{M} \sum^{M}_{m = 1} \bm{B}_j^{\intercal}(t) \hat{\Lambda}^{(m)}_{\hat{\bm{b}}_j} \bm{B}_j(t) = \bm{B}_j^{\intercal}(t) \left \{  \frac{1}{M} \sum^{M}_{m = 1}  \hat{\Lambda}^{(m)}_{\hat{\bm{b}}} \right \}  \bm{B}_j(t) = \bm{B}_j^{\intercal}(t) \bar{\Lambda}_{\hat{\bm{b}}_j, M} \bm{B}_j(t)$, and between-imputation variability given by $B_{j,M}(t)  = \frac{1}{M - 1} \sum^{M}_{m = 1} \left \{ \hat{\beta}_{j}^{(m)}(t) - \bar{\beta}_{j,M}(t) \right \}^2
        = \frac{1}{M - 1} \sum^{M}_{m = 1} \left \{  \bm{B}_j^{\intercal}(t) \left ( \hat{\bm{b}}_j^{(m)} - \bar{\bm{b}}_{j,M}  \right ) \right \}^2 
         \\  = \bm{B}_j^{\intercal}(t) \left \{ \frac{1}{M - 1} \sum^{M}_{m = 1} \left ( \hat{\bm{b}}_j^{(m)} - \bar{\bm{b}}_{j,M}  \right )^{\intercal} \left ( \hat{\bm{b}}_j^{(m)} - \bar{\bm{b}}_{j,M}  \right ) \right \} \bm{B}_j(t)
           = \bm{B}_j^{\intercal}(t)  \bar{\bm{\Omega}}_{j,M}   \bm{B}_j(t),$ where $\bar{\bm{\Omega}}_{j,M}$ is the covariance matrix quantifying the variability in the estimated basis coefficients between imputations. The total variance of $\bar{\beta}_{j,M}(t)$ is then given by $\bar{V}_{j,M}(t) + (1 + \frac{1}{M}) B_{j,M}(t)$, comprising contributions from variability within and between the imputed data sets.  We propose to construct an approximate 95\% confidence interval for $\beta_{j}(t_0)$ using $\bar{\beta}_{j,M}(t_0) \pm 1.96\sqrt{\bar{V}_{j,M}(t_0) + (1 + \frac{1}{M}) B_{j,M}(t_0)}$. 

It is straightforward to construct pooled estimators and corresponding variances for the bivariate coefficient functions, $\rho_k^{(m)}(s,t)$ for $k = 1, \ldots, q$, in model (\ref{fr_mod}).  One can also use these procedures to obtain estimators and corresponding variances for coefficient functions in model (\ref{e_mod}) and use the standard methods proposed in \cite{rubin_1987} for the scalar coefficients.

\section{Simulation Study} \label{s_sim}
Here we investigate the performance of fregMICE algorithm and evaluate the characteristics of the pooled estimators and approximate confidence intervals proposed in the previous section.  The simulation settings were designed to be similar to those encountered in the EMBARC data.  As there are no other methods in the literature to deal with these settings, we compare our fregMICE algorithm to mean imputation and CCA.  
    
\subsection{Data Generation}
\begin{figure}[t]
\centering
  \includegraphics[width=0.75\linewidth]{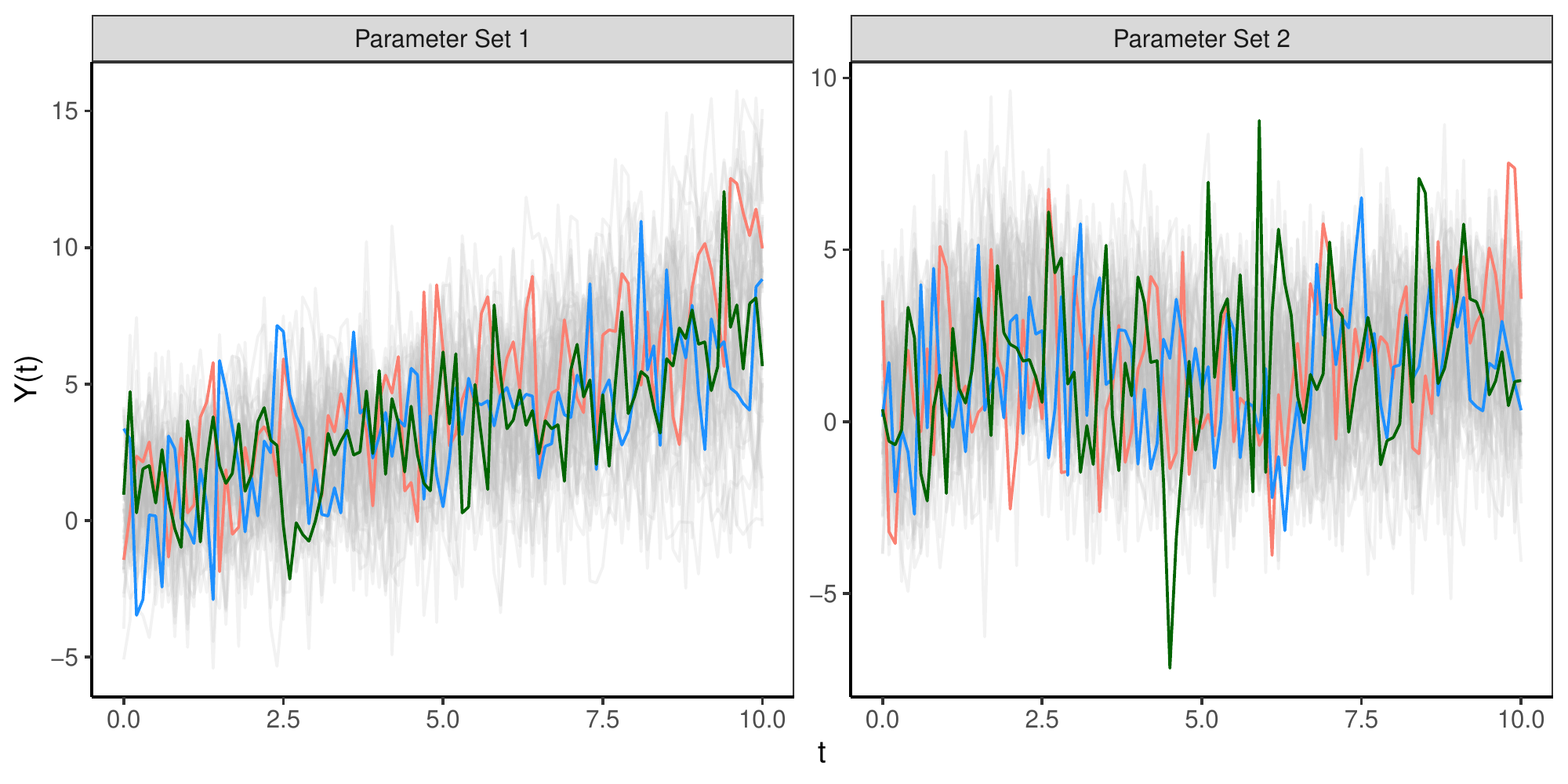}
  \caption{50 simulated responses from Parameter Set 1 with three highlighted observations (\textbf{Left}).  50 simulated responses from Parameter Set 2 with three highlighted observations (\textbf{Right}).}
  \label{fig:sim_Y_fig}
\end{figure}

Our simulation study focuses on settings with a functional response, $Y$, and its association with three scalar predictors, $z_1, z_2,$ and $z_3$.  We generated $z_{1i} \sim Bin(1, 0.4)$ and $(z_{2i}, z_{3i}) \sim N((2, 0), \bigl( \begin{smallmatrix}1 & 0.6\\ 0.6 & 1\end{smallmatrix} \bigr))$ for $i = 1, \ldots, 350$.  We selected $n = 350$ since this is close to the number of subjects in our application.  (Results for $n = 100$ are provided in Appendix Section 2.)  The functional outcome is observed on a grid $\{t_g = \frac{g}{10}: g = 0, 1, \ldots, 100\}$ on the interval $[0, 10]$ and related to the scalar predictors via the equation, 
\begin{equation} \label{sim_mod}
Y_i(t) = \beta_{0}(t) + \beta_{1}(t) z_{1i} + \beta_{2}(t) z_{2i} + \beta_{3}(t) z_{3i} + \varepsilon_i(t),
\end{equation}
where we consider two sets of true coefficient functions.  For the first set, which we refer to as ``Parameter Set 1,'' we have $\beta_{0}(t) = 0.25t$, $\beta_{1}(t) = sin(\frac{\pi t}{10})$, $\beta_{2}(t) = 0.3e^{t/5} $, and $\beta_{3}(t) =  -0.2sin(\frac{\pi t}{10})$.  For the second set, which we refer to as ``Parameter Set 2,'' we have $\beta_{0}(t) = 0.25t$, $\beta_{1}(t) = sin(\frac{\pi t}{5})$, $\beta_{2}(t) = \frac{2}{\sqrt{2 \pi}}e^{-(t - 2)^2 / 2} $, $\beta_{3}(t) = \frac{-1}{\sqrt{2 \pi}}\left \{e^{-(t - 2)^2 / 2} + e^{-(t - 8)^2 / 2}\right\}$.  $\beta_{1}, \beta_{2},$ and $\beta_{3}$ in Parameter Set 2 have more localized features relative to Parameter Set 1 and are therefore more challenging to estimate using the PFFR approach.  In each setting, $\varepsilon_i(t)$ is simulated from a Gaussian process with mean zero and covariance $V(s,t) = 4 \cdot 0.15^{|s-t|} + 0.05^2 \cdot I(s = t)$ where $I(s = t)$ is 1 if $s = t$ and 0 otherwise.  Figure \ref{fig:sim_Y_fig} shows simulated responses generated under Parameter Sets 1 and 2. 

For each set of parameters, we consider two scenarios.  In Scenario (a), only $z_2$ has missing values and in Scenario (b), both $z_2$ and $Y$ have missing values.  In both Scenarios (a) and (b), $z_1$ and $z_3$ are fully observed and the probability that $z_2$ is observed is given by $logit\{P(R_{z_{2i}} = 1)\} = \alpha_0 - 20I\{s_i > \alpha_1 \} + \alpha_2 \frac{1}{1+ e^{-z_{1i}}} + \alpha_3 \frac{1}{1+ e^{-z_{3i}}}$ for observation $i$ where $s_i = \sum_{g=0}^{100}Y_i(t_g)$.  In Scenario (a), where $Y$ is always observed, we set $\alpha_0 = 10, \alpha_2 = \alpha_3 = 0$, and let $\alpha_1$ be the $90th, 80th,$ or $70th$ percentile of $\{s_1, \ldots, s_{350}\}$ so as to achieve exactly 10\%, 20\%, or 30\% missingness in $z_2$ respectively.  In Scenario (b), whether $z_2$ is observed depends only upon the values of $z_1$ and $z_3$ such that $\alpha_1 = max\{s_1, \ldots, s_{350}\}$ (so that missingness is independent of $Y$), $\alpha_2 = 1$, $\alpha_3 = -1$ and $\alpha_0 = 2.1, 1.3,$ or 0.8 to achieve about 10\%, 20\%, or 30\% missingness in $z_2$ respectively.  Also for Scenario (b), the probability that $Y$ is observed is given by $logit\{P(R_{Y_{i}} = 1)\} = \psi_0 + \psi_1 \frac{1}{1+ e^{-z_{1i}}} + \psi_2 \frac{1}{1+ e^{-z_{3i}}}$ where $\psi_1 = -1$, $\psi_2 = 1$, and $\psi_0 = 2.3, 1.5,$ or 0.9 to achieve about 10\%, 20\%, or 30\% missingness in $Y$ respectively.  The mean (sd) proportions of missing values and complete cases realized in the simulation study are provided in Section 2 of the Appendix.  We note that the proportion of complete cases in the EMBARC application is about 61\%, while the proportions of complete cases in our simulation studies range from 49\% to 90\%.  We generated 500 data sets under each parameter set, scenario, and missingness combination.     

In Scenario (a), $z_2$ values are MAR with missingness depending on the response, $Y$, in such a way that observations with missing data have functional responses that tend to have larger values across the domain [0,10].  In this scenario, it is expected that CCA will yield biased estimates.  In Scenario (b), $z_2$ values are MAR with missingness depending only on the other covariates, but not the response.  The response, $Y$, is also MAR.  In this scenario, CCA is not expected to be biased.  

\subsection{Procedures Compared and Performance Measures} \label{sim_procs}
For each simulation setting, we fit the correctly specified model for $Y$ using PFFR where each coefficient function was represented via 20 cubic B-splines and smoothness in the estimated coefficient functions was achieved via a penalty on the magnitude of the second derivative.  Smoothing parameters were estimated via restricted maximum likelihood estimation.

As a benchmark, we used all of the data, prior to imposing missingess on any of the variables to fit model (\ref{sim_mod}).  In the results, we refer to this as ``all no missing'' (ANM).  For mean imputation, we filled in missing $z_2$ values with the mean of the observed $z_2$ values and for missing $Y$ functions, we filled in the point-wise mean function of the observed $Y$ functions.  We then used the mean-imputed dataset to fit model (\ref{sim_mod}).  For CCA, the analysis model was fit on observations with complete data.  For our fregMICE procedure, the imputation model for $z_2$ was $E(z_2 | z_1, z_3, Y) = \gamma_0 + \gamma_1 z_1 + \gamma_2 z_3 + \int Y(t)\omega(t)dt$.  To estimate this model, we used PFR.  In the PFR fitting procedure, functional observations were represented using functional principal components (FPCs) by smoothed covariance \citep{yao_2003} where the number of FPCs was selected to be the minimum number of components explaining at least 99\% of the variance in the functional observations.  The coefficient function, $\omega$, was represented using a basis of 30 thin-plate regression splines and the fitting procedure penalized the magnitude of the second derivative.  In scenarios where $Y$ had missing values, the imputation model for $Y$ was the same as (\ref{sim_mod}).  To estimate this model, we used the correctly specified analysis model fit via PFFR, using 20 cubic B-splines to represent each coefficient function and penalized the magnitude of the second derivative.  We ran the fregMICE procedure for 20 iterations and constructed 5 imputed data sets. We fit model (\ref{sim_mod}) on each imputed data set and used the extension of Rubin's rules described in Section \ref{sec:MIanalysis} to pool estimates from the 5 data sets.  Regardless of the method used to handle missing data, we used the \verb+pffr+ function from the \verb+refund+ package \citep{refund} to fit the analysis model.  This function provides different estimates for the covariance matrix that are useful for constructing confidence intervals.  In our simulations and application, we employ the Bayesian posterior covariance matrix \citep{ruppert2003} which was also used in \cite{ivanescu2015}.   

For each coefficient function ($\beta_j(t)$ for $j = 0, 1, 2, $ and 3), we show point-wise standardized bias (pwSB) plots.  The pwSB was calculated as $SB(t_g) = \frac{\bar{\hat{\beta}}_{j}(t_g) - \beta_{j}(t_g)}{sd\{\hat{\beta}_{j}(t_g)\}}$, where $\bar{\hat{\beta}}_{j}(t_g)$ is the average of the 500 estimates of $\beta_{j}(t_g)$ and $sd\{\hat{\beta}_{j}(t_g)\}$ is the Monte-Carlo standard deviation of the estimates.  We also provide plots of across-the-function mean point-wise coverage (pwCov) and mean point-wise width (pwWidth) for the estimated 95\% confidence bands for each coefficient function by taking the mean coverage and width, respectively, over all $t_g$ values and then averaging over the 500 simulation runs.  Additional results are presented in Appendix Section 2. 

\subsection{Results}
\begin{figure}[h]
\centering
  \includegraphics[width=0.49\linewidth]{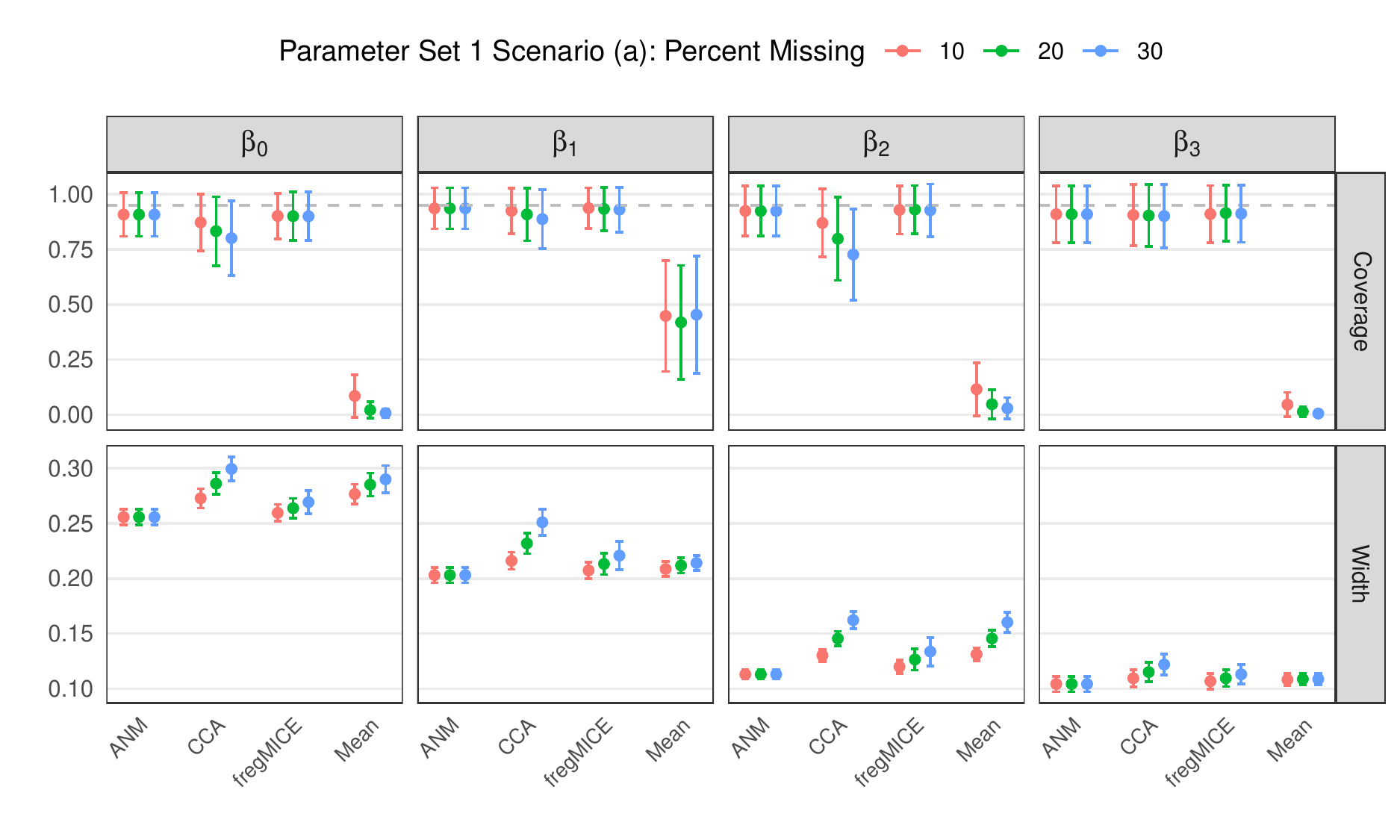}
  \includegraphics[width=0.49\linewidth]{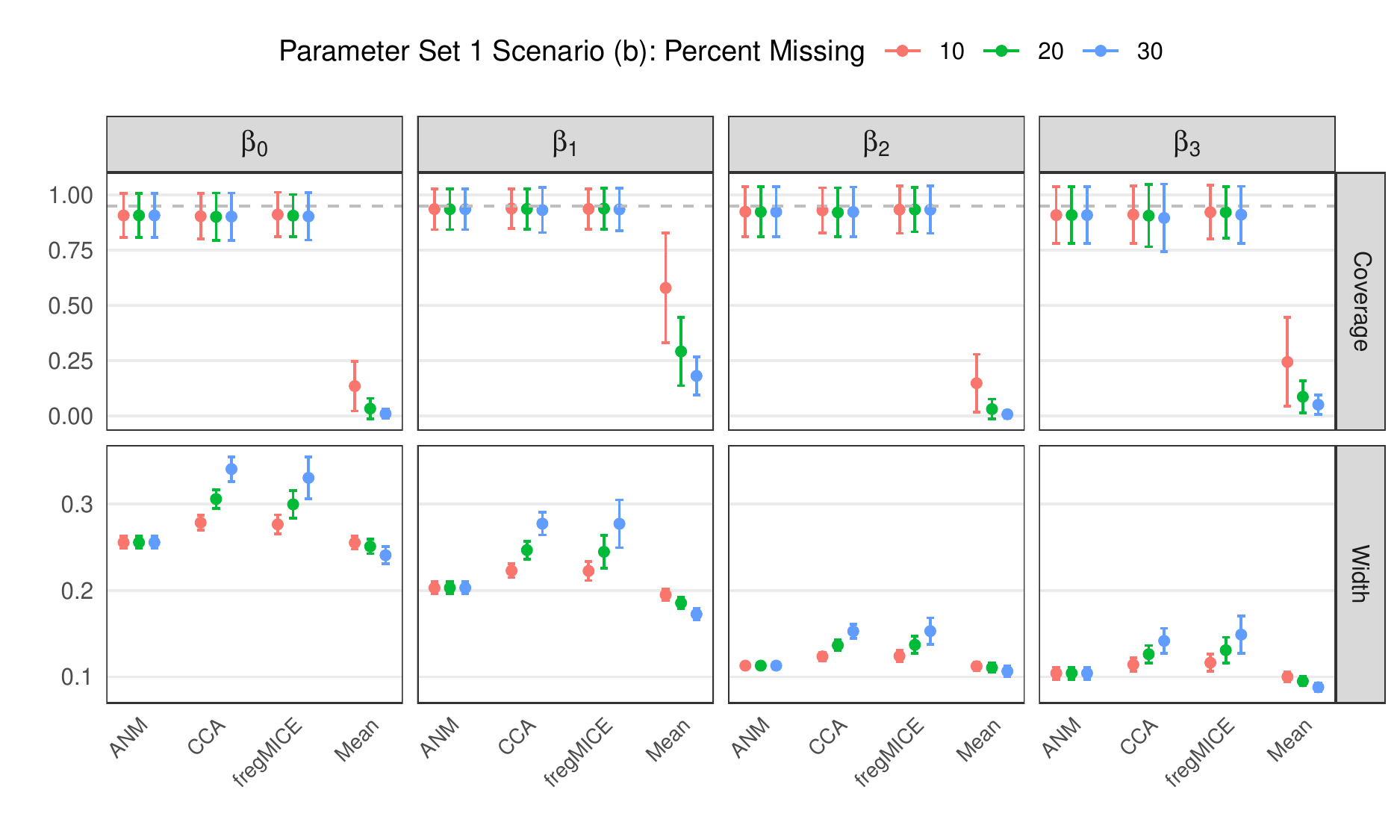}
  \includegraphics[width=0.49\linewidth]{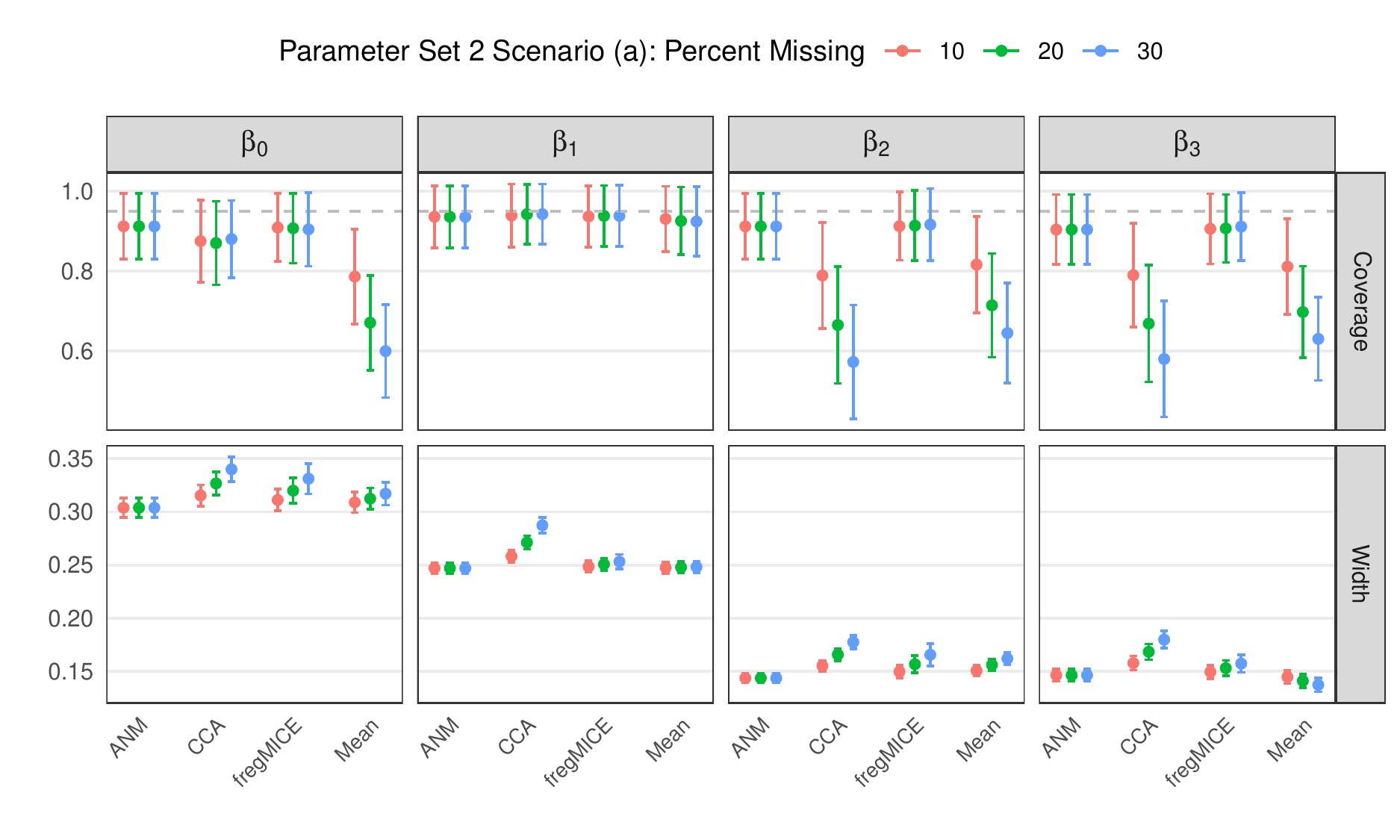}
  \includegraphics[width=0.49\linewidth]{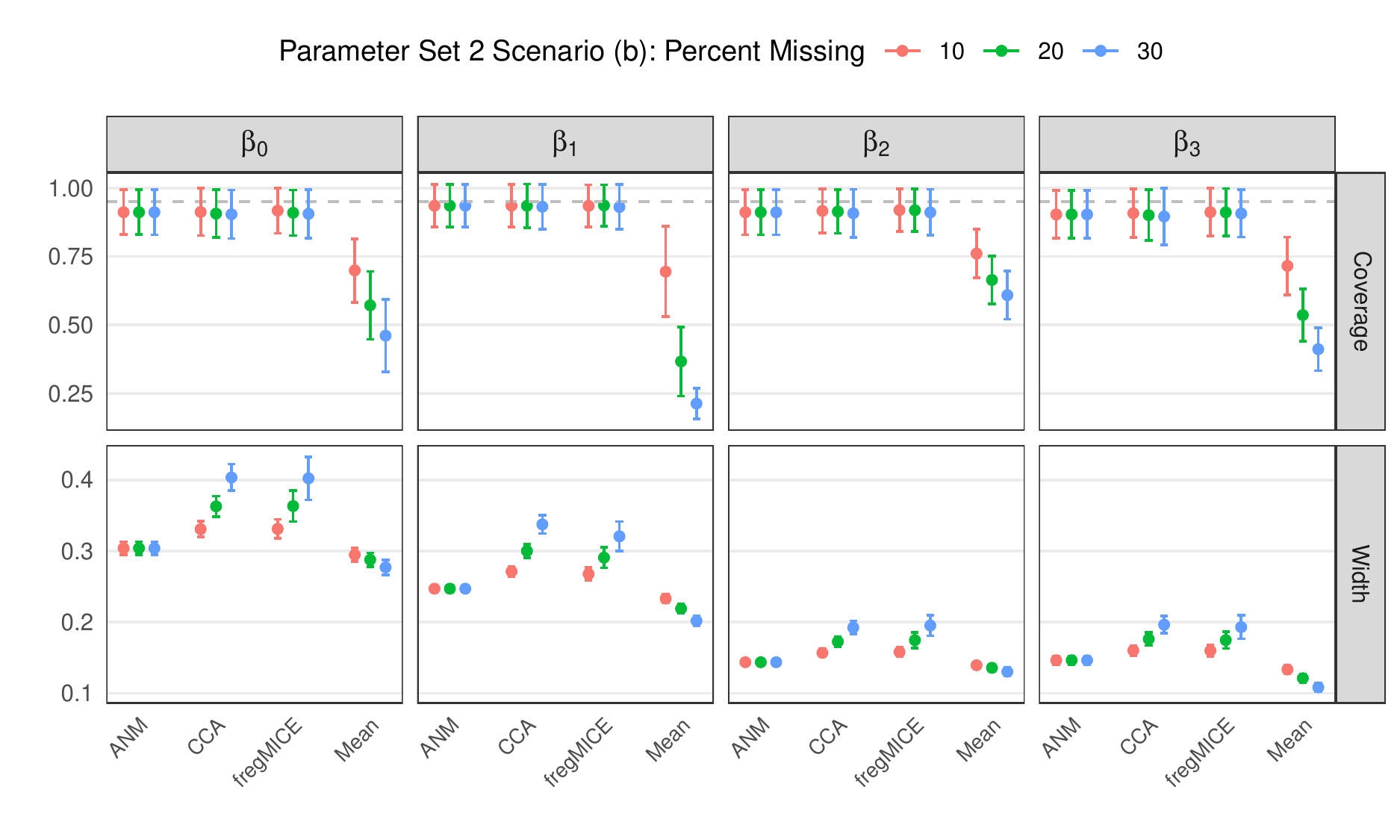}
  \caption{Across-the-function mean point-wise 95\% confidence interval coverage and width for all combinations of Parameter Settings 1 \& 2, Scenarios (a) \& (b), and missingness. Error bars are $\pm$ Monte Carlo standard deviation.}
  \label{fig:sim_covwidthplots_all}
\end{figure}

\textbf{Parameter Set 1 Results}: Figure \ref{fig:sim_pwm_1ab350} (Top) shows that, for Scenario (a), pwSB for fregMICE and ANM estimates are similar while estimates based on CCA and mean imputation show considerable bias for each coefficient function.  The degree of bias is considerably greater for mean imputation. Bias for both mean imputation and CCA increases as the amount of missing data increases, whereas bias is relatively stable for fregMICE.  Mean pwCov and pwWidth, for Scenario (a), are shown in the top left of Figure \ref{fig:sim_covwidthplots_all}.  Mean pwCov for fregMICE and ANM are similar, though the intervals tend to be slightly wider for fregMICE, especially as the amount of missing data increases.  Coverage decreases substantially and width increases slightly for intervals for $\beta_2$ from CCA and mean imputation as the amount of missing data increases.  Coverage is poor for intervals derived via the mean imputation procedure.             

Figure \ref{fig:sim_pwm_1ab350} (Bottom) shows that, for Scenario (b), pwSB for fregMICE and ANM estimates are similar.  CCA estimates also perform similarly to ANM.  This is expected since missingness in both $Y$ and $z_2$ depends on the completely observed covariates, $z_1$ and $z_3$.  pwSB is large for estimates based on mean imputation.  Mean pwCov and pwWidth, for Scenario (b), are shown in the top right of Figure \ref{fig:sim_covwidthplots_all}.  ANM, fregMICE, and CCA are similar with respect to pwCov but both fregMICE and CCA have greater pwWidths that increase with larger amoutns of missing data.  Again, coverage is poor for intervals based on mean imputed data.

\begin{figure}
\centering
  \includegraphics[width=0.70\linewidth]{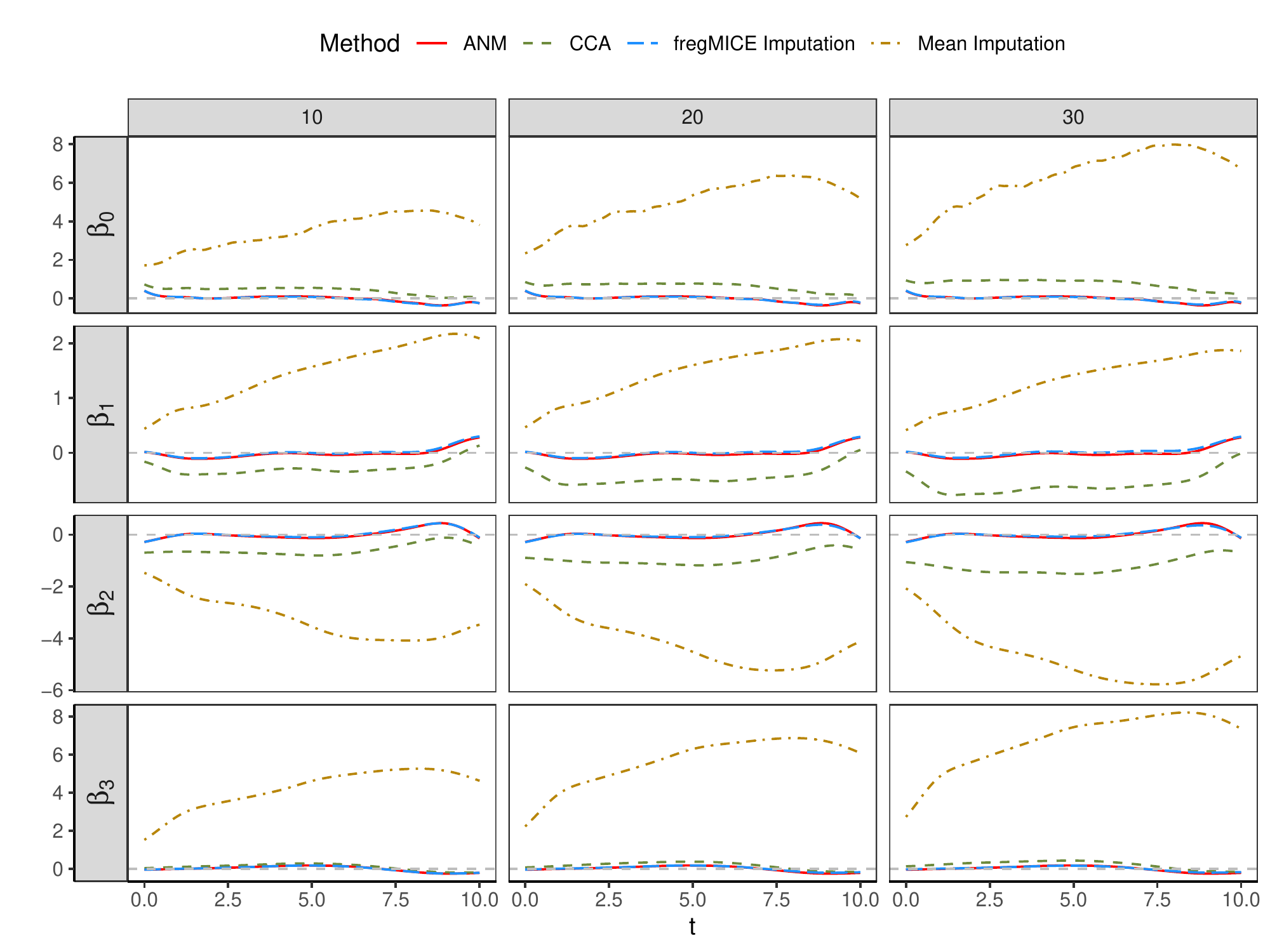}
  \includegraphics[width=0.70\linewidth]{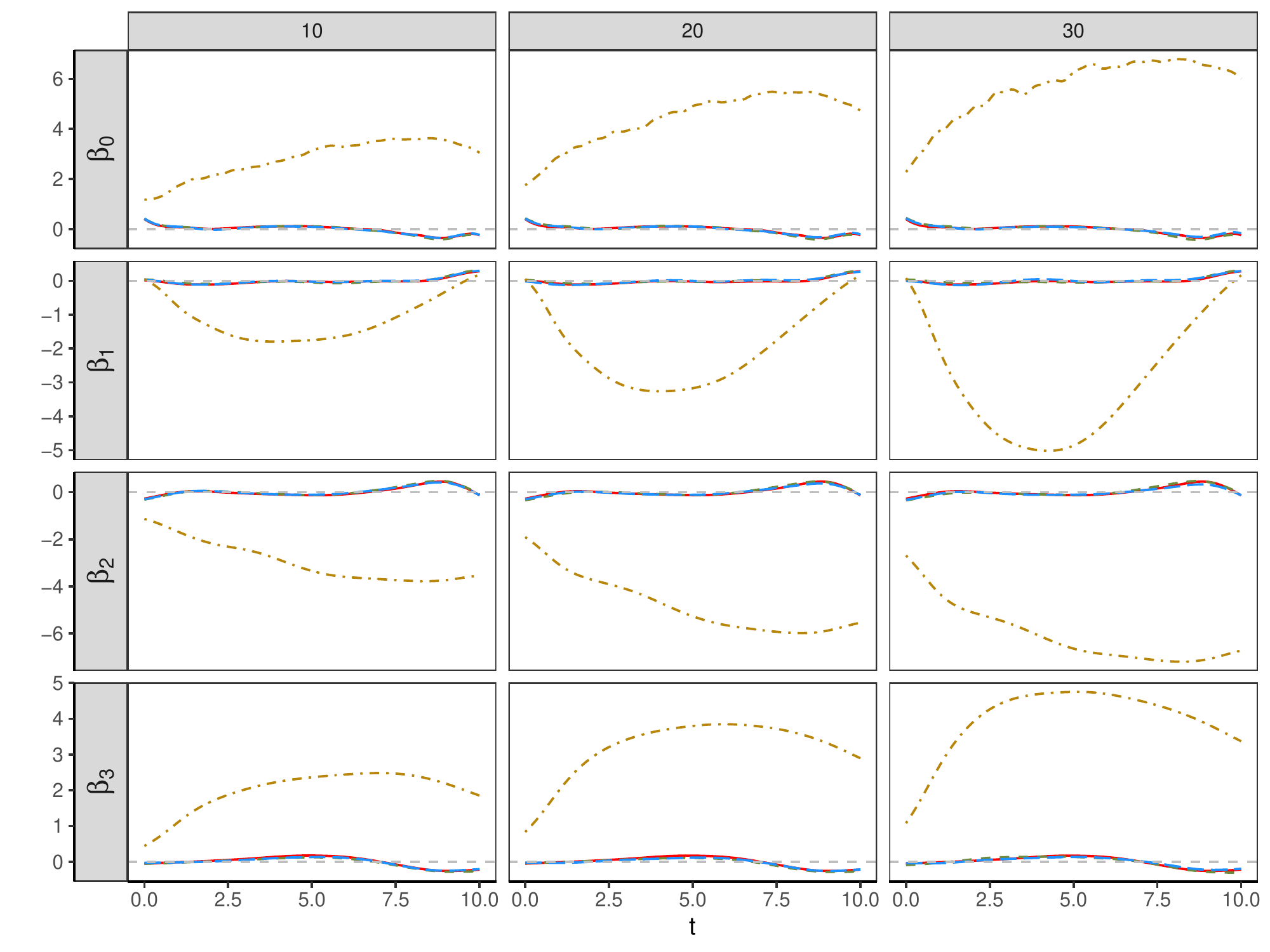}
  \caption{Point-wise standardized bias curves in Setting 1 Scenario (a) (\textbf{Top}) and Setting 1 Scenario (b) (\textbf{Bottom}).  Columns (left to right) correspond to 10, 20, and 30\% missing data.  Rows (top to bottom) correspond to functional parameters $\beta_0, \beta_1, \beta_2,$ and $\beta_3$.}  
  \label{fig:sim_pwm_1ab350}
\end{figure}

\textbf{Parameter Set 2 Results}: Figure \ref{fig:sim_pwm_2ab350} (Top) shows that, for Scenario (a), pwSB for fregMICE and ANM estimates are similar with slight increases for fregMICE with increasing amounts of missing data.  CCA and mean imputation-based estimates show considerable bias for each coefficient function across all amounts of missingness.  Mean pwCov and pwWidth, for Scenario (a), are shown in the bottom left of Figure \ref{fig:sim_covwidthplots_all}.  Mean pwCov for fregMICE and ANM are similar, though the intervals tend to be wider for fregMICE, especially as the amount of missing data increases.  Mean pwCov of $\beta_1$ for CCA is similar to ANM, but coverage of the other functional coefficients is lower, especially for $\beta_2$ and $\beta_3$.  Coverage is poor for intervals based on mean imputed data.

Figure \ref{fig:sim_pwm_2ab350} (Bottom) shows that, for Scenario (b), pwSB for fregMICE and ANM estimates are similar.  CCA estimates also perform similarly to ANM.  As noted above, this is expected since missingness in both $Y$ and $z_2$ depends only on the completely observed covariates, $z_1$ and $z_3$.  pwSB is large for estimates based on mean imputation.  Mean pwCov and pwWidth, for Scenario (b), are shown in the bottom right of Figure \ref{fig:sim_covwidthplots_all}.  While coverage is similar for ANM, CCA, and fregMICE, the mean pwWidths for CCA and fregMICE are larger and increase with larger amounts of missing data.  Again, coverage is poor for intervals based on mean imputed data.  

\begin{figure}
\centering
  \includegraphics[width=0.70\linewidth]{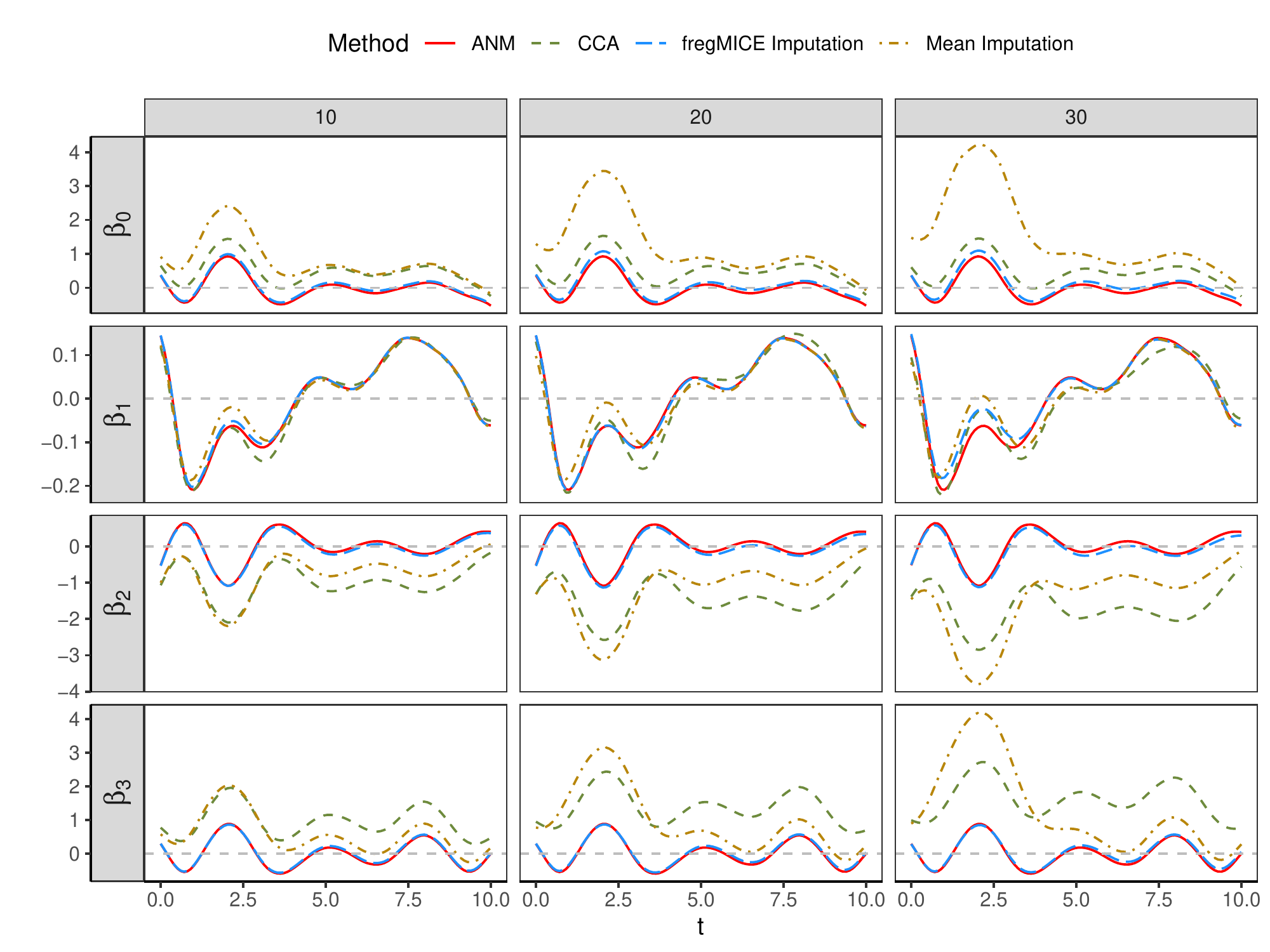}
   \includegraphics[width=0.70\linewidth]{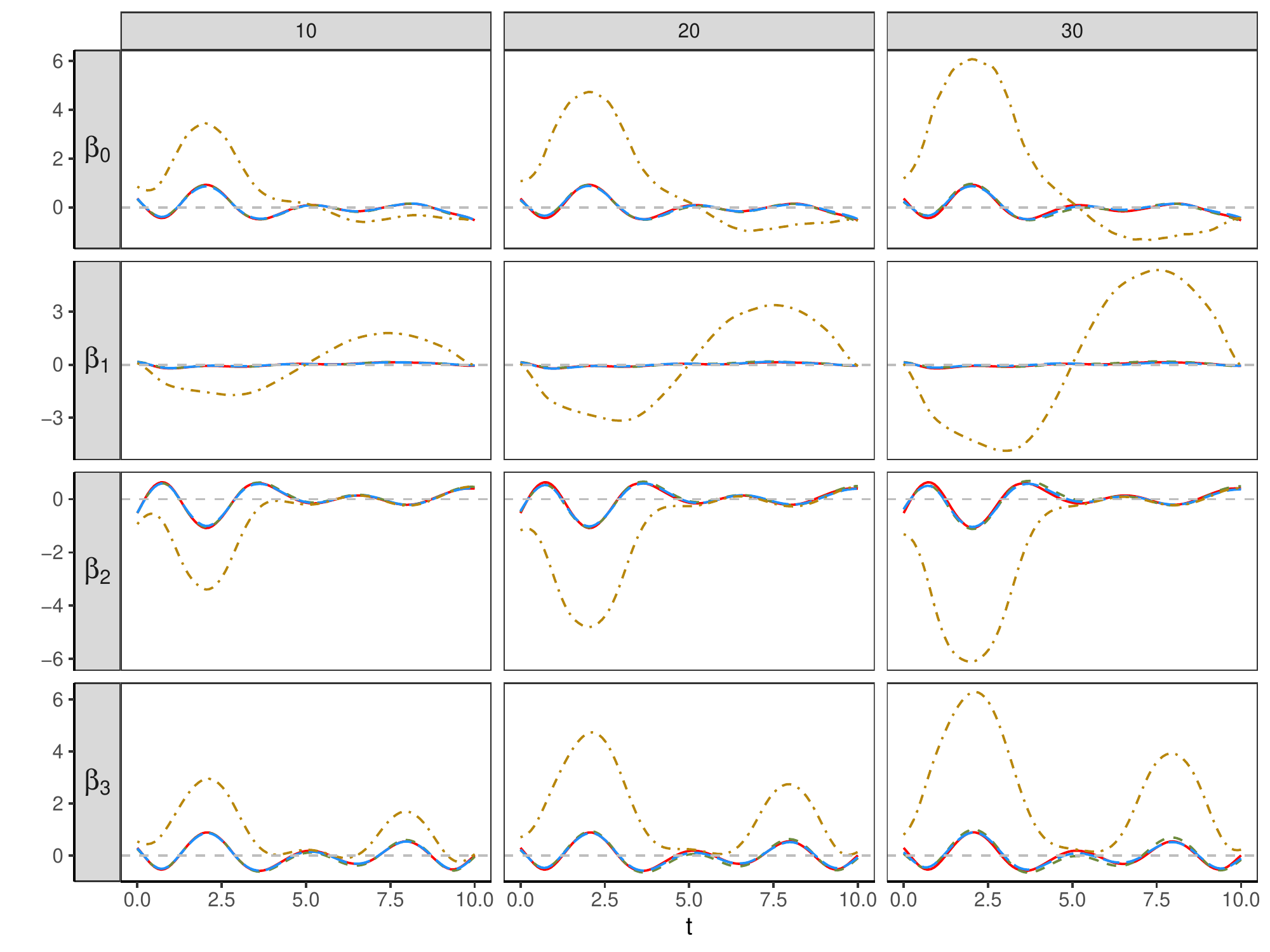}
  \caption{Point-wise standardized bias curves in Setting 2 Scenario (a) (\textbf{Top}) and Setting 2 Scenario (b) (\textbf{Bottom}).  Columns (left to right) correspond to 10, 20, and 30\% missing data.  Rows (top to bottom) correspond to functional parameters $\beta_0, \beta_1, \beta_2,$ and $\beta_3$.}
  \label{fig:sim_pwm_2ab350}
\end{figure}

\textbf{Results Summary}: The fregMICE procedure performs simialrly to the best case scenario where all data are available. It also performs at least as well as or much better than CCA and mean imputation.  Though CCA should be unbiased in settings where missingness is independent of the outcome, fregMICE still tends to perform as well or slightly better on the reported performance measures.  Across all settings, mean imputation performs poorly in all respects and we recommend against its use.  We provide additional results for these settings, as well as for settings where $n = 100$, in Appendix Section 2.  There, we also discuss results from a different set of simulations where the analysis model has a scalar outcome and functional and scalar predictors.        

\section{Application to EMBARC Data} \label{s_app}

In Section \ref{s_intro}, we stated that our goal is to use the EMBARC data to characterize the association between FA and MDD status. \cite{vanderVinne17} suggest that analysis of FA should adjust for both age and gender and consider their potential modifying effects. \cite{kaiser2018} suggest controlling for handedness (left vs. right) and cognition (mental ability).  We follow these suggestions in formulating the functional response analysis model:
\begin{align} \label{analysis_mod} 
\small \nonumber
FA_i(t) = \beta_{0}(t)  & + AGE_i \beta_{1}(t) + EHI_i \beta_{2}(t) + WASIV_i \beta_{3}(t) + MDD_i \beta_{4}(t)  \\
 & + SEX_i \beta_{5}(t) + MDD_i \times SEX_i \beta_{6}(t) + MDD_i \times AGE_i \beta_{7}(t) + \varepsilon_i(t),
\end{align}
where $\varepsilon_i(t) \sim N(0,\sigma^2)$.  In model (\ref{analysis_mod}), $FA_i$ is the normalized CSD asymmetry curve (see Figure \ref{fig:raw}) for subjects having EEG data with ``Good'' or ``Acceptable'' quality designations.  $AGE_i$ is the mean-cetered age in years (i.e., $AGE_i$ = 0 corresponds to the mean age in the sample of 37.16 years),  $EHI_i$ is the Edinburgh Handedness Inventory score (ranging from $-100$ to $100$; completely left to right-handed, respectively), $WASIV_i$ is the raw score for the verbal component of the Wechsler Abbreviated Scale of Intelligence (a measure of cognitive ability with higher values indicating better performance; values range from $20$ to $80$), $MDD_i$ indicates disease status (1 = MDD, 0 = HC), and $SEX_i$ indicates sex ($1$ = Female, $0$ = Male).  (Note that we are breaking with our notation convention and using words with non-bold capital letters to represent scalar quantities.)    

Table \ref{tab:summary_stats}  shows summary statistics by diagnostic group.  $WASIV$ is missing for 49 subjects. EEG data are missing for 11 subjects.  Among the remaining 324 subjects with EEG data available, 88 have data that are ``Marginal'' or ``Unacceptable.''  We treat these EEG data as missing.  

\rowcolors{2}{white}{white}
\begin{table}
\caption{\label{tab:summary_stats} Summary statistics for variables in the analysis model by disease status. } 
\centering
\scriptsize
\begin{tabular}{lcccc}
\hline
              &  \multicolumn{2}{c}{HC ($n = 40$)}            & \multicolumn{2}{c}{MDD ($n = 295$)} \\
Variable & $n$ Available & Mean (SD) or $n$ ($\%$) & $n$ Available & Mean (SD) or $n$ ($\%$) \\
\hline
AGE      & 40  & 37.62 (14.85) & 295 & 37.10 (13.29)  \\
EHI        & 40  & 70.76 (51.43) & 295 & 71.58 (48.23) \\
WASIV  & 35 & 67.40 (7.29)  & 251 & 63.62 (9.59) \\
QIDS    & 40 & 1.40 (1.30) & 295 & 18.08 (2.81) \\
SEX (Female)      & 40   & 25 (62.5) & 295 & 193 (65.4) \\
FA                   & 39 &                  & 285 & \\
\ \ Good          &      &   17 (42.5) & & 126 (42.7) \\ 
\ \ Acceptable     &      &  5 (12.5) & & 88 (29.8) \\          
\ \ Marginal        &      &  14 (35.0) & & 44 (14.9)\\
\ \ Unacceptable     &      &  3 (7.5) & & 27 (9.2)\\           
 \ \ Missing           &      & 1 (2.5) & & 10 (3.4)\\           
\hline
\end{tabular}
\end{table}

A CCA approach for fitting model (\ref{analysis_mod}) uses 204 (60.9\%) complete observations.  As an alternative, we used our fregMICE method to impute the missing $WASIV$ scores and  $FA$ functions.  To allow for interaction between diagnostic group and each covariate in the imputation models, we imputed the missing values separately within the HC ($n_{HC} = 40$) and MDD ($n_{MDD} = 295$) groups, thus making the imputation models more general than the analysis model.  The imputation model for the $WASIV$ variable had the same form given in (\ref{sc_mod}) with age, sex, EHI, diagnostic status, and FA curves as predictors as well as QIDS score, a measure of depressive symptomatology available for all subjects.  We employed PFR to fit the $WASIV$ imputation model with settings similar to those outlined in Section \ref{sim_procs}.  The only difference is that we chose to represent the unknown coefficient functions with a set of 30 B-spline basis functions.  The imputation model for the FA variable was similar to model (\ref{analysis_mod}) with the additional QIDS score predictor.  We employed PFFR to fit this model and used the same settings outlined in Section \ref{sim_procs}.  We generated 20 imputed data sets, running the fregMICE Algorithm for 20 iterations to obtain each imputed data set.     

\begin{figure}[t]
 \center
  \includegraphics[width=0.70\linewidth]{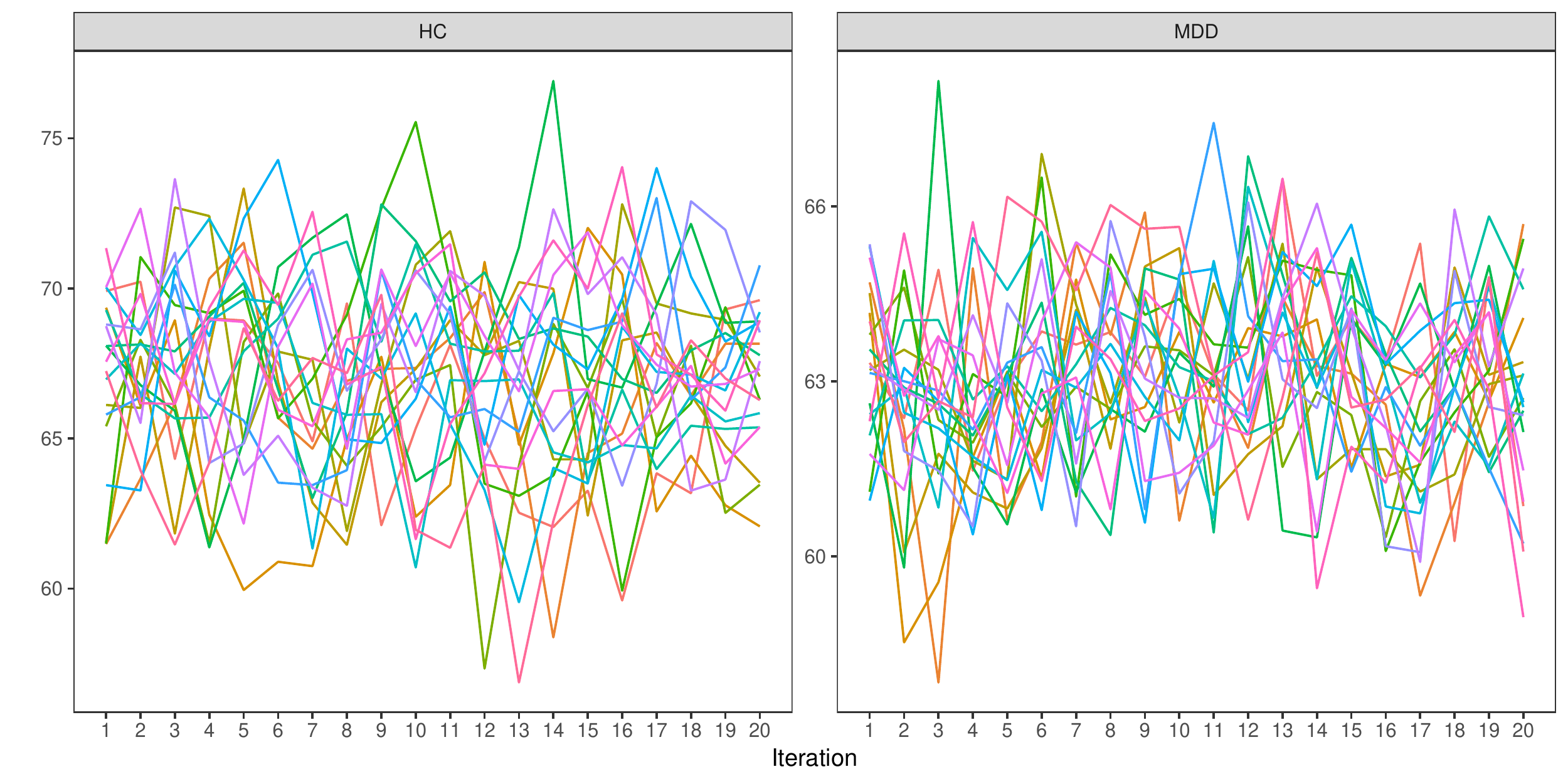}
  \caption{Scalar convergence plots: Mean of the imputed WASIV values for HC and MDD subjects.  The different colors correspond to the different streams.}
  \label{fig:scalar_conv_wasiv}
\end{figure}

\begin{figure}[t]
  \center
  \includegraphics[width=0.70\linewidth]{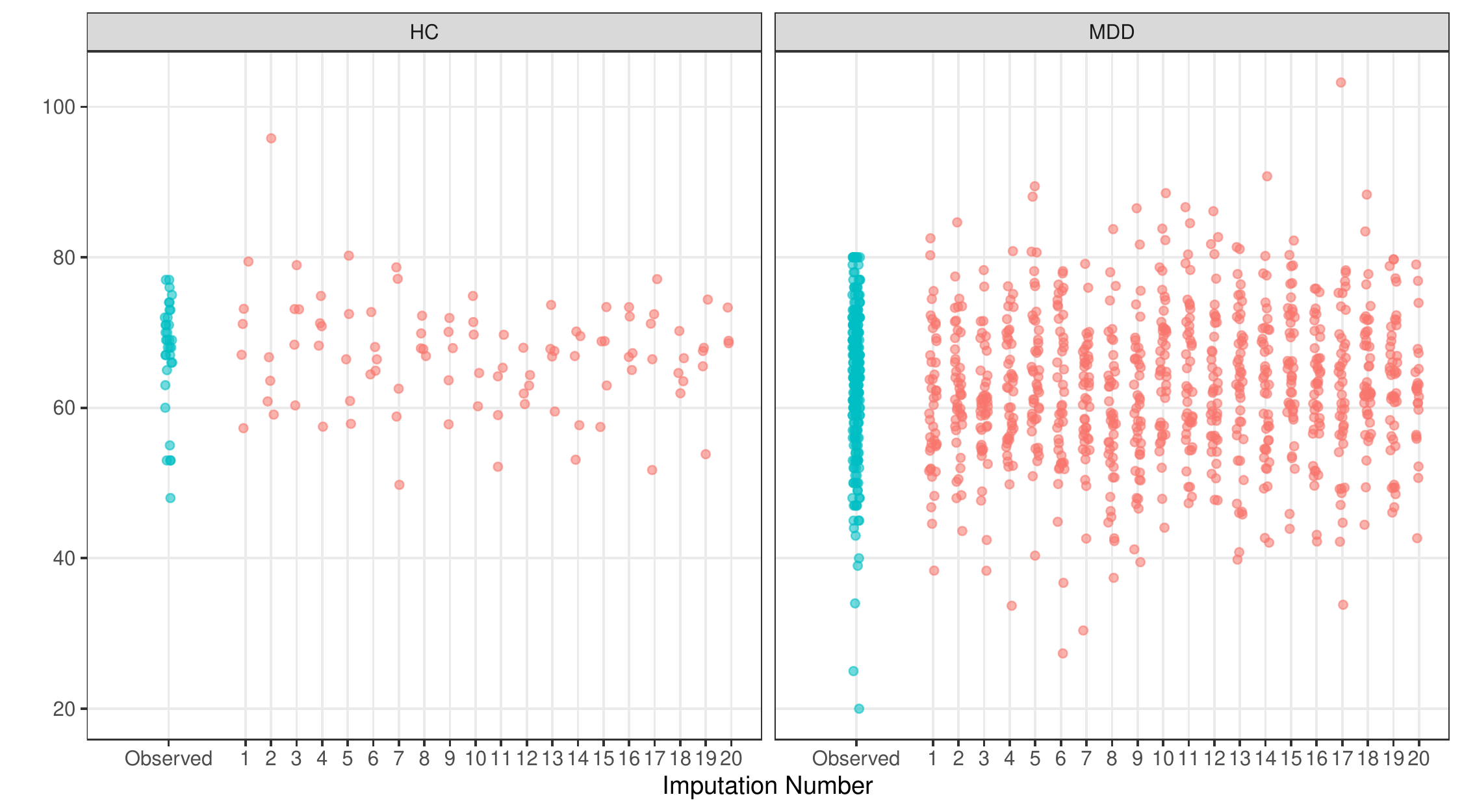}
  \caption{Strip plots of WASIV values.  Horizontal axis shows imputation number ($m$). Vertical axis shows WASIV value.  Observed values are blue and imputed values are red.}
  \label{fig:scalar_strip_wasiv}
\end{figure}

Figure \ref{fig:scalar_conv_wasiv} shows diagnostic plots of the mean value of the imputed $WASIV$ scores at each iteration of the fregMICE algorithm for each of the 20 streams (various line colors) in the HC and MDD subsets.  We see a fair amount of mixing in the streams and no discernible patterns that would suggest convergence issues.  Strip plots showing the imputed and observed values for the the $WASIV$ scores from the 20 imputed data sets are provided in Figure \ref{fig:scalar_strip_wasiv}.  The plots reveal several instances when the imputed $WASIV$ scores were higher than the maximum possible score of 80.  These scores were set to 80 prior to fitting the analysis model.   
 
\begin{figure}[th]
 \center
  \includegraphics[width=1.0\linewidth]{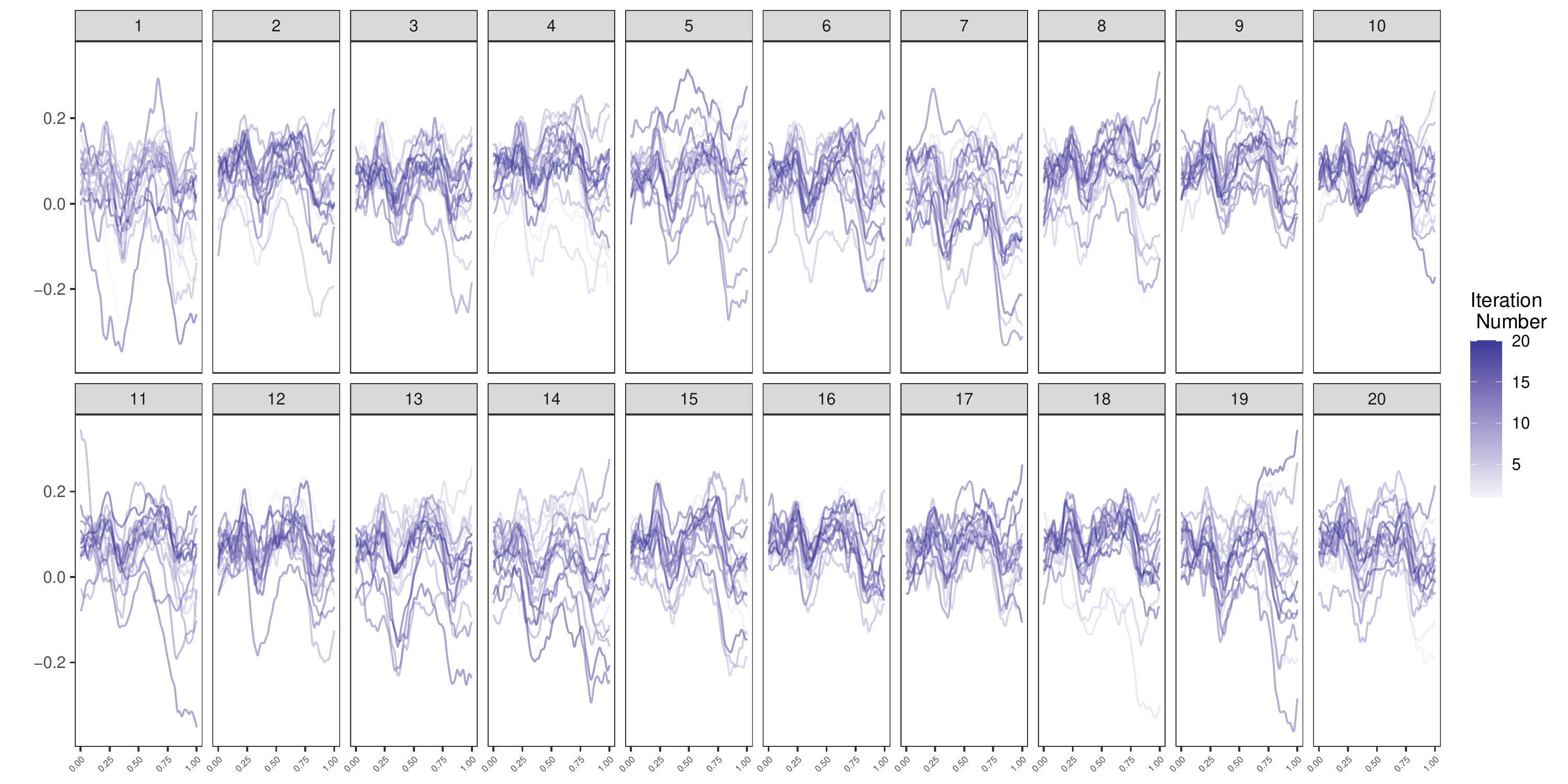}
  \caption{Functional convergence plots. Point-wise mean of the imputed function values for HC subjects.  Each panel corresponds to one imputation stream.  Dark colors correspond to later iterations.}
  \label{fig:func_conv_HC_MDD}
\end{figure}

Figure \ref{fig:func_conv_HC_MDD} shows plots of the means from the imputed FA curves at each iteration of the fregMICE algorithm for each of the 20 streams in the HC subset.  Note that we have scaled the horizontal axis to be between 0 and 1.  If we look across the 20 panels, we see a fair amount of overlap in the streams and no discernible patterns to suggest convergence issues with the imputed FA values.  A similar pattern arose for the MDD subset (plot given in Appendix Section 3).  Strip plots showing the imputed and observed values for the the FA functions from each of the 20 imputed data sets are provided in Appendix Section 3.  These plots show that the imputed FA functions tend to fall within the range of values of the observed FA functions and have similar characteristics.  
 
We combined the $m^{th}$ imputed data sets for the HC and MDD subsets for $m = 1, \ldots 20$ to obtain 20 complete data sets.  We fit model (\ref{analysis_mod}) on each of the 20 complete imputed data sets via PFFR using the same settings as the imputation model for the FA curves described above.  Results were pooled and approximate 95\% point-wise confidence bands were calculated according to Section \ref{sec:MIanalysis}.  As in Section \ref{sim_procs}, we used the Bayesian posterior covariance matrix of the estimated basis coefficients to obtain point-wise standard errors for the estimated coefficient functions.

\begin{figure}[th] 
\center
  \includegraphics[width=0.85\linewidth]{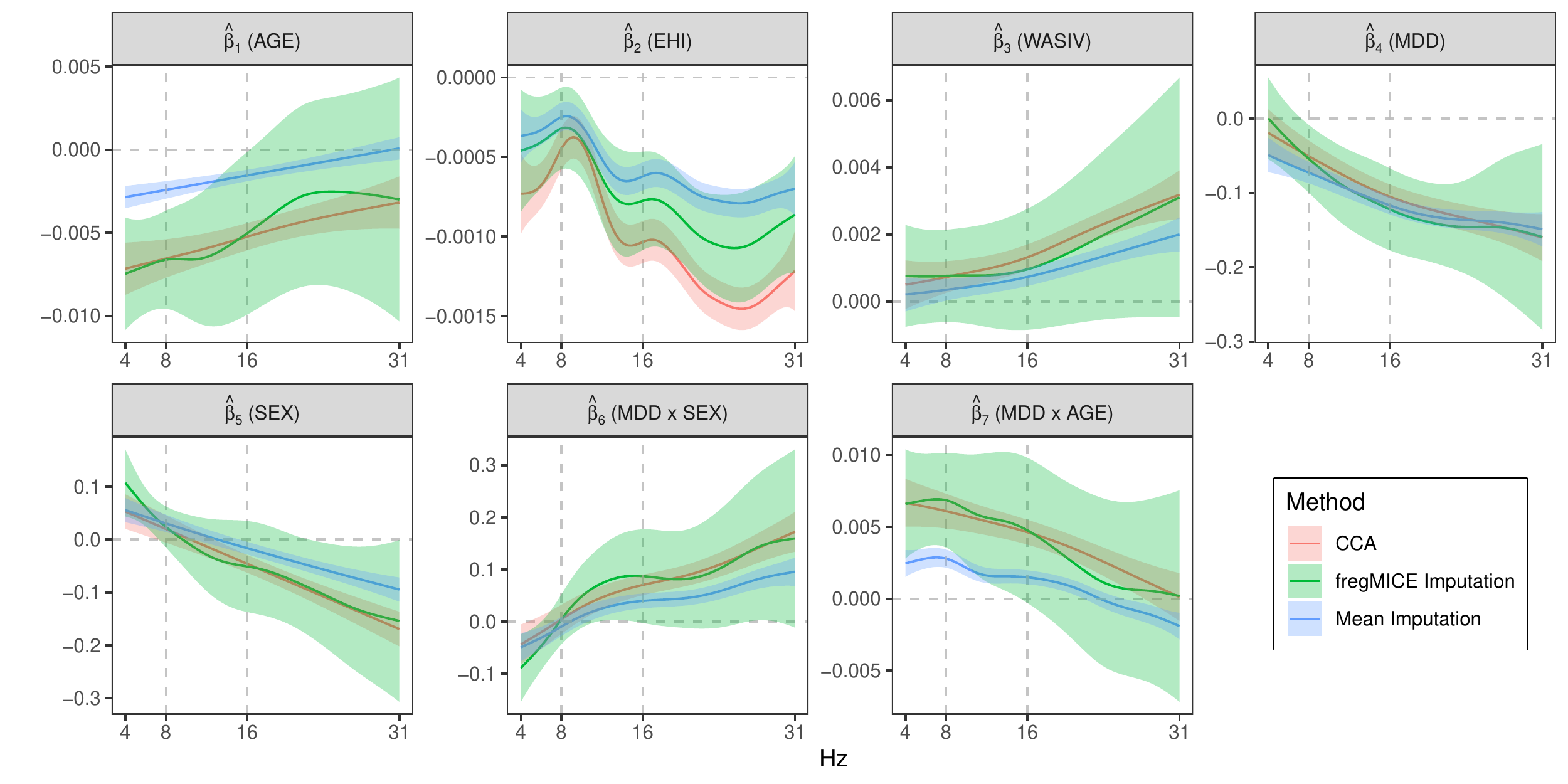}
  \caption{Coefficient function estimates from CCA, fregMICE, and mean imputation.}
  \label{fig:coefs_pffr}
\end{figure}

Figure \ref{fig:coefs_pffr} shows the pooled functional coefficient estimates and corresponding 95\% point-wise confidence bands for model (\ref{analysis_mod}) as well as estimates and confidence bands derived from CCA and from the mean-imputed data.  Most fregMICE coefficient estimates are similar to those derived from CCA, but with considerably wider point-wise confidence bands.  Coefficient estimates from the mean-imputed data tend be closer to 0 across most functions in comparison to the fregMICE and CCA estimates.  Inspection of the estimates from each of the 20 imputed data sets (not shown here) reveals that the wide widths of the confidence bands around the fregMICE estimates are due to the relatively large amount of between imputation variance.  This is not surprising considering the HC sample was small with only 22 subjects having ``Good'' or ``Acceptable'' quality EEG data.             

From the coefficient plots in Figure \ref{fig:coefs_pffr} we see evidence of differences in the FA curves between the MDD and HC groups that may depend on both sex and age.  The dependence on sex and age is more pronounced with the CCA estimates whose point-wise confidence bands are narrower.  To better understand how the differences between the MDD and HG groups depend on both sex and age, we constructed plots of the model-based mean FA curves for different combinations of the predictor values.  We provide a Shiny app, available in the Supplementary Materials, to do this.  

Figure \ref{fig:pred_pffr} shows one set of plots in which the mean FA curves for MDD and HC subjects are stratified by sex for three different age values corresponding to the mean age in the sample (37.16), one standard deviation below (23.70), and one standard deviation above (50.62).  For all plots, EHI and WASIV were set to their sample mean values of 71.48 and 64.08 respectively.  Both the CCA and fregMICE estimates show that, among females, HC subjects tend to have greater FA than MDD subjects at a younger age across most frequency values, but that the difference between the groups decreases with age and ultimately the difference reverses direction with older female MDD subjects having greater FA, primarily in the theta and alpha frequency bands.  For males, the plots for both the CCA and fregMICE estimates show that HC subjects tend to have greater FA than MDD subjects at a younger age and that difference persists, though slightly diminished, in the late alpha and beta frequency bands at older ages.  In the theta and early alpha bands, the difference diminishes with increasing age and ultimately reverses with older male MDD subjects showing greater FA in the theta and early alpha bands than older male HC subjects.  In contrast to CCA and fregMICE, the coefficient estimates derived from mean imputation show similar patterns of difference between the HC and MDD groups for both males and females and across different ages.  This follows from the fact that the coefficient estimates for the interaction terms ($\hat{\beta}_6$ and $\hat{\beta}_7$) based on mean imputation are closer to zero than estimates from either CCA or fregMICE.    

\begin{figure}[t]
\center
  \includegraphics[width=0.49\linewidth]{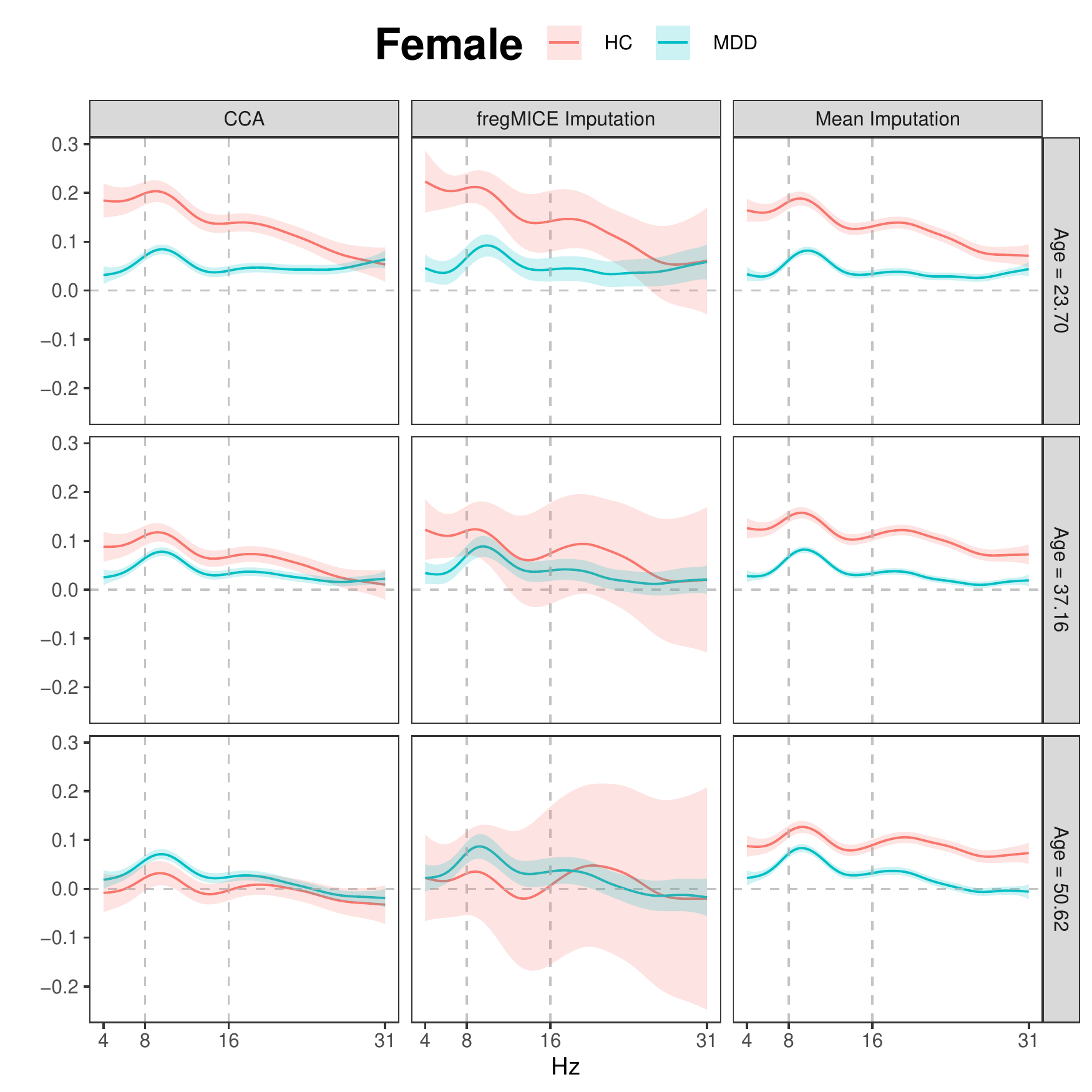}
  \includegraphics[width=0.49\linewidth]{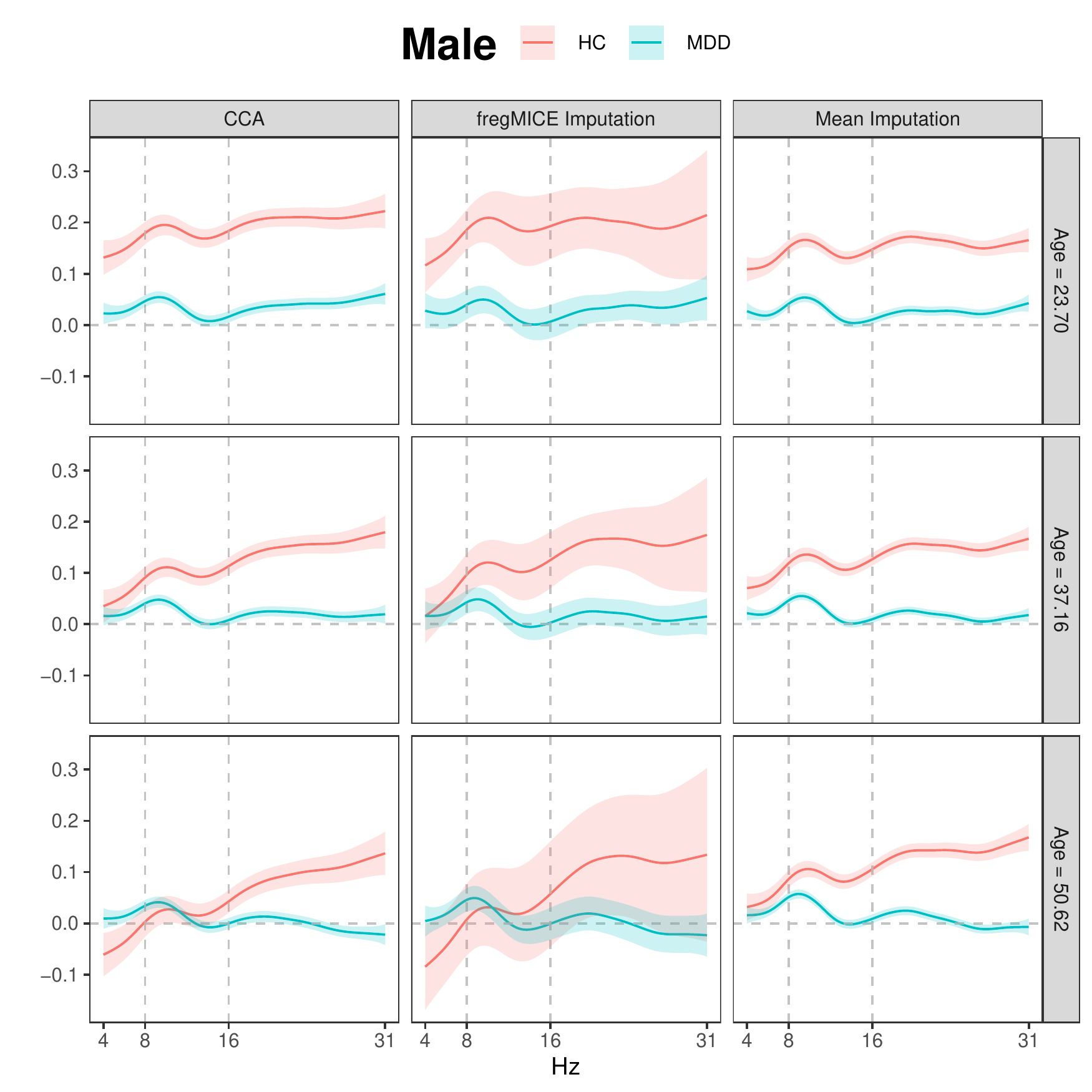}
  \caption{Model-based mean FA curves from CCA, fregMICE, and mean imputation.  Curves are given for three AGE values (mean age in the sample (37.16), one standard deviation below (23.70), and one standard deviation above (50.62)) with EHI = 71.48 and WASIV = 64.08 (the sample means for each variable).}
  \label{fig:pred_pffr}
\end{figure}

\section{Discussion} \label{s_disc}
Research on handling missing data in FDA is extremely limited \citep{nipals, func_mi_sim, febrerobande19}.  To our knowledge, CCA is the current state-of-the-art when entire functional values are missing.  As in the purely scalar setting, CCA is not a universally acceptable approach.  In this article, we extended the MICE algorithm to perform multiple imputation with missing data that are scalar or function-valued and we extended Rubin's Rules to conduct valid inference.  Our simulations show that, in some settings, CCA leads to greater bias in the parameter estimates and correspondingly poor coverage for point-wise confidence bands when compared to the fregMICE procedure.  They also show that mean imputation can perform extremely poorly.  

We applied the proposed extensions to conduct an analysis of the association between fontal CSD power asymmetry and depression status in subjects from the EMBARC study, some having missing values on relevant variables.  Using a functional response model, we found that the relationship between fontal CSD power asymmetry and depression status depends on both gender and age.  The CCA and the analysis based on fregMICE yielded similar results, though the variability in pooled estimates based on fregMICE was greater.   

Though motivated by our need to fit a functional response model on EEG data, the fregMICE method, paired with the extension of Rubin's Rules to functional data, comprises a set of tools that can be applied to both functional and scalar response models with both scalar and/or functional predictors.  Here, we chose to employ PFR and PFFR for fitting the imputation models, but we could have used any other suitable functional regression methods.  One benefit of using PFR and PFFR is that both can be extended to handle sparsely sampled functional data, functional data that are not observed on the same grid-points, or functional data that are observed with noise.  We refer the reader to \cite{GoldsmithPenalized} and \cite{ivanescu2015} for more information.   

Our proposed extensions to existing multiple imputation methods do have several limitations.  First, as in the completely scalar case, the MAR assumption should hold in order for the proposed methods to yield unbiased estimates of the parameters of interest.  \cite{Potthoff_2006} discuss this issue and propose techniques for assessing the MAR assumption. For MNAR cases, more complex imputation models, which include joint modeling of data and missingness, are needed. Second, it is clear that the imputation models should be specified so as to provide high-quality imputations.  This becomes a complex task in settings with many scalar and functional variables.  New robust methods will need to be developed to handle missing data in such settings.  Third, the fregMICE procedure is computationally intensive.  Though the computational burden does not prohibit its use in settings with a handful of variables (e.g., 5 to 10 scalar and functional variables), it may not be practically employed in settings with many scalar and functional variables with high rates of missingness.  Such settings will likely arise as biomedical and public health research studies collect greater amounts of both scalar and functional variables.  New approaches that increase computational efficiency will need to be developed.  Fourth, as we noted in Section \ref{sub_conv}, checking convergence of the standard MICE procedure tends to rely on ad hoc approaches like inspection of various convergence plots of summary measures.  While we employed a similar approach in our application via plotting the point-wise mean of the imputed functions at each iteration, it may be instructive to consider other summary measures (e.g., cross-covariance, measures of smoothness, etc.) or methods to assess convergence.  Lastly, we applied normal approximations to construct confidence bands for the functional parameters.  Such approximations are common in the functional regression literature \citep{GoldsmithPenalized, ivanescu2015}, but confidence bands show better performance if they are based on t-distributions with the appropriate degrees of freedom. \\   
 
\noindent \textbf{Supplementary Materials} \\
\verb+fregMICE_Appendix.pdf+ provides additional information and simulation results.  The zip file \verb+Model_Based_FA_Shiny_App.zip+ contains the Shiny app described in Section \ref{s_app}.  \verb+R+ code for running the simulations is available in the zip file \verb+fregMICE_R_Code.zip+.
 
 \newpage
 \appendix

 \begin{center}
 \textbf{Appendix to: Multiple imputation in functional regression with \\ applications to EEG data from a depression study}
 \end{center}

\singlespace 
 
 \section{Overview of Fitting Procedures for Complete Data}

\subsection{PFFR for Fitting Functional Response Models (FRMs)}

Briefly, the PFFR fitting procedure is carried out as follows.  First, each coefficient function is represented by a set of basis functions (e.g., B-splines, cubic-regression splines, thin-plate splines, etc.).  That is, we let $\beta_j(t) = \bm{B}_j^{\intercal}(t) \bm{b}_j$ for $j = 0, \ldots, p$, where $\bm{B}_j(t) = (B_{j,1}(t), \ldots, B_{j,L_j}(t))^{\intercal}$ is the vector of known basis functions selected for $\beta_j(t)$ and $\bm{b}_j = (b_{j,1}, \ldots, b_{j,L_j})^{\intercal}$ is the corresponding vector of unknown basis coefficients. Similarly, we let $\rho_{k}(s,t) = \bm{U}_{k}^{\intercal}(s,t)\bm{u}_{k}$ for $k = 1, \ldots, q$ where $\bm{U}_{k}(s,t) = (U_{k,1}(s,t), \ldots, U_{k,L_k}(s,t))^{\intercal}$ is the vector of known bivariate basis functions (e.g., bivariate thin-plate splines) selected for $\rho_k(s,t)$ and $\bm{u}_k = (u_{k,1}, \ldots, u_{k,L_k})^{\intercal}$ is the corresponding vector of unknown basis coefficients.  The numbers of basis functions used in the representations ($L_j$ and $L_k$) are typically selected to be larger than is assumed necessary to adequately represent a given function.  The integral terms are approximated as 
$$\int X_{i,k}(s)\rho_{k}(s,t)ds \approx \sum_{d = 1}^{D_k} \Delta_d \rho(s_d, t)X_{i,k}(s_d)
                                             = \left\{ \sum_{d = 1}^{D_k} \bm{U}_{k}^{\intercal}(s_d,t) X_{i,k}(s_d)\Delta_d \right\} \bm{u}_{k}, $$
where $s_d$ are the grid points over which $X_{i,k}$ is observed and $\Delta_d$ is the corresponding interval length between grid points.  Substituting the integral approximation and basis representations into model (1) from the main article gives 
$$g\{\mu_{i}(t) \}  = \bm{B}_0^{\intercal}(t) \bm{b}_0 + \sum_{j = 1}^{p} z_{i,j}\bm{B}_j^{\intercal}(t) \bm{b}_j +
                              \sum_{k = 1}^{q} \left \{ \sum_{d = 1}^{D_k} \bm{U}_{k}^{\intercal}(s_d,t) X_{i,k}(s_d)\Delta_d \right \} \bm{u}_{k},$$
and the log-likelihood function is given by 
$$l(\bm{b}_0, \ldots, \bm{b}_p, \bm{u}_1, \ldots, \bm{u}_q) = \sum_{g = 1}^{N_t} \sum_{i = 1}^n log\left [ EF \{ \mu_{i}(t_g | \bm{b}_0, \ldots, \bm{b}_p, \bm{u}_1, \ldots, \bm{u}_q), \eta_t) \} \right ],$$
where $\{t_g; g = 1, \ldots, N_t\}$ are the grid points at which $Y_i$ is observed and we explicitly write the mean as a function of the vectors of parameters $\bm{b}_j$ and $\bm{u}_k$ for $j = 0, \ldots, p$ and $k = 1, \ldots, q$.  As $L_j$ and $L_k$ are chosen to be large, smoothness in the estimated coefficient functions is achieved by including penalties on the estimated basis coefficients when maximizing the log-likelihood.  Hence the objective function to be maximized is 
$$l(\bm{b}_0, \ldots, \bm{b}_p, \bm{u}_1, \ldots, \bm{u}_q) - \sum_{j = 0}^p \lambda_{\bm{b}_j}P_{\bm{b}_j}(\bm{b}_j) - \sum_{k = 1}^q \lambda_{\bm{u}_k}P_{\bm{u}_k}(\bm{u}_k),$$ 
where $P_{\bm{b}_j}(\bm{b}_j)$ and $P_{\bm{u}_k}(\bm{u}_k)$ are known penalty functions and $\lambda_{\bm{b}_j}$ and $\lambda_{\bm{u}_k}$ are the respective non-negative tuning parameters that control the amount of smoothness.  Using quadratic penalties of the form $P_{\bm{b}_j}(\bm{b}_j) = \bm{b}_j^{\intercal} \bm{D}_{\bm{b}_j} \bm{b}_j$ and $P_{\bm{u}_k}(\bm{u}_k) = \bm{u}_k^{\intercal} \bm{D}_{\bm{u}_k} \bm{u}_k$, where $\bm{D}_{\bm{b}_j}$ and $\bm{D}_{\bm{u}_k}$ are known penalty matrices allows to use a mixed-effects model framework, with the coefficient functions viewed as random effects.  In this setting, one can use restricted maximum likelihood (REML) to simultaneously select the tuning parameters and estimate the basis coefficients corresponding to the coefficient functions.  Furthermore, approximate confidence intervals for the basis coefficients can be obtained.  The estimated basis coefficients and their approximate intervals can then be substituted into the basis representations to obtain the estimated coefficient functions and point-wise confidence bands.  Full details can be found in \cite{ivanescu2015}.

\subsection{PFR for Fitting Scalar Response Models (SRMs)}

Briefly, PFR is carried out as follows.  First the functional covariates are represented using a truncated Karhunen-Lo\'eve decomposition, $X_{ij}(t) = \sum_{k = 1}^{K_j}c_{ijk} \psi_{jk}(t) =  \bm{\psi}_{j}(t)\bm{c}_{ij}$, where $\psi_{jk}(t)$ for $k = 1, \ldots, K_j$ are the first $K_j$ eigenfunctions (functional principal components (FPCs)) of the smoothed covariance operator corresponding to $cov\{X_{ij}(s), X_{ij}(t)\}$, $c_{ijk} = \int X_{ij}(t)\psi_{jk}(t) dt$ are the FPC scores,  $\bm{\psi}_{j}(t) = (\psi_{j1}(t), \ldots, \psi_{jK_j}(t))$ is a $1 \times K_j$ vector of the eigenfunctions, and $\bm{c}_{ij}^{\intercal} = (c_{ij1}, \ldots, c_{ijK_j})$ is a $1 \times K_j$ vector of FPC scores representing the $j$-th functional covariate for the $i$-th subject.    Next, each coefficient function, $\beta_j$, is represented using a suitable set of basis functions given by $\bm{\phi}_{j}(t) = (\phi_{j1}(t), \ldots, \phi_{jL_j}(t))$ such that $\beta_{j}(t) = \sum_{\ell = 1}^{L_j} b_{j \ell}\phi_{j \ell}(t) = \bm{\phi}_{j}(t) \bm{b}_{j}$, where $\bm{b}_{j}^{\intercal} = (b_{j1}, \ldots, b_{jL_j})$.  Using these representations, the right-hand side of the equation (2) from the main article can be re-expressed as $\theta_{0} + \bm{z}_{i} \bm{\theta} + \sum_{j = 1}^{q}  \tilde{\bm{X}}_{ij}\bm{b}_j $ where $\tilde{\bm{X}}_{ij} = \bm{c}_{ij}^{\intercal} \bm{G}_{j}$ and $\bm{G}_{j}$ is a $K_j \times L_j$ dimensional matrix with entry $(u, v)$ given by $\int \psi_{ju}(t)\phi_{jv}(t)dt$.  

As with the PFFR method to estimate the FRM given in model (1) from the main article, PFR uses a penalized log-likelihood objective function with similar penalties on the basis coefficients and employs a mixed effects model framework to obtain estimates for the basis coefficients.  Using this framework, the basis coefficients are viewed as random effects in a mixed effects model and turning parameters are estimated via REML.  As with PFFR, since PFR employs a mixed effects framework, it is possible to obtain approximate confidence intervals for the scalar coefficients and approximate point-wise confidence intervals for the coefficient functions.  The reader is referred to \cite{GoldsmithPenalized} for the complete details of the PFR estimation procedure.  $K_j$ and $L_j$ for $j = 1, \ldots, q$ are tuning parameters that need to be chosen prior to estimation.  Selection can be based on a data-driven approach such as cross-validation, but this approach may be computationally intensive.  Alternatively, \cite{GoldsmithPenalized} note that as long as they are chosen ``large enough,'' then their specific values have minor impact on the quality of the estimates.

\section{Additional Simulation Results}

In the main article, we present results for simulations in which the analysis model is a FRM with 3 scalar predictors.  In those simulations, the sample size is $n = 350$.  Here we present additional results for the same settings and scenarios, but for a smaller sample size of $n = 100$. In addition, we include plots for the point-wise coverages and point-wise widths of the 95\% confidence intervals for both $n = 350$ and $n = 100$.  In the main article, we provide tables with across-the-function mean point-wise coverage and point-wise width. We hope that the plots included here provide additional insight into the performance of the different methods for handling missing data in the context of fitting a FRM.  These additional results are provided in Sections \ref{sim_sec_01} and \ref{sim_sec_02}.

We also introduce a new simulation study in Section \ref{sim_sec_03} where the analysis model is a SRM with scalar and functional predictors.  The complete case, mean imputation, and fregMICE approaches for handling missing data are applied and evaluated.
  
\subsection{Proportion of Missing Observations} \label{sim_sec_prop_miss}

Tables \ref{p_mis_tab_s1_n100} and \ref{p_mis_tab_s1_n350} below show mean (sd) proportion of missing observations in the data sets generated under the simulation settings described in the main article for sample sizes of $n = 100$ and $n = 350$ respectively.

\begin{table}[h!]
\caption{\label{p_mis_tab_s1_n100} Mean (SD) proportion of missing observations with $n = 100$ for Settings 1 and 2. } 
\centering
\scriptsize
\begin{tabular}{llccc}
\hline
 & & 10\% Intended & 20\% Intended & 30\% Intended \\
\hline
Scenario (a) & Prop. Miss. $z_2$        & 0.10 (0.00) & 0.20 (0.00) & 0.30 (0.00)  \\
Scenario (b) & Prop. Miss. $z_2$        & 0.10 (0.03) & 0.20 (0.04) & 0.29 (0.05)  \\
                     & Prop. Miss. $Y$          & 0.10 (0.03) & 0.20 (0.04) & 0.31 (0.05)  \\
                     & Prop. Miss. $z_2$ or $Y$ & 0.19 (0.04) & 0.36 (0.05) & 0.51 (0.05)  \\
\hline
\end{tabular}
\end{table}

\begin{table}[h!]
\caption{\label{p_mis_tab_s1_n350} Mean (SD) proportion of missing observations with $n = 350$ for Settings 1 and 2. } 
\centering
\scriptsize
\begin{tabular}{llccc}
\hline
 & & 10\% Intended & 20\% Intended & 30\% Intended \\
\hline
Scenario (a) & Prop. Miss. $z_2$        & 0.10 (0.00) & 0.20 (0.00) & 0.30 (0.00)  \\
Scenario (b) & Prop. Miss. $z_2$        & 0.10 (0.02) & 0.20 (0.02) & 0.29 (0.03)  \\
              & Prop. Miss. $Y$          & 0.10 (0.02) & 0.20 (0.02) & 0.31 (0.03)  \\
              & Prop. Miss. $z_2$ or $Y$ & 0.20 (0.02) & 0.36 (0.02) & 0.52 (0.03)  \\
\hline
\end{tabular}
\end{table}

Note that, for Scenario (a), for either sample size, the missing proportions are exactly 10\%, 20\%, or 30\% when they are intended to take on these values.  This is because $z_2$ is set to missing if the corresponding functional response, $Y$, has an average value above the 90th percentile (for 10\% missing), 80th percentile (for 20\% missing), or 70th percentile (for 30\% missing). 

In Scenario (b), both $z_2$ and/or $Y$ can be missing for an observation.  The bottom rows of Tables \ref{p_mis_tab_s1_n100} and \ref{p_mis_tab_s1_n350} show the mean (sd) proportions of incomplete observations in a data set.  

\subsection{Setting 1 Additional Results} \label{sim_sec_01}

\subsubsection{Point-wise Mean Plots for $n = 350$}

In the main article, we provided point-wise standardized bias plots for each setting and scenario for $n = 350$.  Figures \ref{fig:sim_pwm_1ab350} and \ref{fig:sim_pwm_2ab350} show the point-wise means over the 500 simulation runs.

\begin{figure}
\centering
  \includegraphics[width=0.85\linewidth]{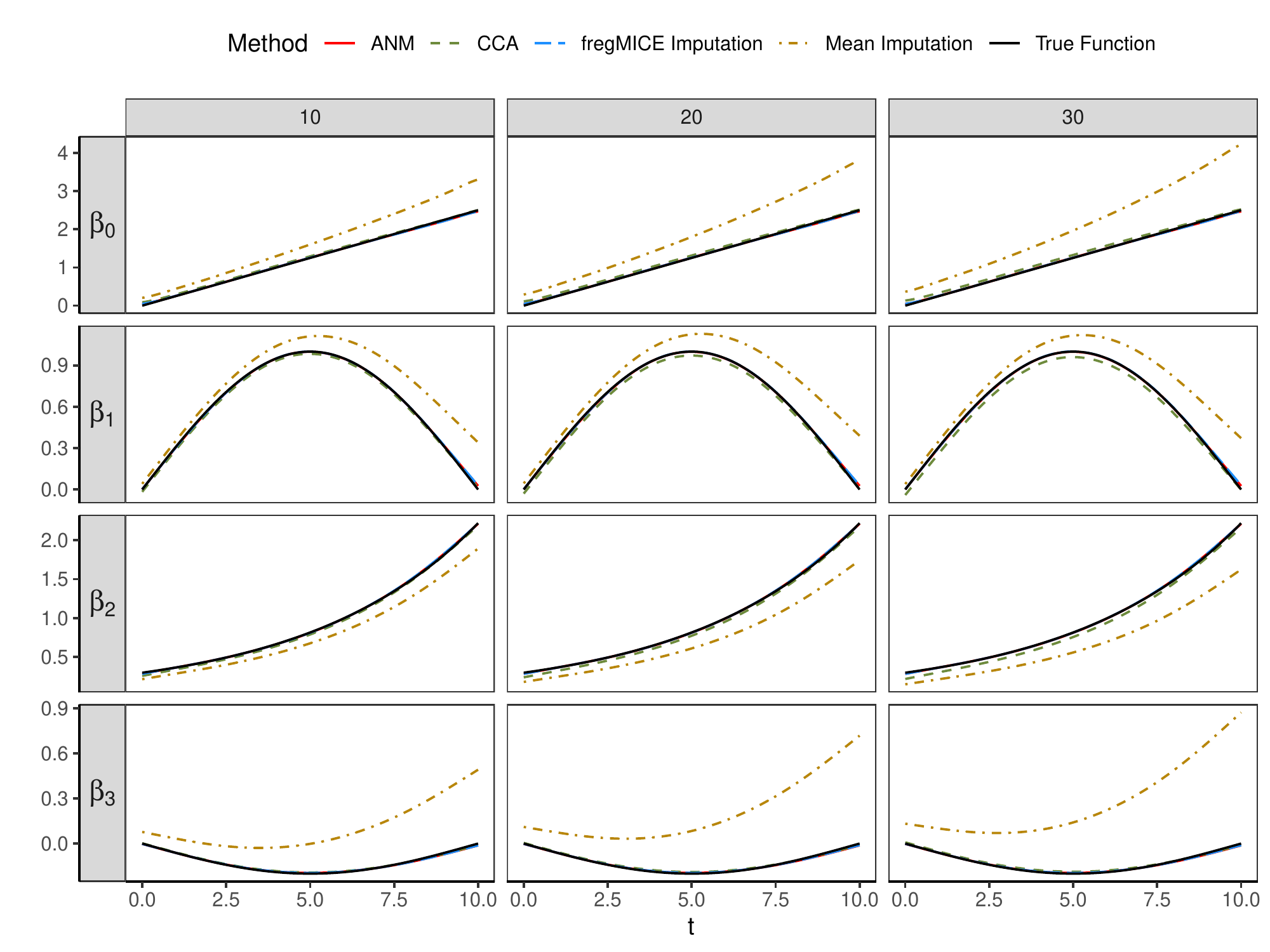}
  \includegraphics[width=0.85\linewidth]{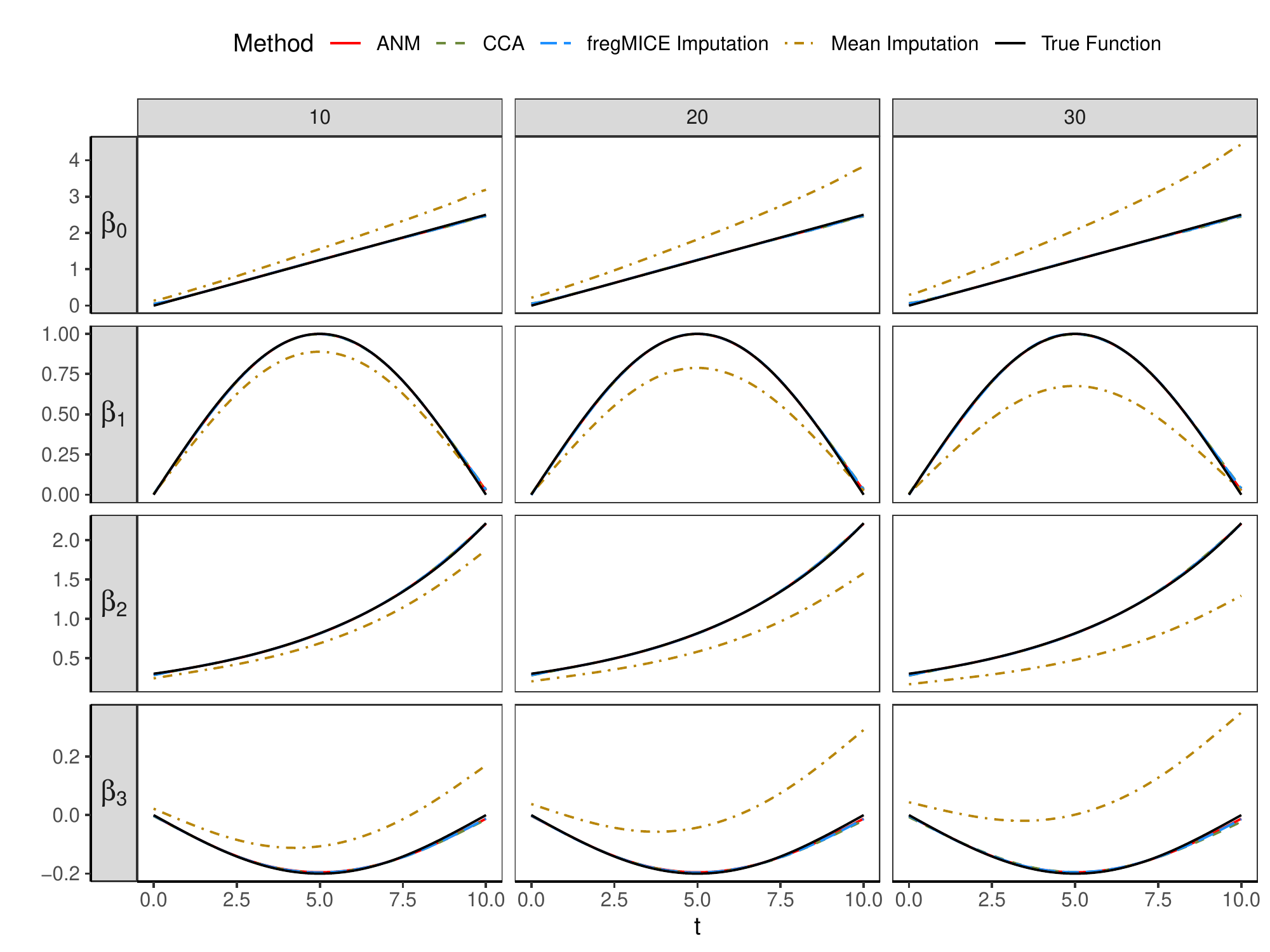}
  \caption{Point-wise mean curves in Setting 1 Scenario (a) (\textbf{Top}) and Setting 1 Scenario (b) (\textbf{Bottom}).  Columns (left to right) correspond to 10, 20, and 30\% missing data.  Rows (top to bottom) correspond to functional parameters $\beta_0, \beta_1, \beta_2,$ and $\beta_3$.}  
  \label{fig:sim_pwm_1ab350}
\end{figure}

\begin{figure}
\centering
  \includegraphics[width=0.85\linewidth]{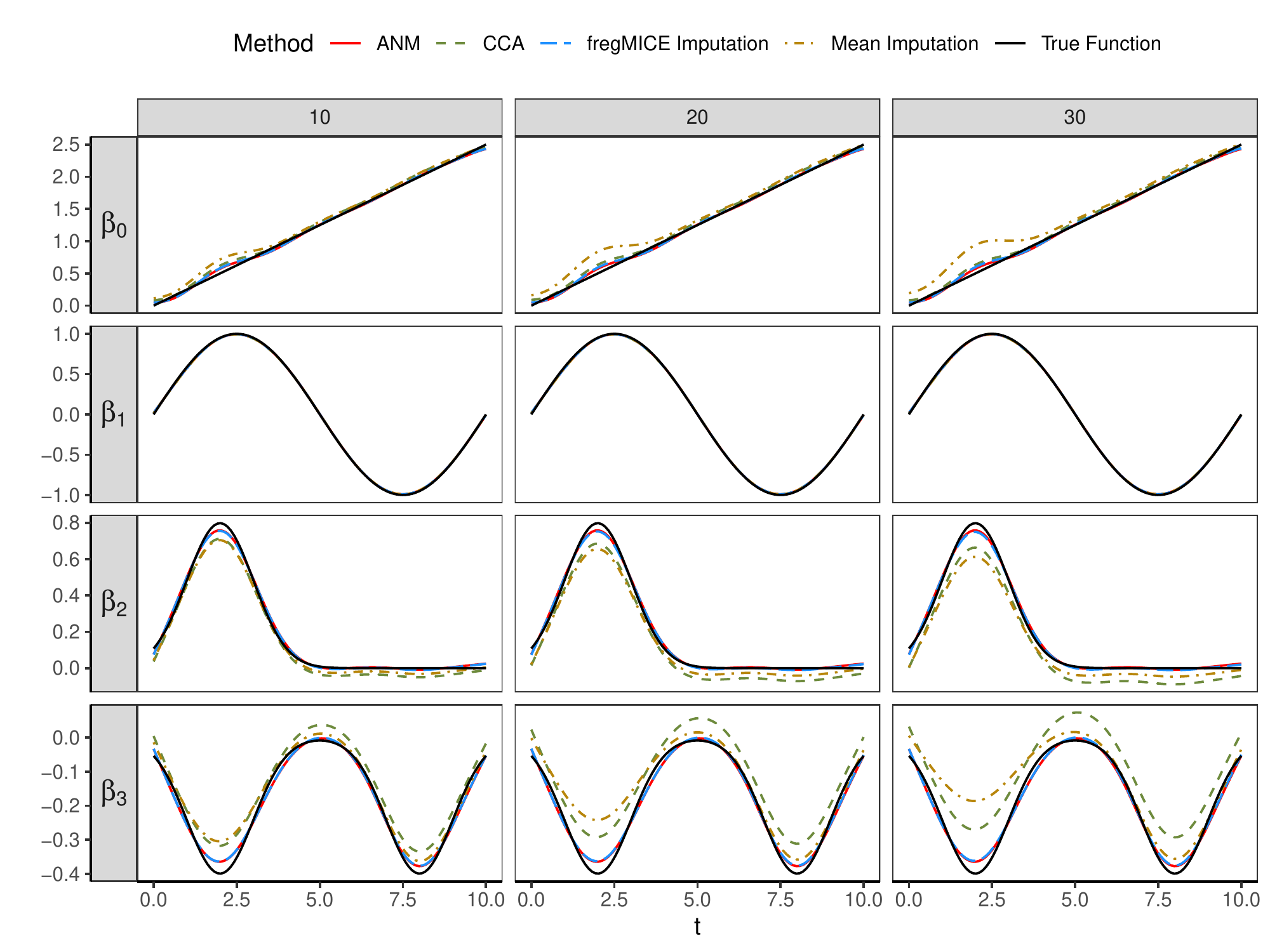}
  \includegraphics[width=0.85\linewidth]{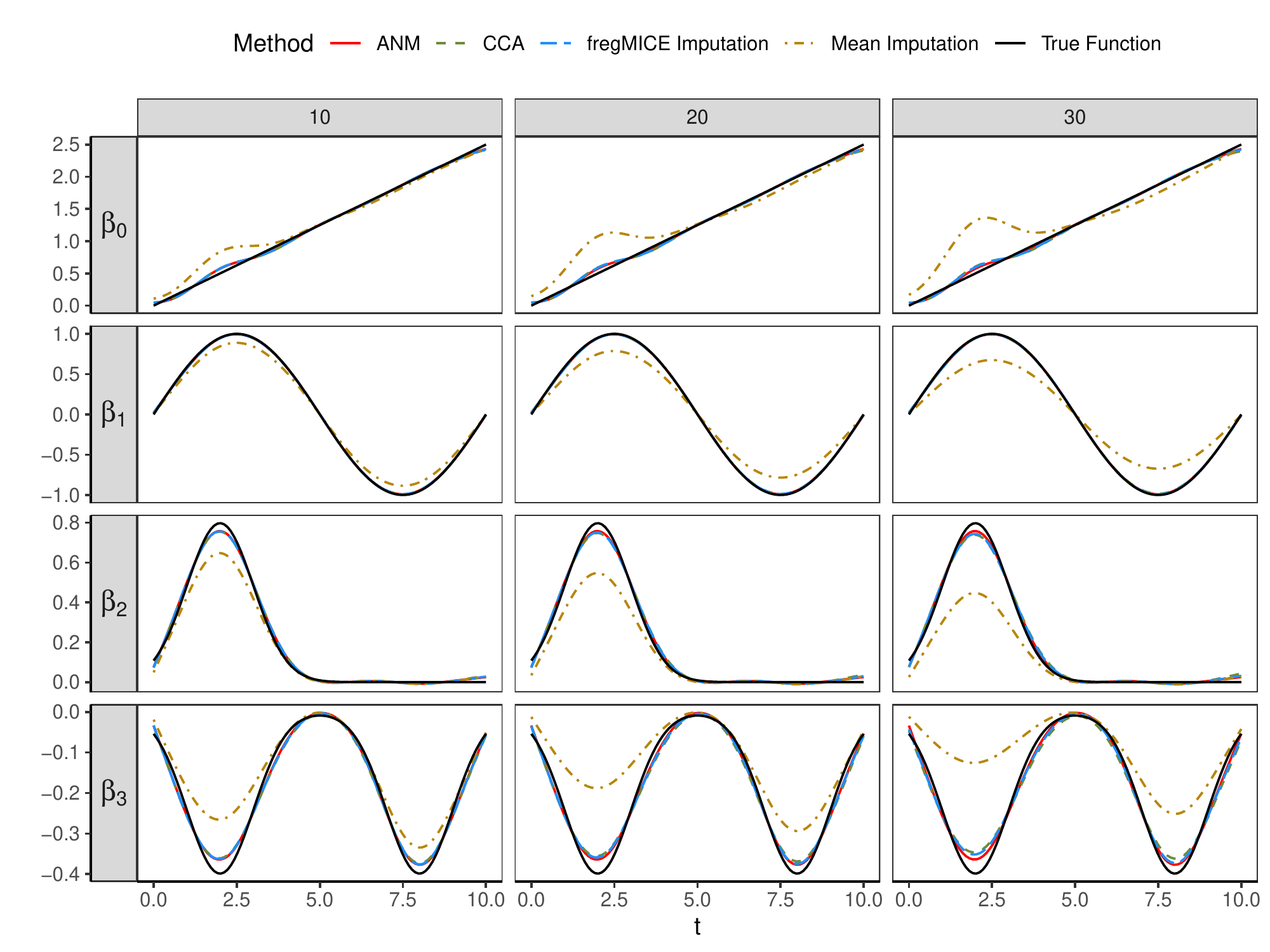}
  \caption{Point-wise mean curves in Setting 2 Scenario (a) (\textbf{Top}) and Setting 2 Scenario (b) (\textbf{Bottom}).  Columns (left to right) correspond to 10, 20, and 30\% missing data.  Rows (top to bottom) correspond to functional parameters $\beta_0, \beta_1, \beta_2,$ and $\beta_3$.}  
  \label{fig:sim_pwm_2ab350}
\end{figure}

\subsubsection{Point-wise Mean and Point-wise Standardized Bias Plots for $n = 100$}

\begin{figure}
\centering
  \includegraphics[width=0.85\linewidth]{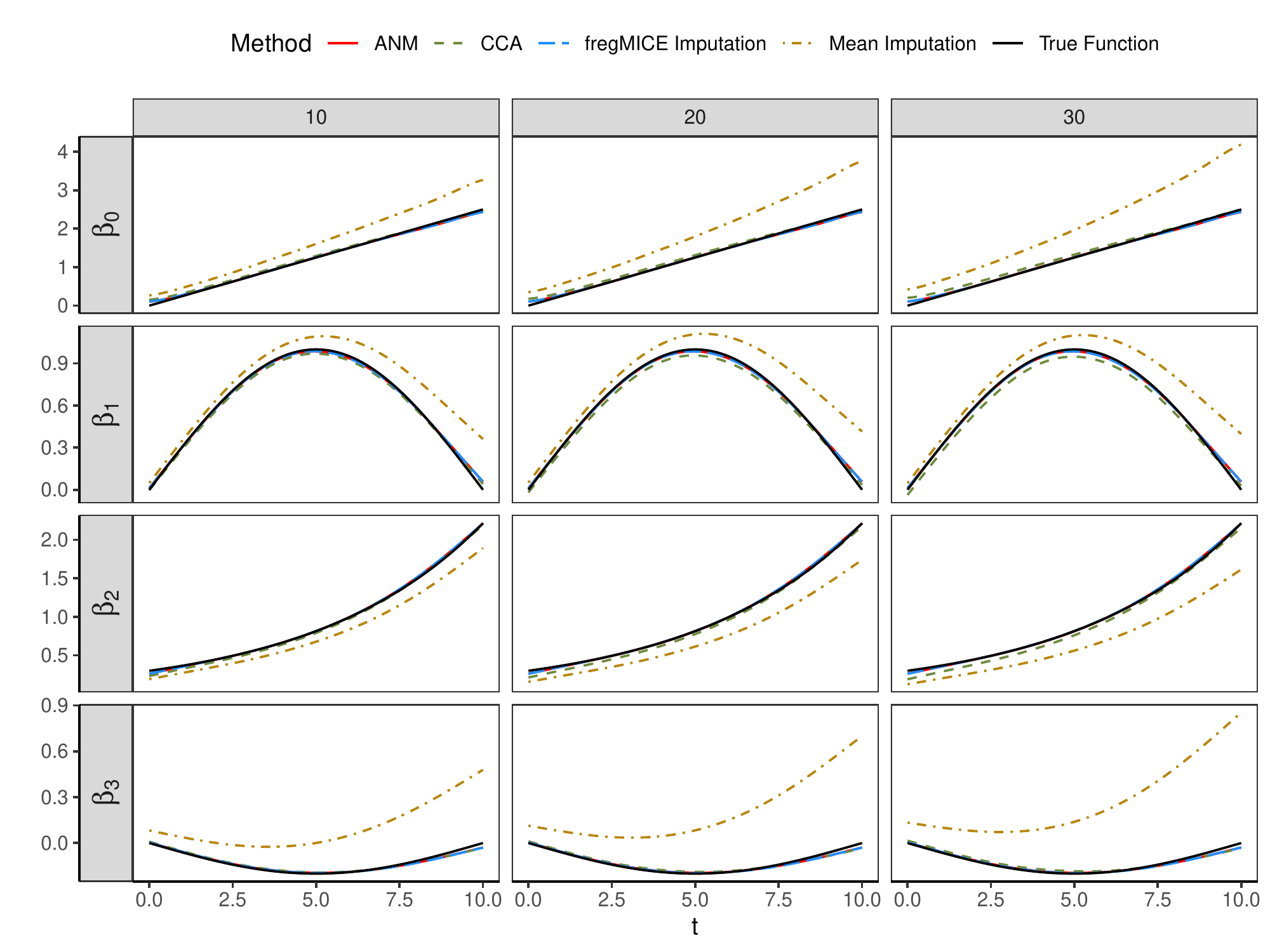}
  \includegraphics[width=0.85\linewidth]{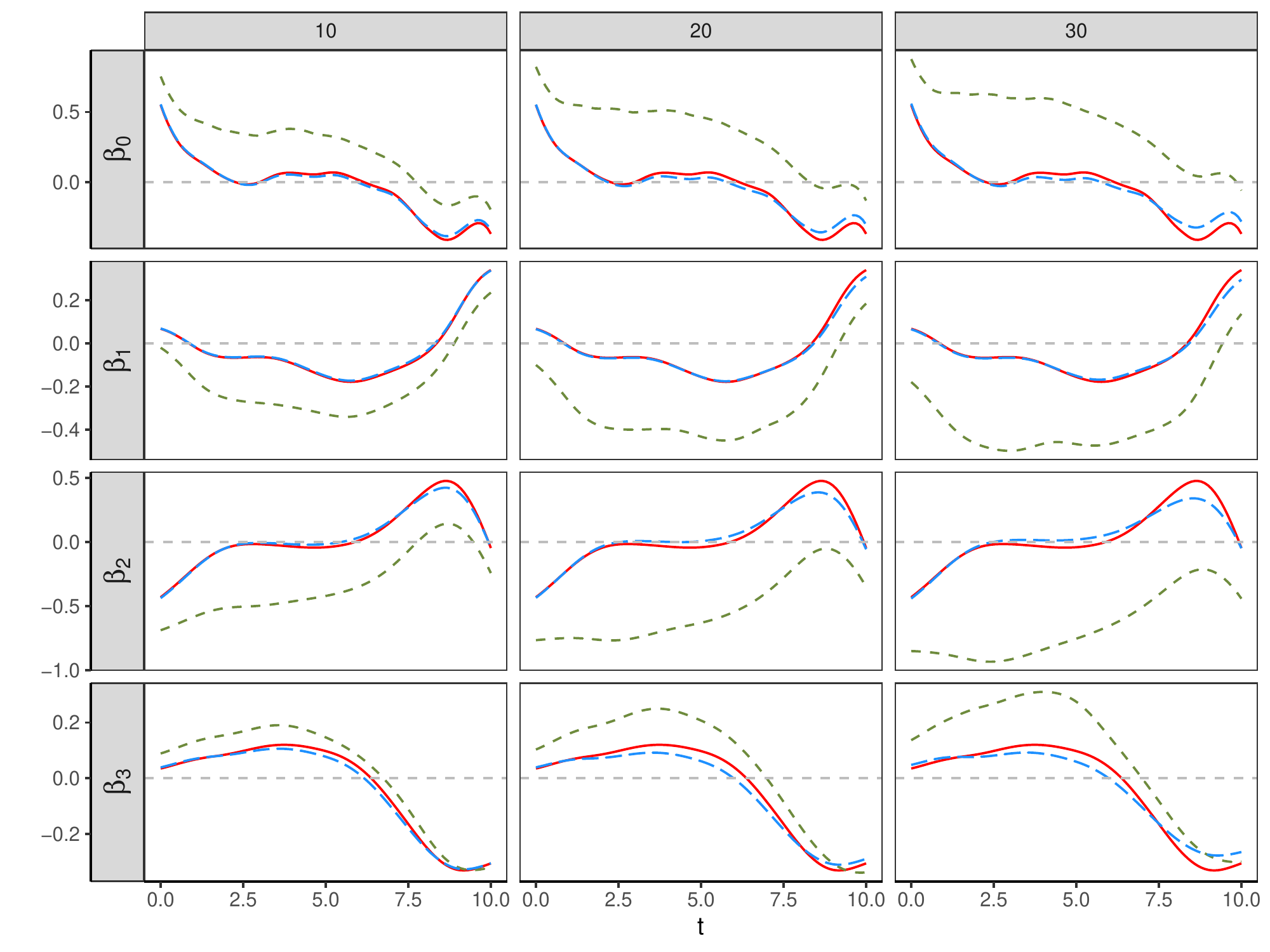}
  \caption{(\textbf{Top}) Point-wise mean curves and (\textbf{Bottom}) point-wise standardized bias in Setting 1 Scenario (a) for $n = 100$.  Columns (left to right) correspond to 10, 20, and 30\% missing data.  Rows (top to bottom) correspond to functional parameters $\beta_0, \beta_1, \beta_2,$ and $\beta_3$.  Point-wise standardized bias curve for mean imputation is removed to better compare estimates from ANM, CCA, and fregMICE.}
  \label{fig:sim_pwm_1a100}
\end{figure}

\begin{figure}
\centering
  \includegraphics[width=0.85\linewidth]{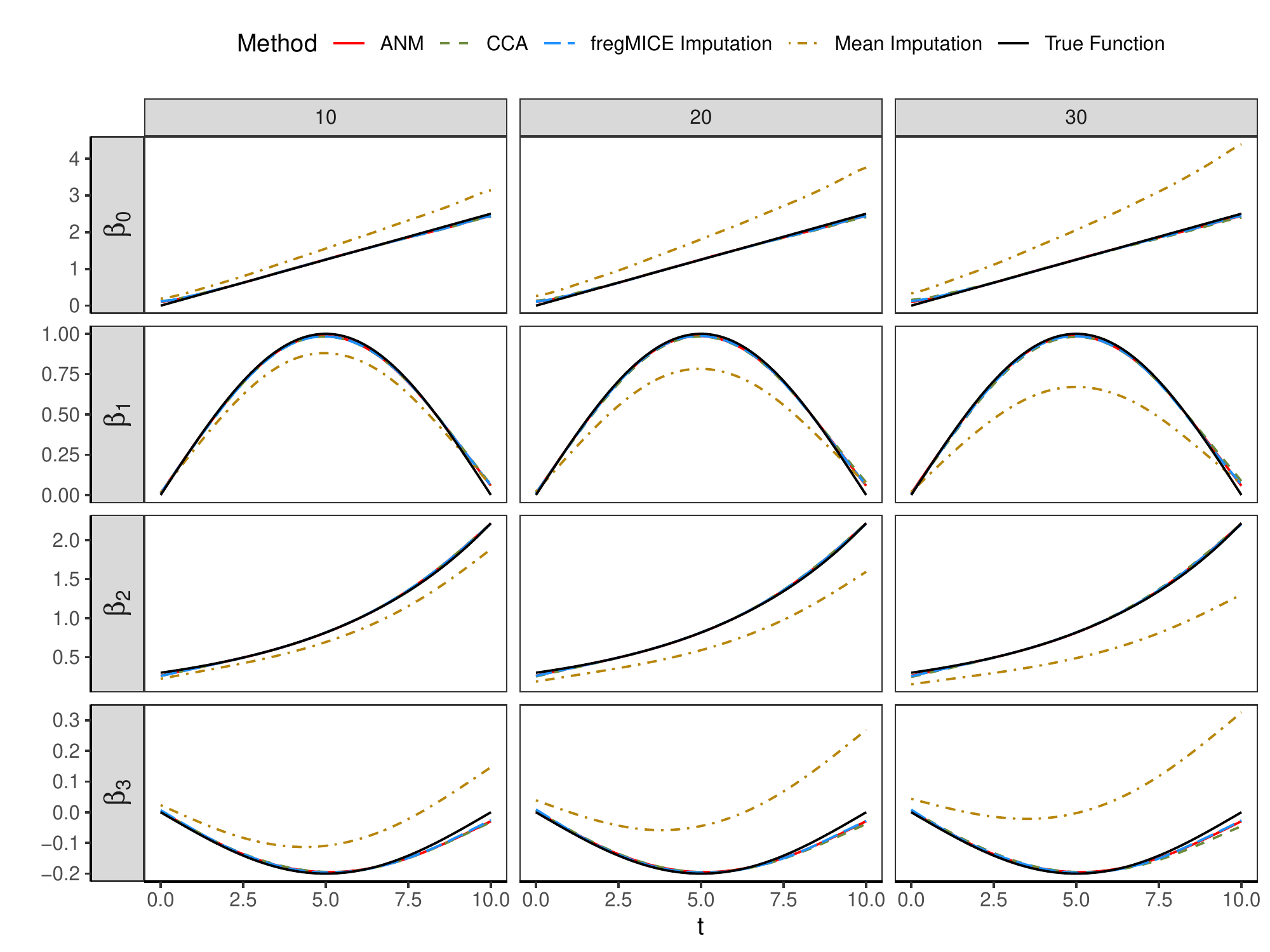}
  \includegraphics[width=0.85\linewidth]{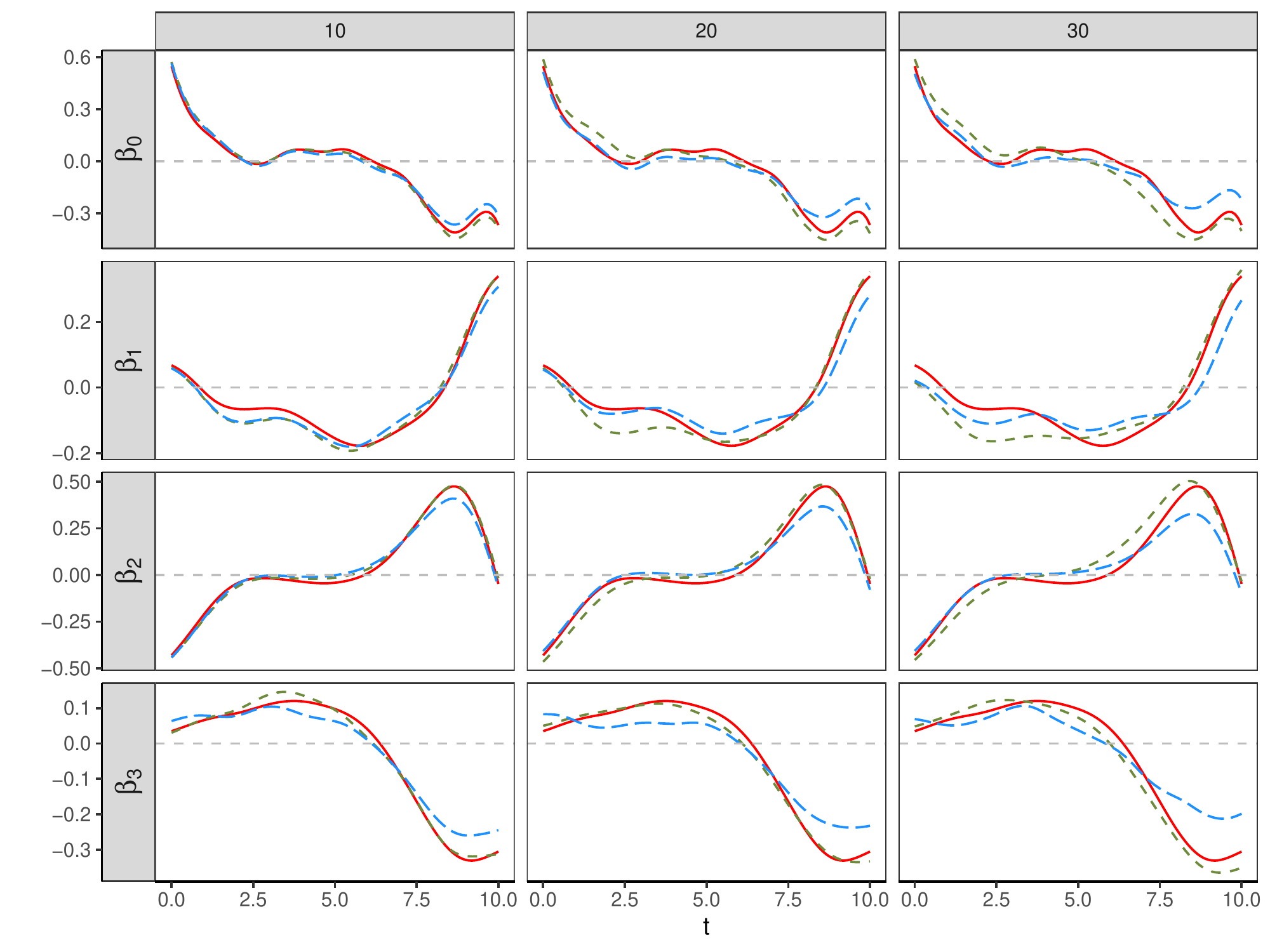}
  \caption{(\textbf{Top}) Point-wise mean curves and (\textbf{Bottom}) point-wise standardized bias in Setting 1 Scenario (b) for $n = 100$.  Columns (left to right) correspond to 10, 20, and 30\% missing data.  Rows (top to bottom) correspond to functional parameters $\beta_0, \beta_1, \beta_2,$ and $\beta_3$.  Point-wise standardized bias curve for mean imputation is removed to better compare estimates from ANM, CCA, and fregMICE.}
  \label{fig:sim_pwm_1b100}
\end{figure}

Figures \ref{fig:sim_pwm_1a100} and \ref{fig:sim_pwm_1b100} show point-wise mean and point-wise standardized bias plots for simulation settings presented in the main article, but with a smaller sample size of $n = 100$.  (ANM = all no missing; CCA = complete case analysis; fregMICE = functional regression MICE)  Overall, we see that the performance of the estimates derived from the fregMICE-imputed data, relative the the other approaches for handling missing data, is similar to that seen in the main article for the larger sample size of $n = 350$.   

\subsubsection{Point-wise 95\% Confidence Band Coverage and Width}

\begin{figure}[!htbp]
\centering
  \includegraphics[width=0.85\linewidth]{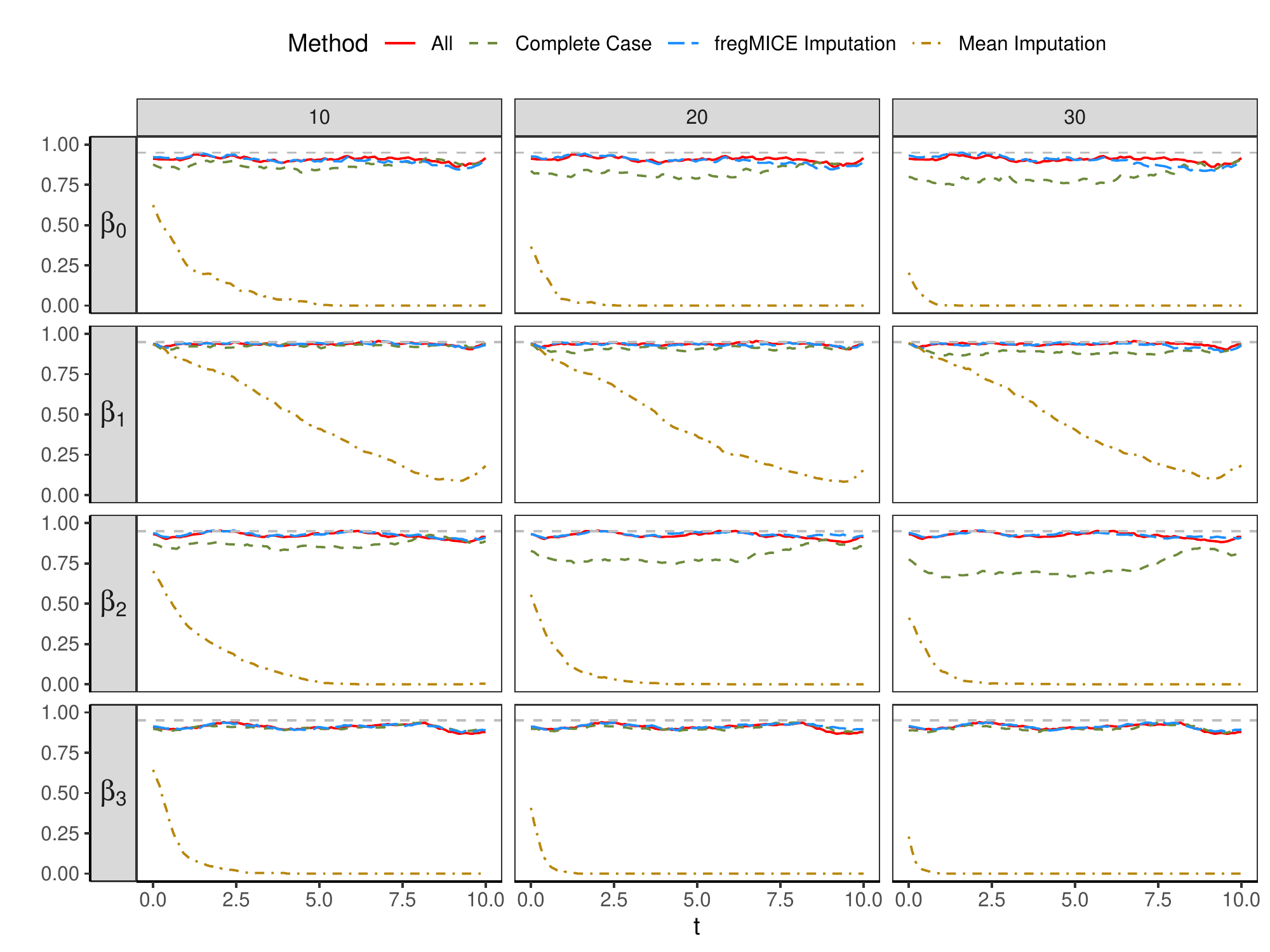}
  \includegraphics[width=0.85\linewidth]{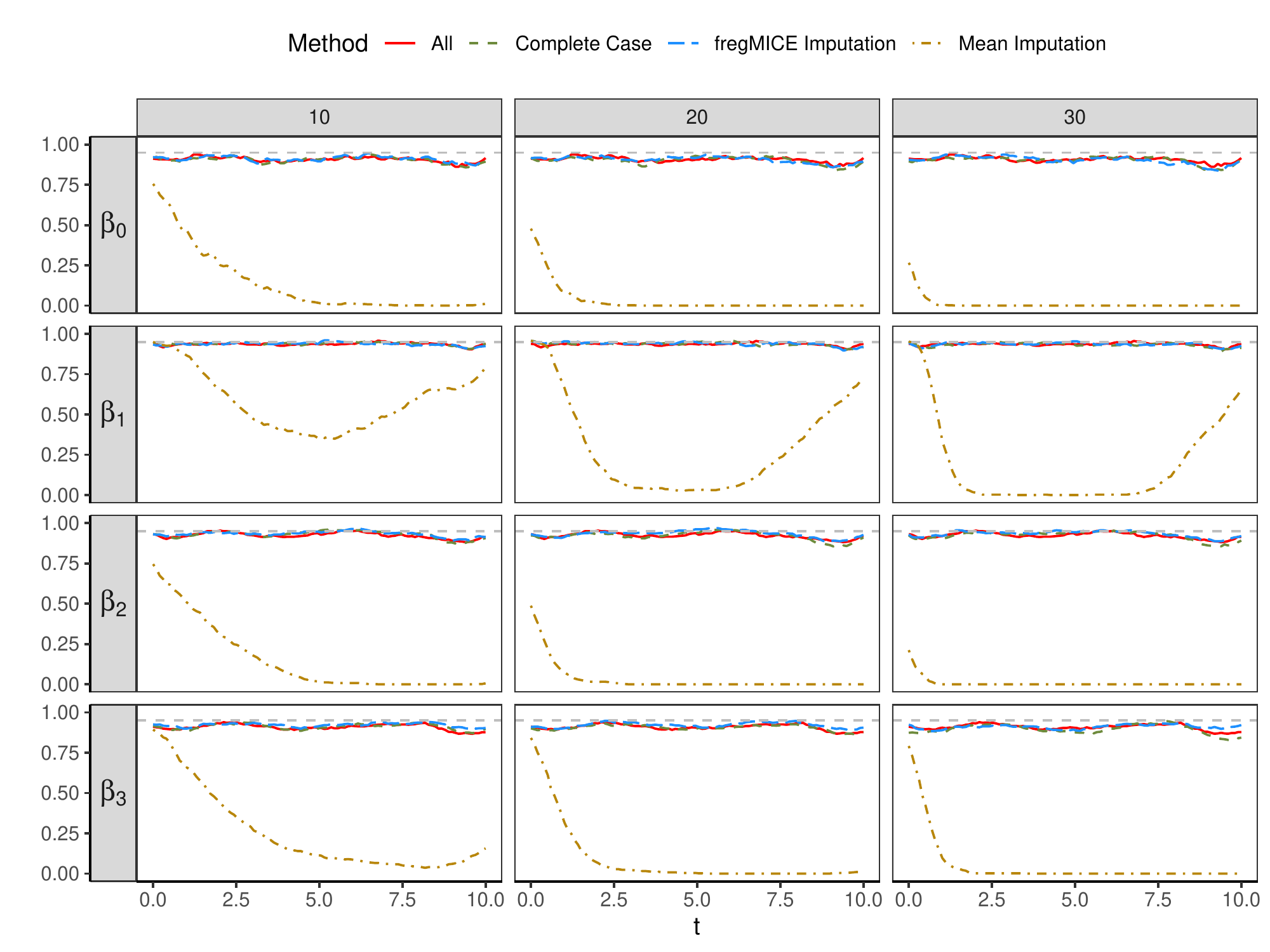}
  \caption{Point-wise 95\% confidence band coverage for Setting 1 Scenarios (a) (\textbf{Top}) and (b) (\textbf{Bottom}) with $n = 350$.  Columns (left to right) correspond to 10, 20, and 30\% missing data.  Rows (top to bottom) correspond to functional parameters $\beta_0, \beta_1, \beta_2,$ and $\beta_3$.}
  \label{fig:pwcov_2ab350}
\end{figure}

\begin{figure}[!htbp]
\centering
  \includegraphics[width=0.85\linewidth]{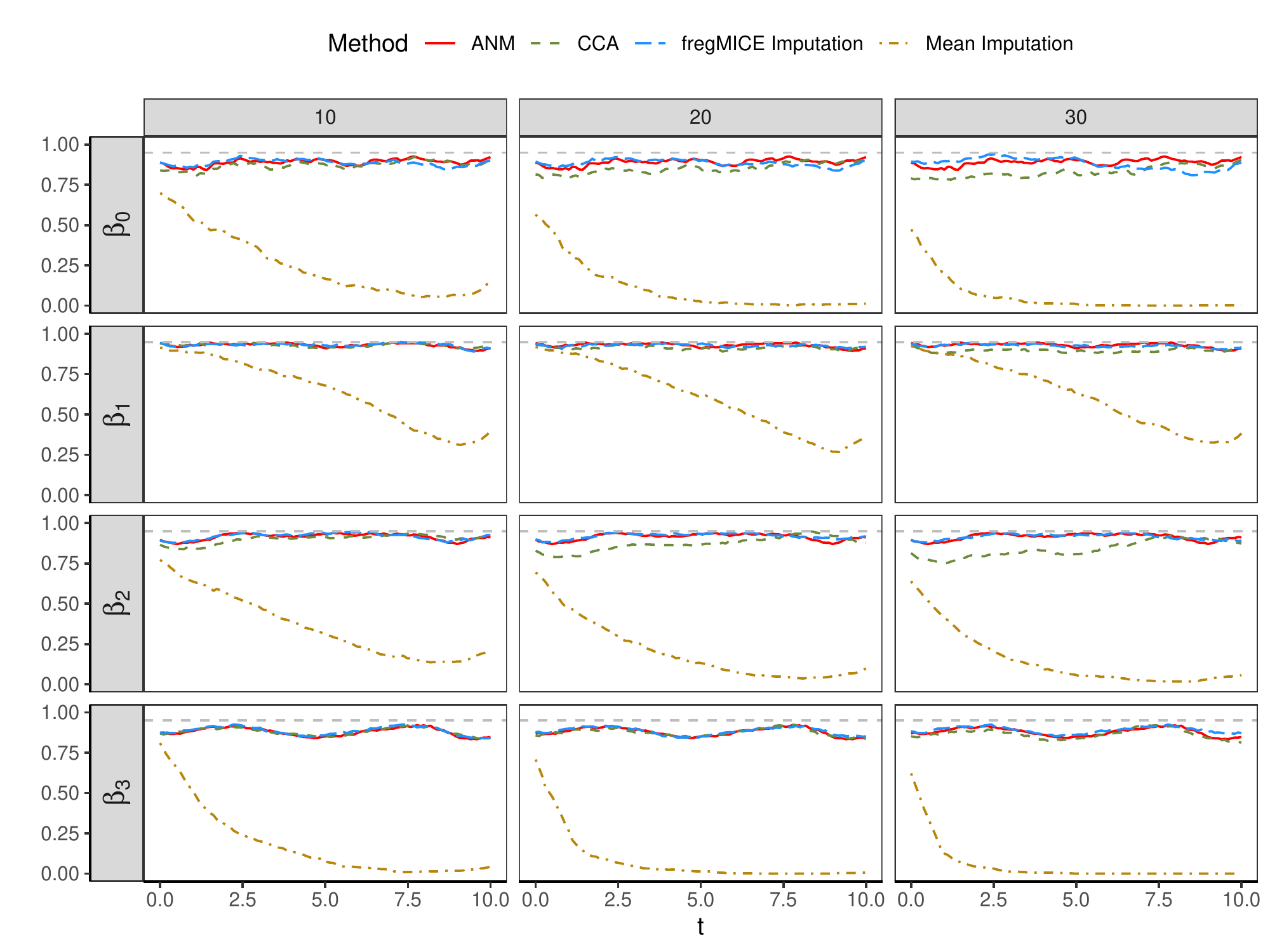}
  \includegraphics[width=0.85\linewidth]{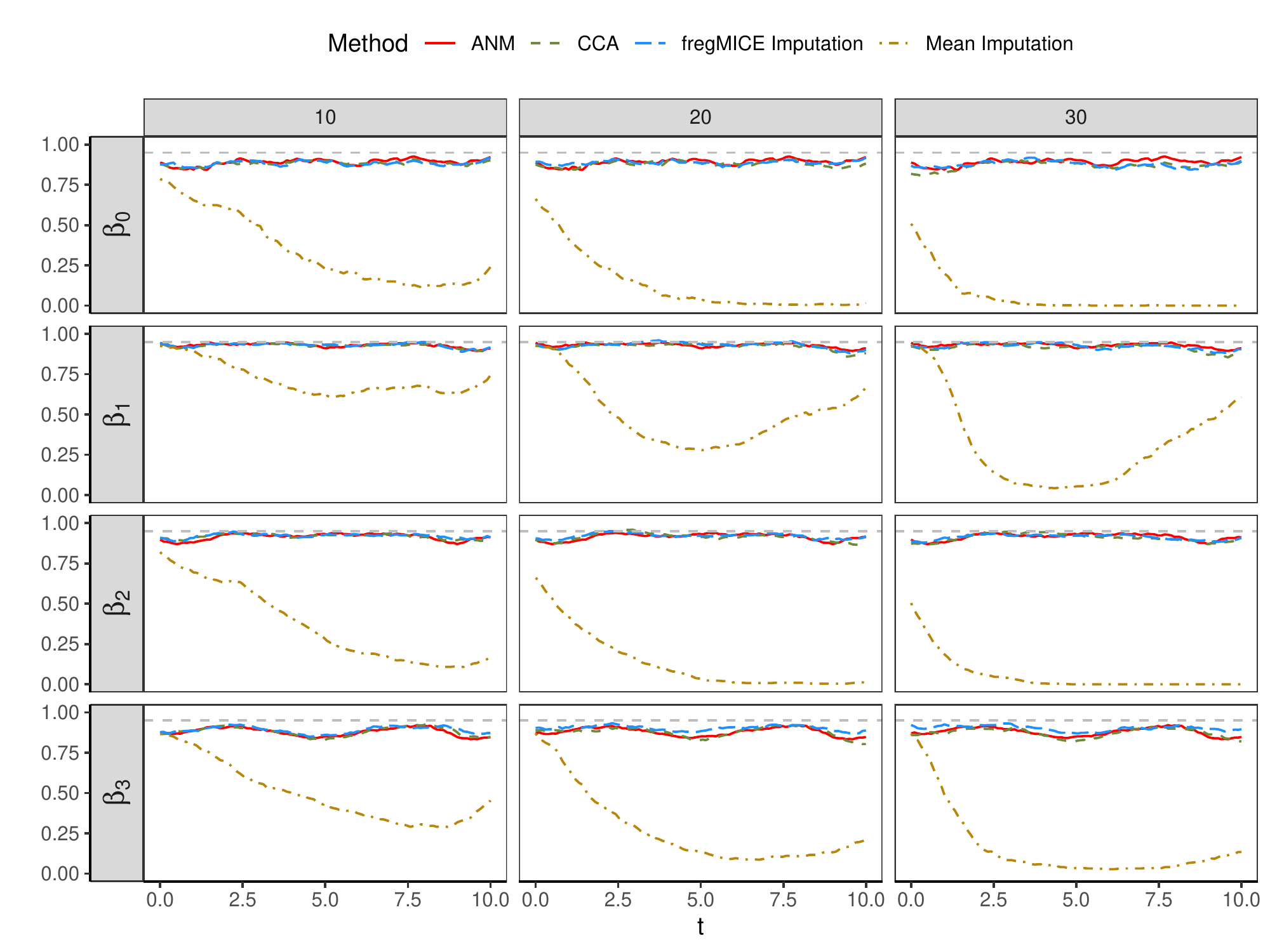}
  \caption{Point-wise 95\% confidence band coverage for Setting 1 Scenarios (a) (\textbf{Top}) and (b) (\textbf{Bottom}) with $n = 100$.  Columns (left to right) correspond to 10, 20, and 30\% missing data.  Rows (top to bottom) correspond to functional parameters $\beta_0, \beta_1, \beta_2,$ and $\beta_3$.}
  \label{fig:pwcov_2ab100}
\end{figure}

\begin{figure}[!htbp]
\centering
  \includegraphics[width=0.85\linewidth]{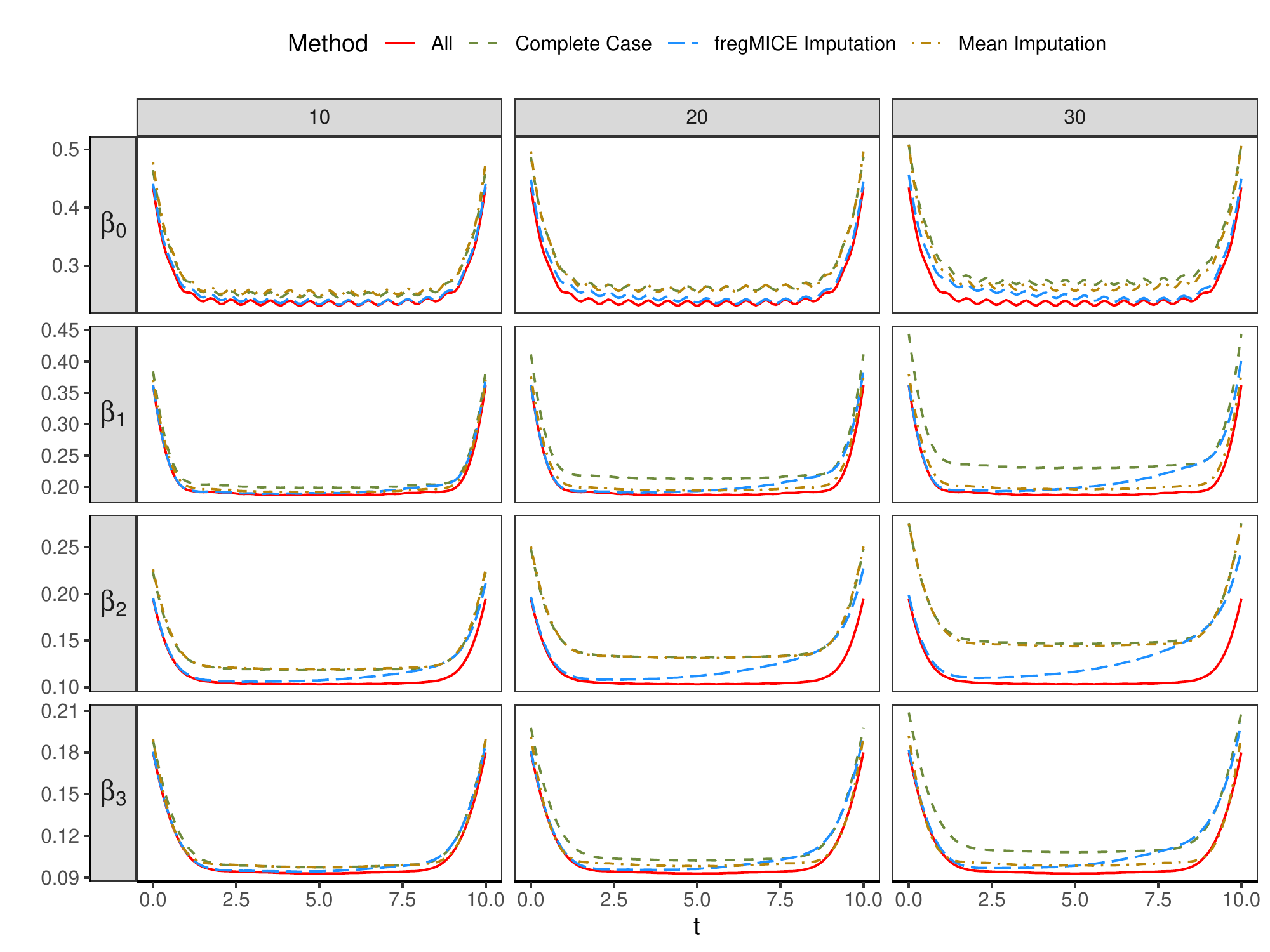}
  \includegraphics[width=0.85\linewidth]{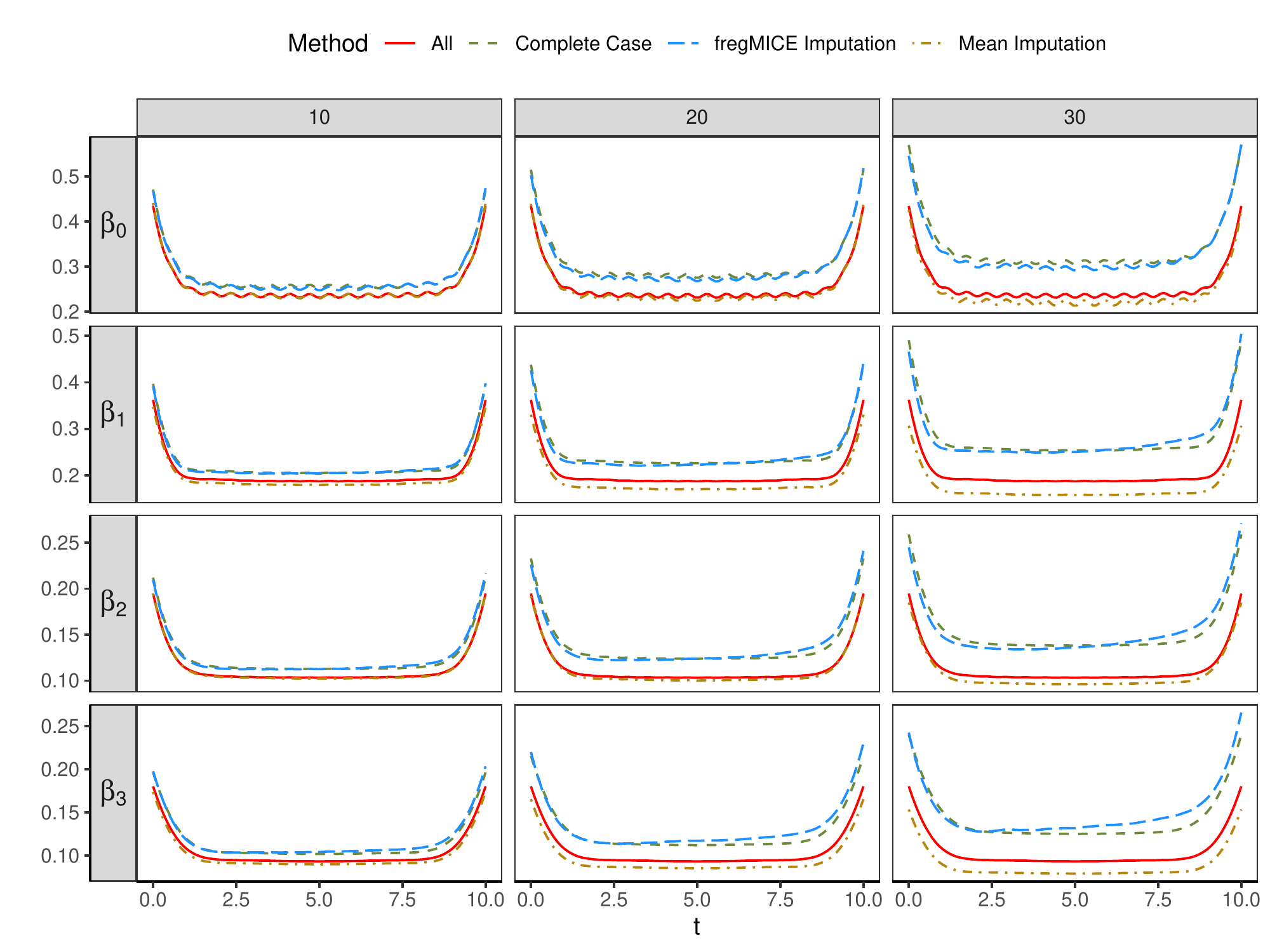}
  \caption{Point-wise 95\% confidence band width for Setting 1 Scenarios (a) (\textbf{Top}) and (b) (\textbf{Bottom}) with $n = 350$.  Columns (left to right) correspond to 10, 20, and 30\% missing data.  Rows (top to bottom) correspond to functional parameters $\beta_0, \beta_1, \beta_2,$ and $\beta_3$.}
  \label{fig:pww_2ab350}
\end{figure}

\begin{figure}[!htbp]
\centering
  \includegraphics[width=0.85\linewidth]{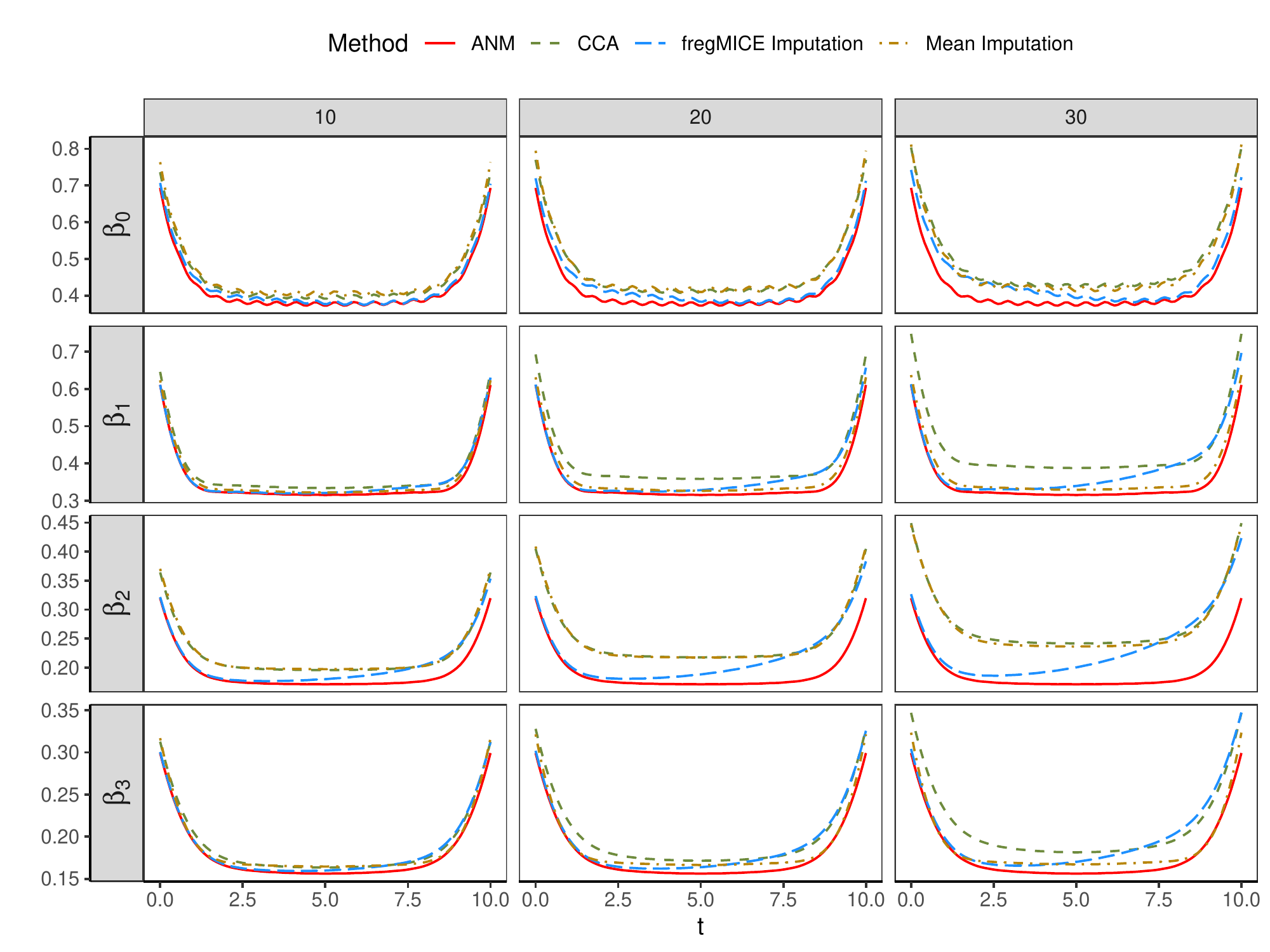}
  \includegraphics[width=0.85\linewidth]{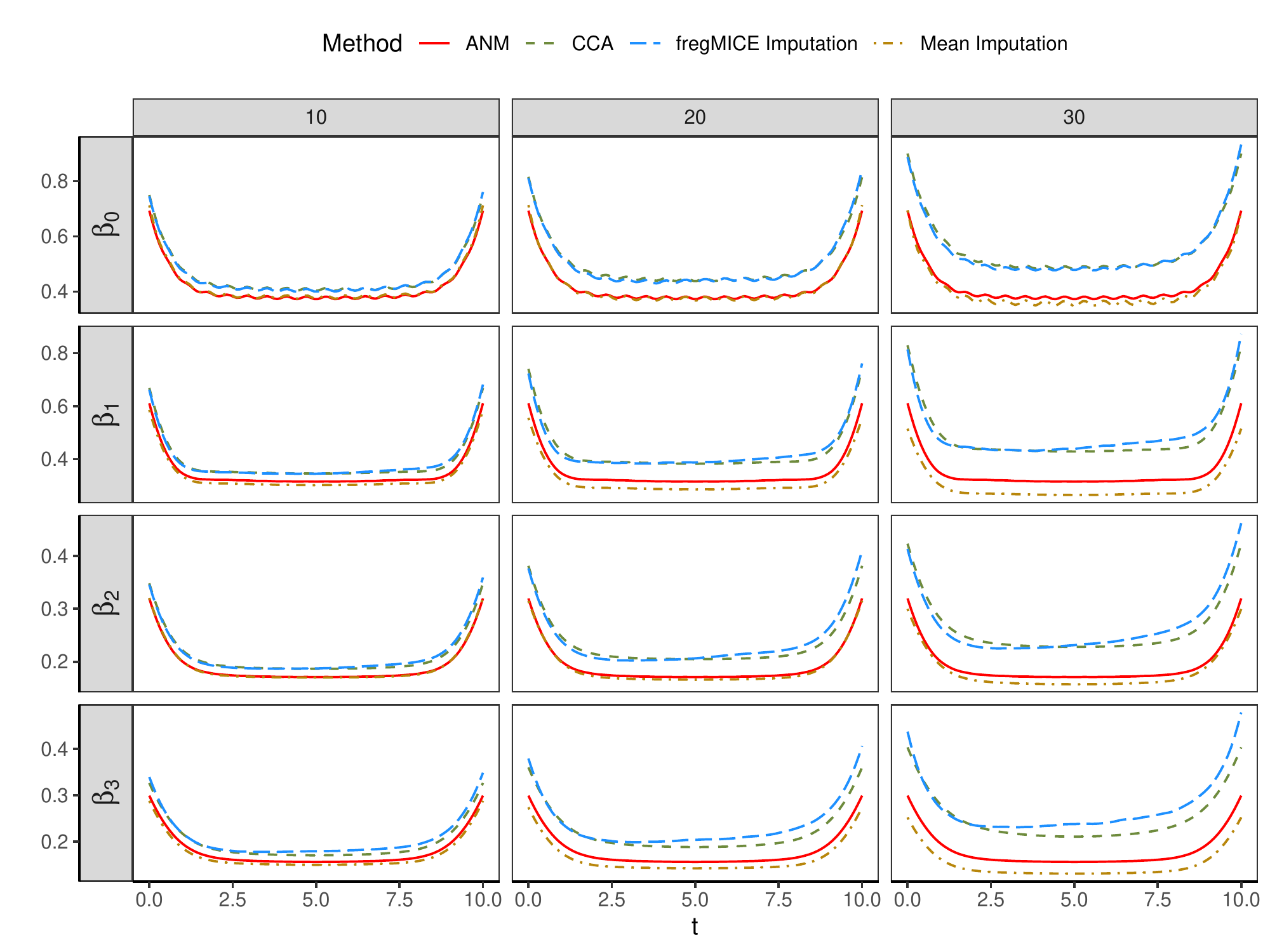}
  \caption{Point-wise 95\% confidence band width for Setting 1 Scenarios (a) (\textbf{Top}) and (b) (\textbf{Bottom}) with $n = 100$.  Columns (left to right) correspond to 10, 20, and 30\% missing data.  Rows (top to bottom) correspond to functional parameters $\beta_0, \beta_1, \beta_2,$ and $\beta_3$.}
  \label{fig:pww_2ab100}
\end{figure}

Figures \ref{fig:pwcov_2ab350} and \ref{fig:pwcov_2ab100} show the coverage probability for the estimated 95\% confidence bands at each point on $[0,10]$ for Setting 1 Scenarios (a) and (b) with $n = 350$ and $n = 100$ respectively, over the 500 simulation runs for each method of handling the missing data.  Figures \ref{fig:pww_2ab350} and \ref{fig:pww_2ab100} show the corresponding mean point-wise widths for the 95\% confidence bands.

\newpage 

\subsection{Setting 2 Additional Results}  \label{sim_sec_02}

\subsubsection{Point-wise Mean and Point-wise Standardized Bias Plots for $n = 100$}

\begin{figure}
\centering
  \includegraphics[width=0.85\linewidth]{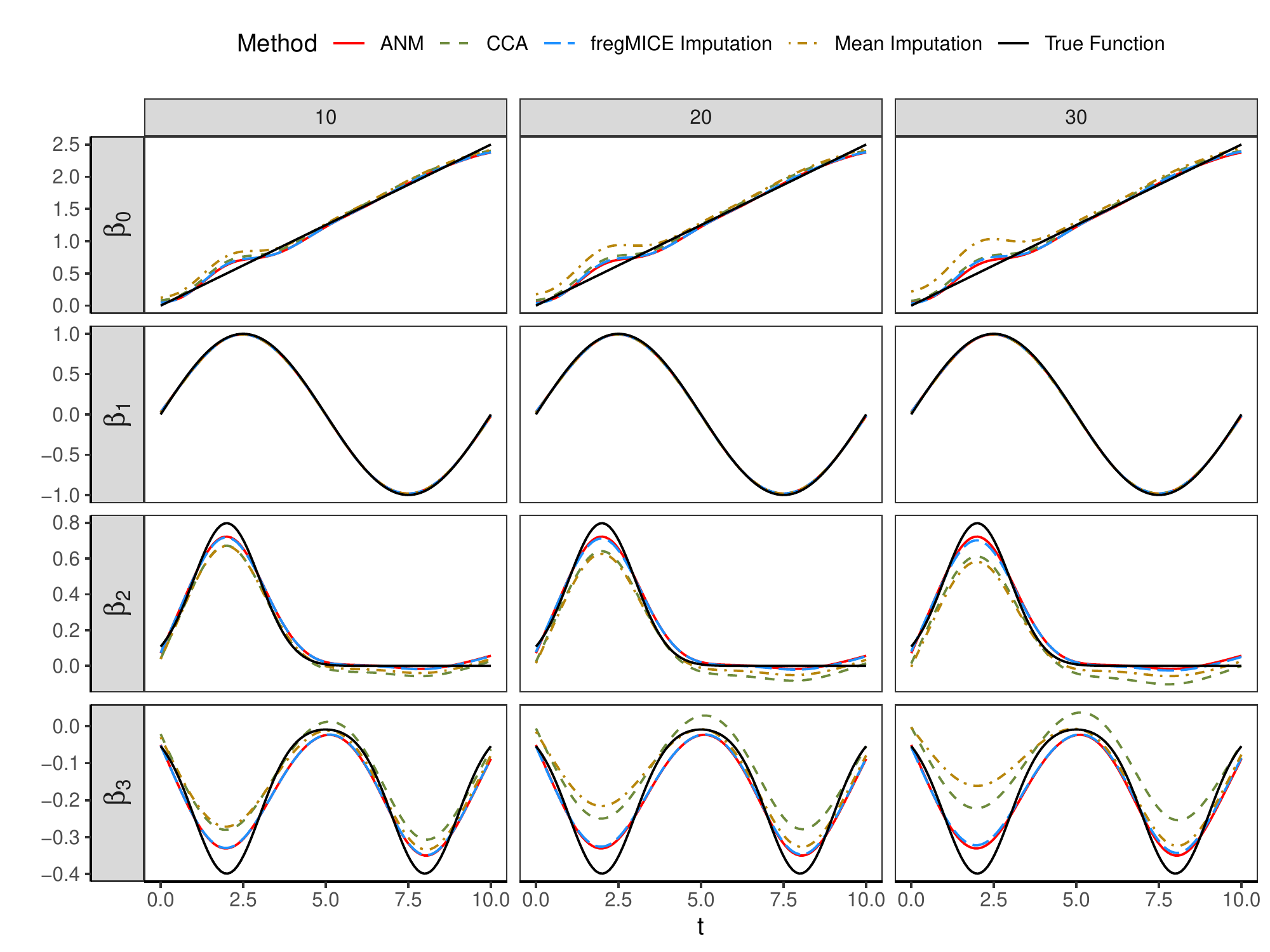}
  \includegraphics[width=0.85\linewidth]{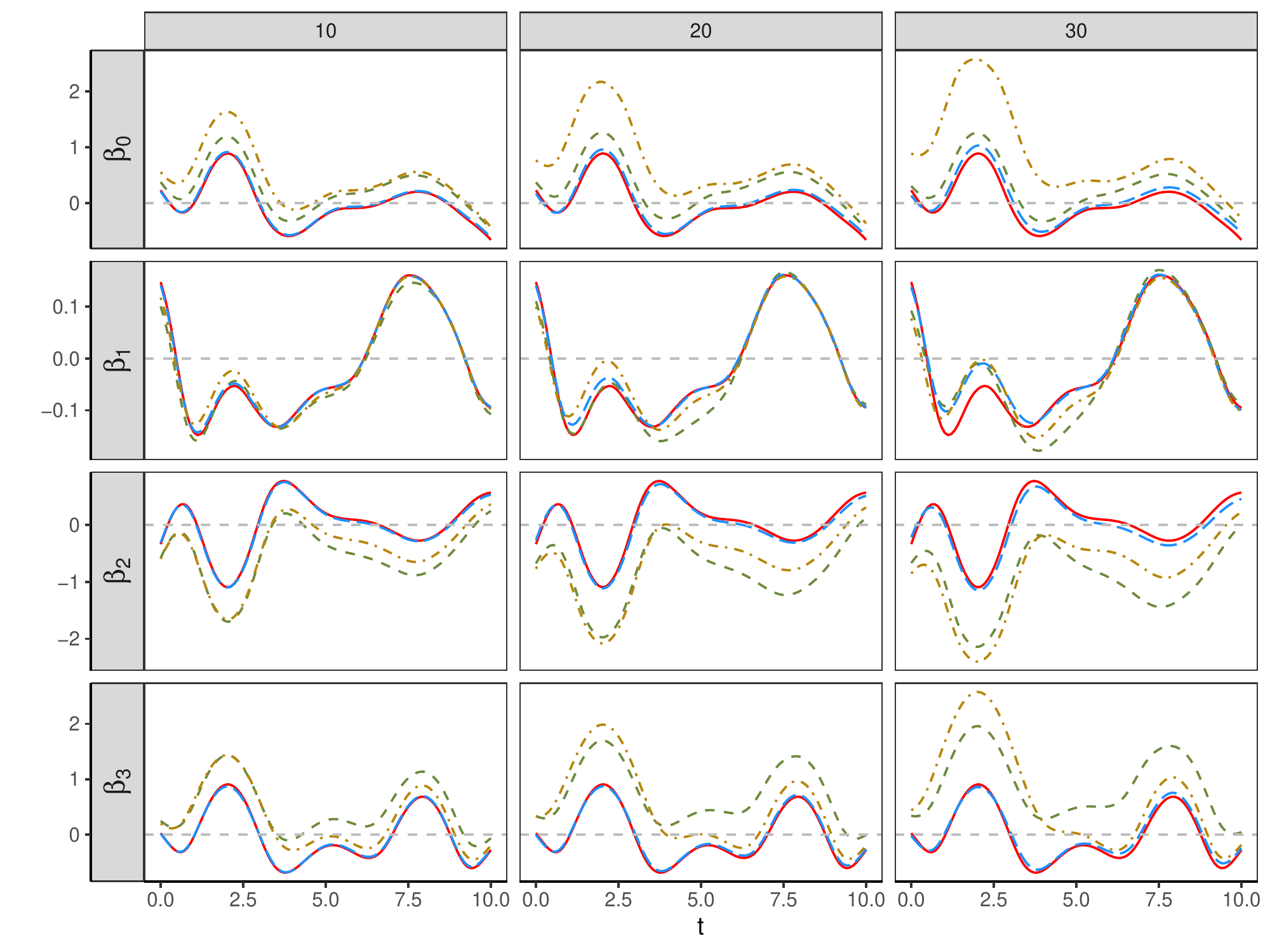}
  \caption{(\textbf{Top}) Point-wise mean curves and (\textbf{Bottom}) point-wise standardized bias in Setting 2 Scenario (a) for $n = 100$.  Columns (left to right) correspond to 10, 20, and 30\% missing data.  Rows (top to bottom) correspond to functional parameters $\beta_0, \beta_1, \beta_2,$ and $\beta_3$.  Point-wise standardized bias curve for mean imputation is removed to better compare estimates from ANM, CCA, and fregMICE.}
  \label{fig:sim_pwm_2a100}
\end{figure}

\begin{figure}
\centering
  \includegraphics[width=0.85\linewidth]{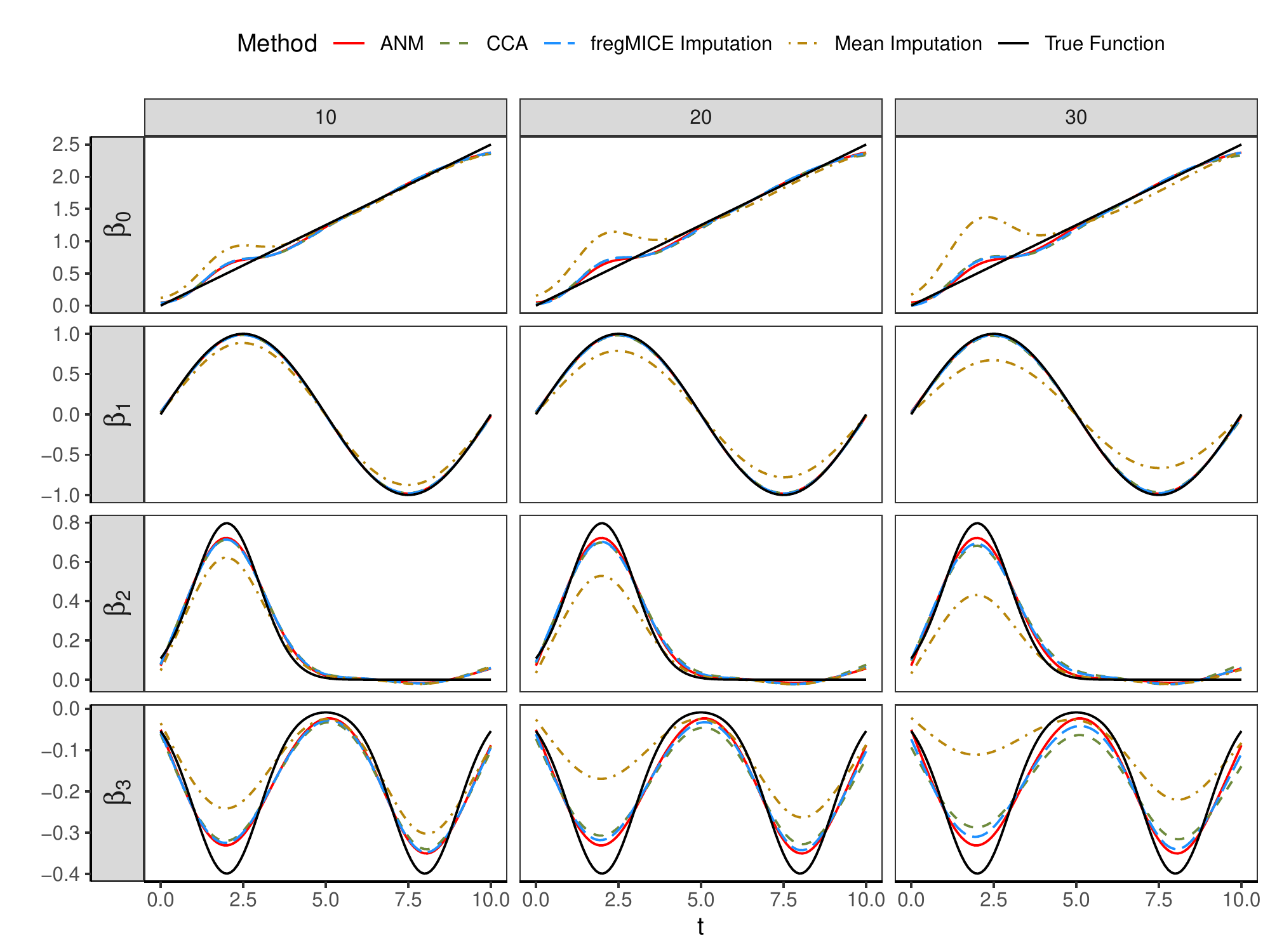}
  \includegraphics[width=0.85\linewidth]{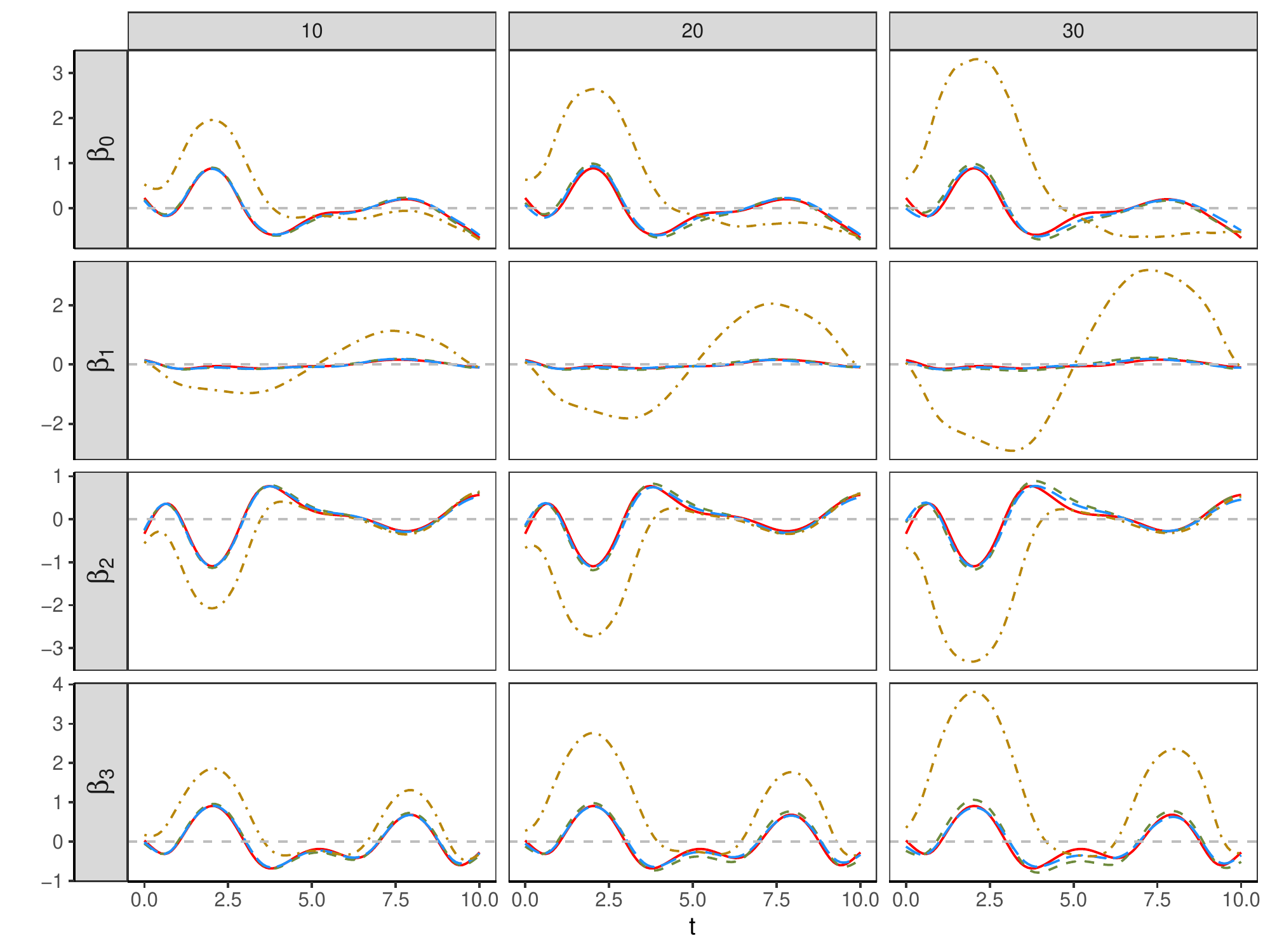}
  \caption{(\textbf{Top}) Point-wise mean curves and (\textbf{Bottom}) point-wise standardized bias in Setting 2 Scenario (b) for $n = 100$.  Columns (left to right) correspond to 10, 20, and 30\% missing data.  Rows (top to bottom) correspond to functional parameters $\beta_0, \beta_1, \beta_2,$ and $\beta_3$.  Point-wise standardized bias curve for mean imputation is removed to better compare estimates from ANM, CCA, and fregMICE.}
  \label{fig:sim_pwm_2b100}
\end{figure}

Figures \ref{fig:sim_pwm_2a100} and \ref{fig:sim_pwm_2b100} show point-wise mean and point-wise standardized bias plots for simulation settings presented in the main article, but with a smaller sample size of $n = 100$.  Overall, we see that the performance of the estimates derived from the fregMICE-imputed data, relative the the other approaches for handling missing data, is similar to that seen in the main article for the larger sample size of $n = 350$.   

\subsubsection{Point-wise 95\% Confidence Band Coverage and Width}

\begin{figure}[!htbp]
\centering
  \includegraphics[width=0.85\linewidth]{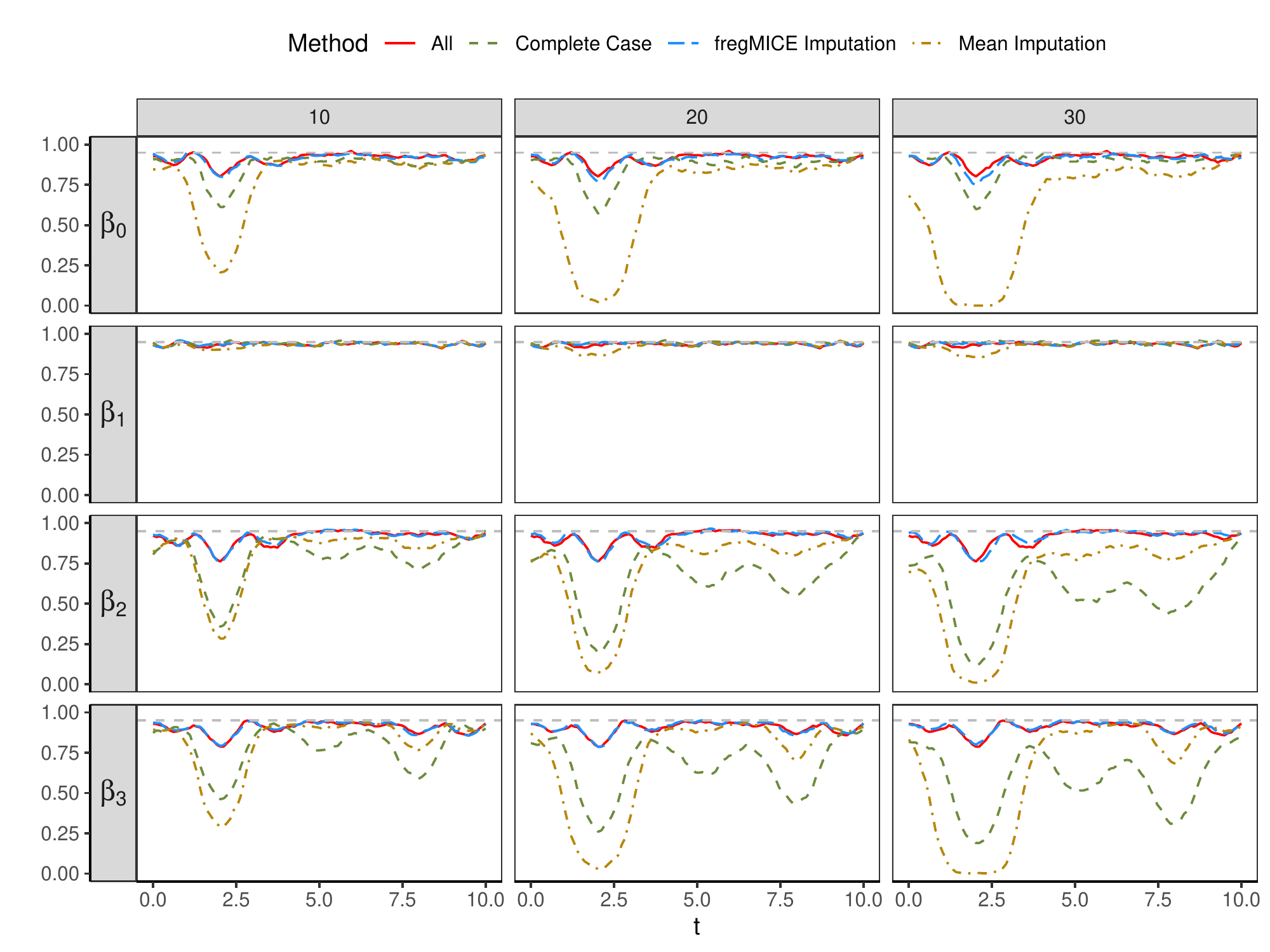}
  \includegraphics[width=0.85\linewidth]{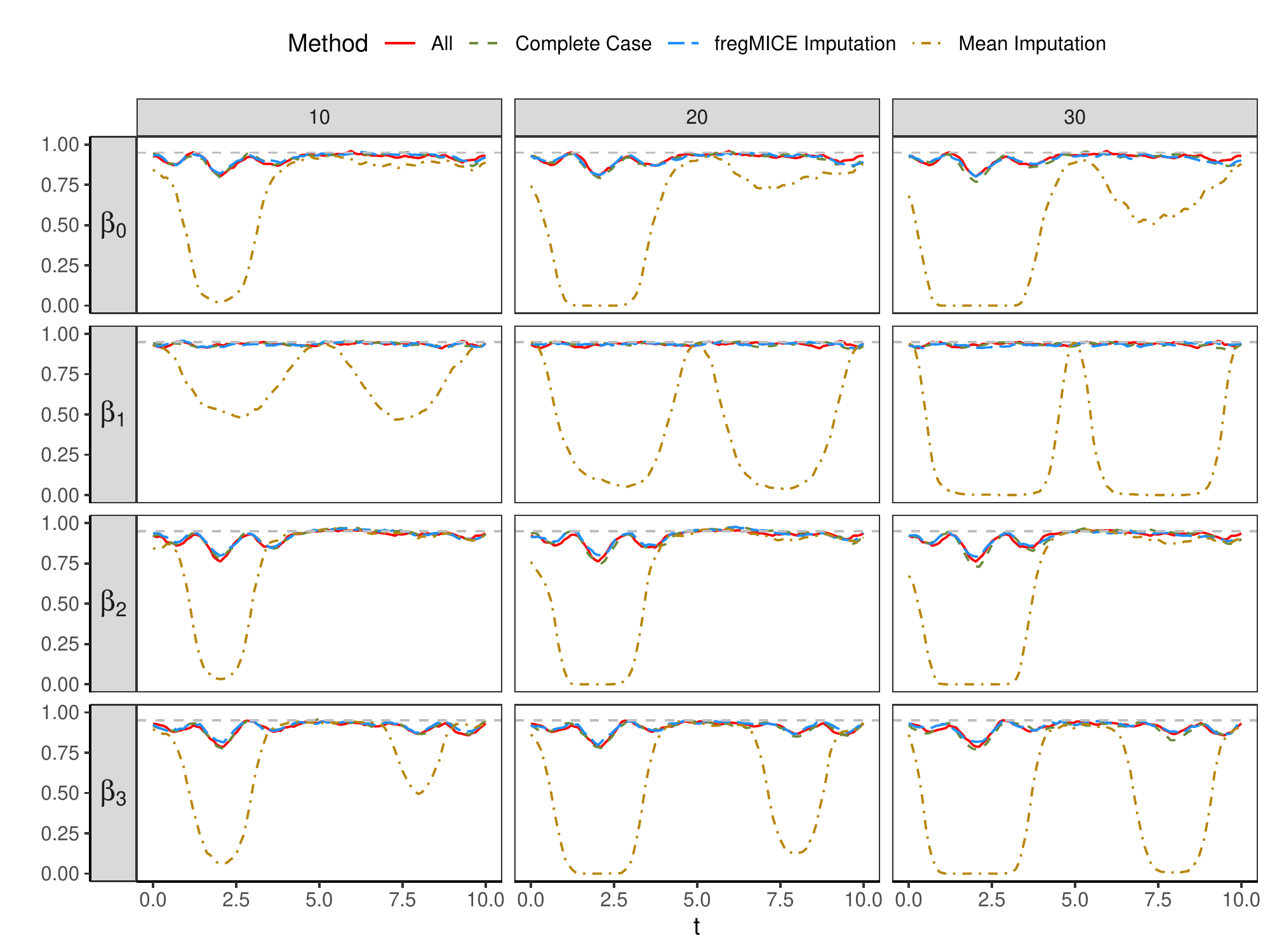}
  \caption{Point-wise 95\% confidence band coverage for Setting 2 Scenarios (a) (\textbf{Top}) and (b) (\textbf{Bottom}) with $n = 350$.  Columns (left to right) correspond to 10, 20, and 30\% missing data.  Rows (top to bottom) correspond to functional parameters $\beta_0, \beta_1, \beta_2,$ and $\beta_3$.}
  \label{fig:pwcov_1ab350}
\end{figure}

\begin{figure}[!htbp]
\centering
  \includegraphics[width=0.85\linewidth]{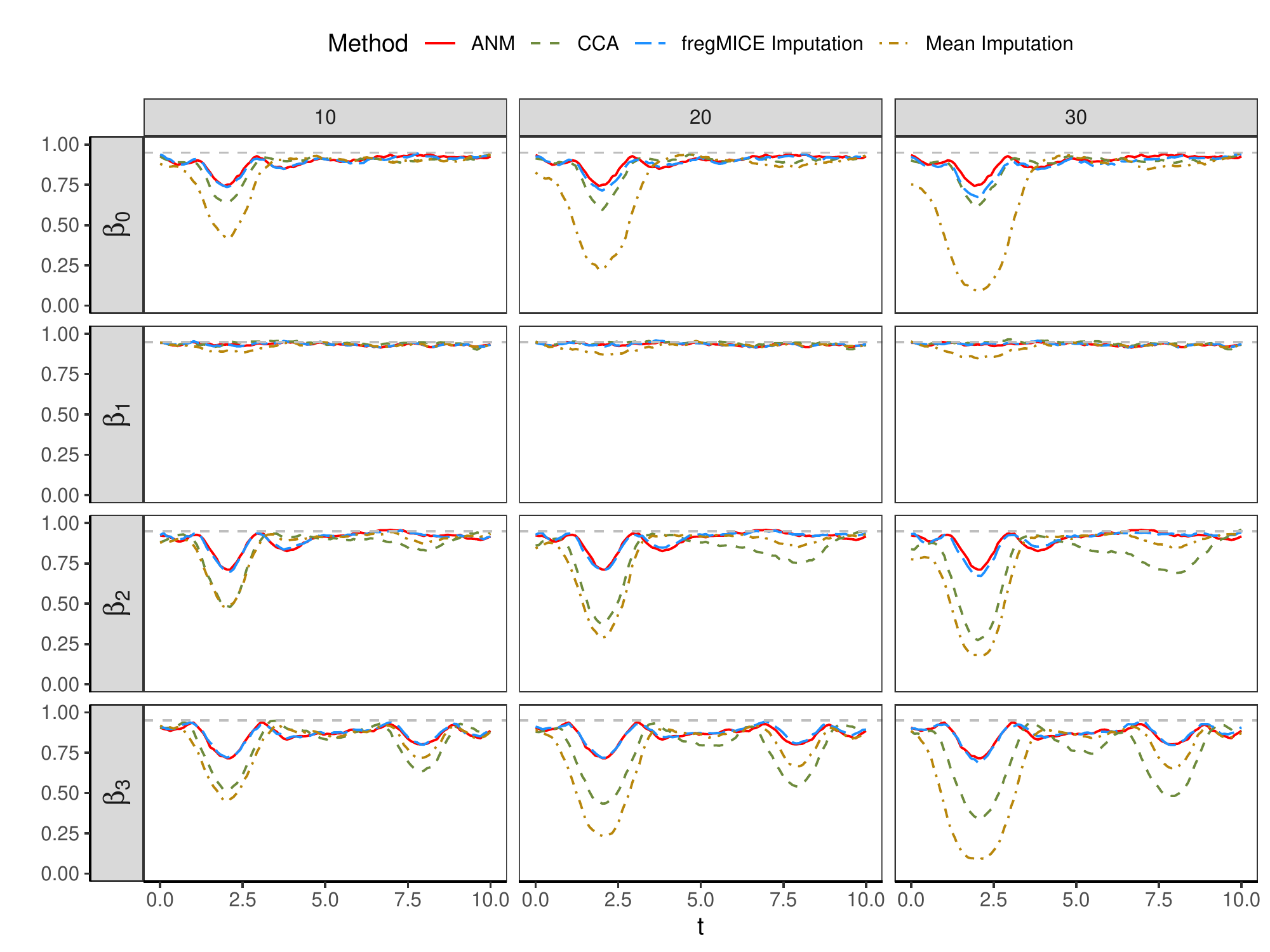}
  \includegraphics[width=0.85\linewidth]{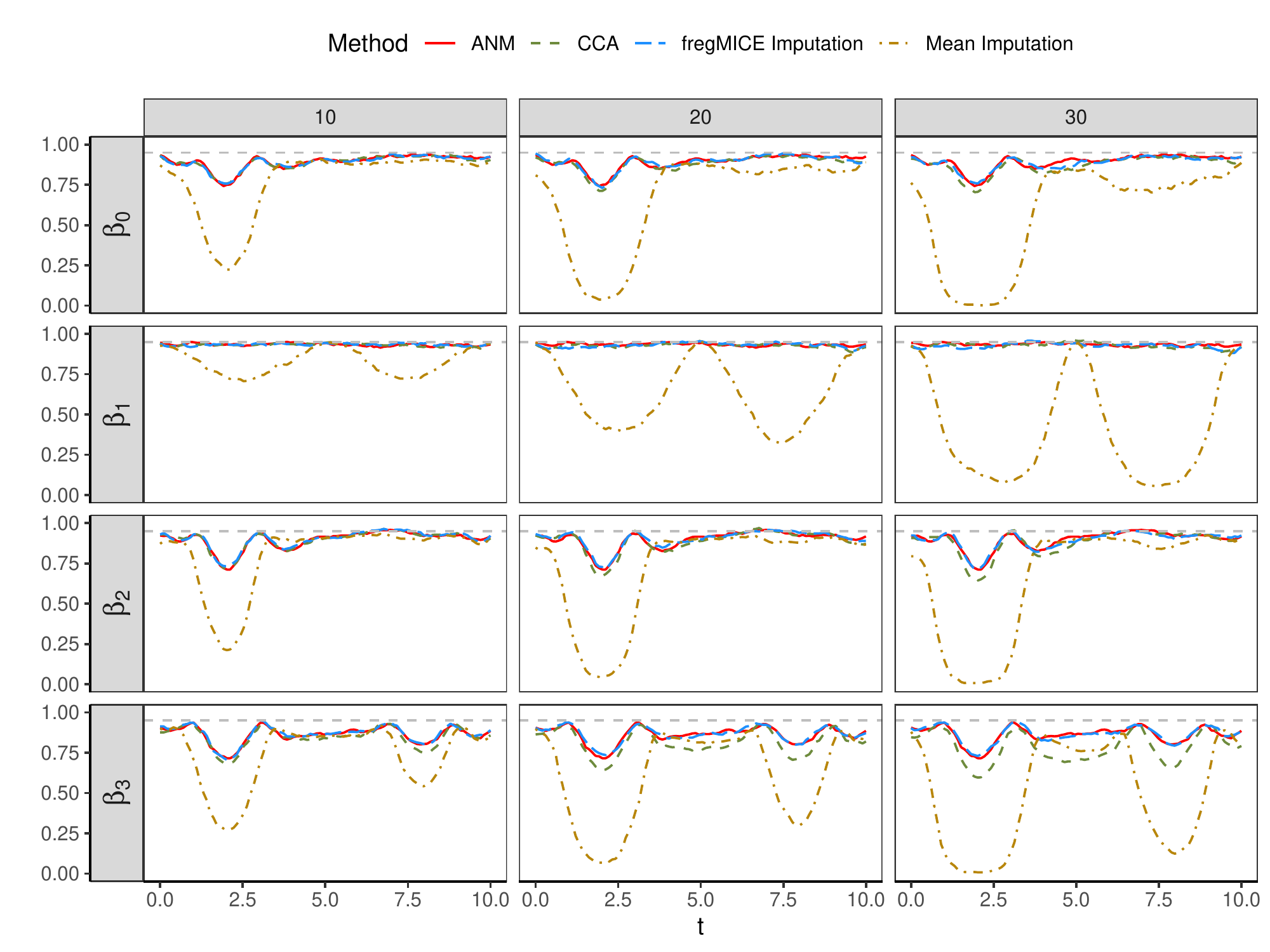}
  \caption{Point-wise 95\% confidence band coverage for Setting 2 Scenarios (a) (\textbf{Top}) and (b) (\textbf{Bottom}) with $n = 100$.  Columns (left to right) correspond to 10, 20, and 30\% missing data.  Rows (top to bottom) correspond to functional parameters $\beta_0, \beta_1, \beta_2,$ and $\beta_3$.}
  \label{fig:pwcov_1ab100}
\end{figure}

\begin{figure}[!htbp]
\centering
  \includegraphics[width=0.85\linewidth]{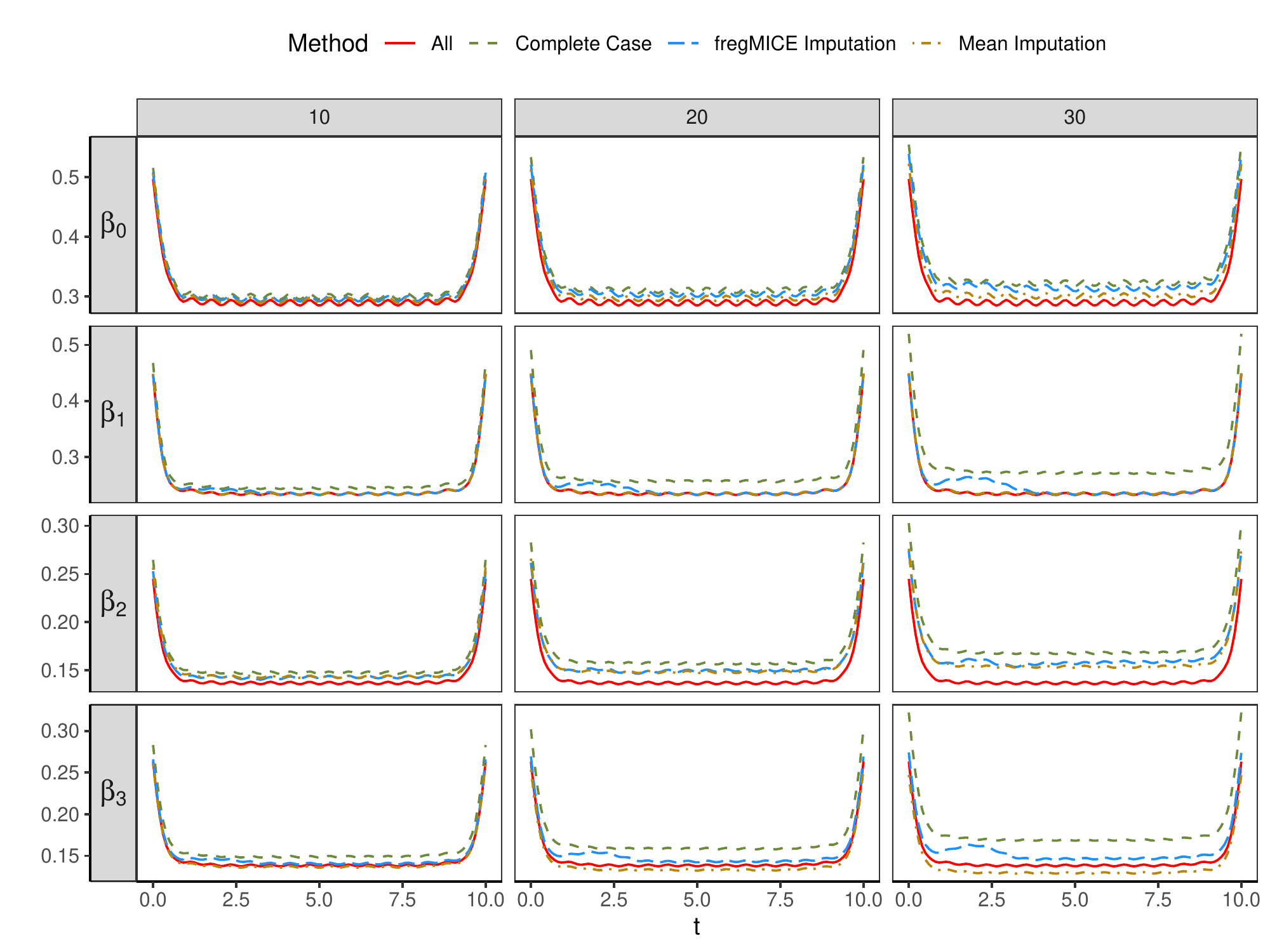}
  \includegraphics[width=0.85\linewidth]{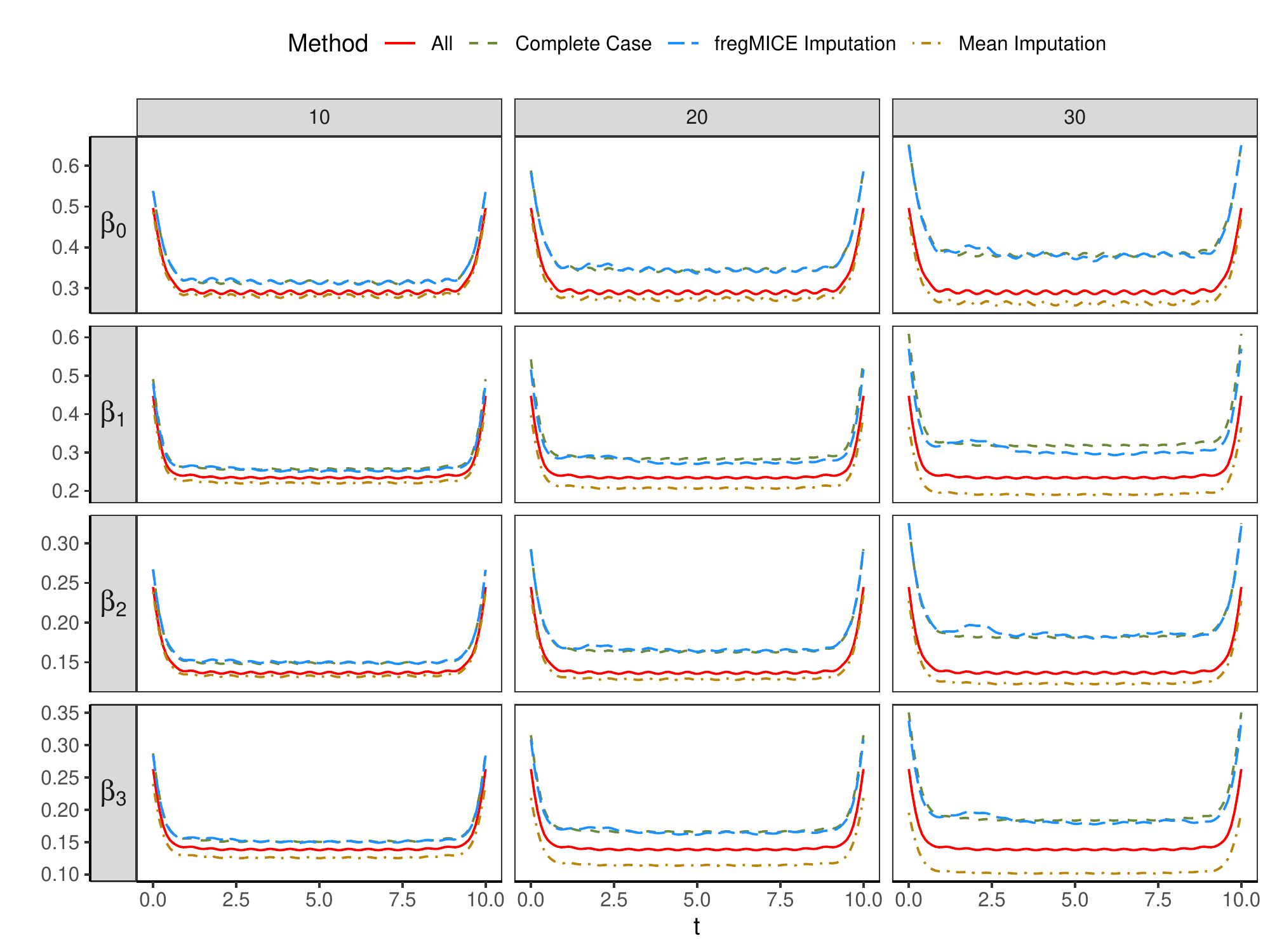}
  \caption{Point-wise 95\% confidence band width for Setting 1 Scenarios (a) (\textbf{Top}) and (b) (\textbf{Bottom}) with $n = 350$.  Columns (left to right) correspond to 10, 20, and 30\% missing data.  Rows (top to bottom) correspond to functional parameters $\beta_0, \beta_1, \beta_2,$ and $\beta_3$.}
  \label{fig:pww_1ab350}
\end{figure}

\begin{figure}[!htbp]
\centering
  \includegraphics[width=0.85\linewidth]{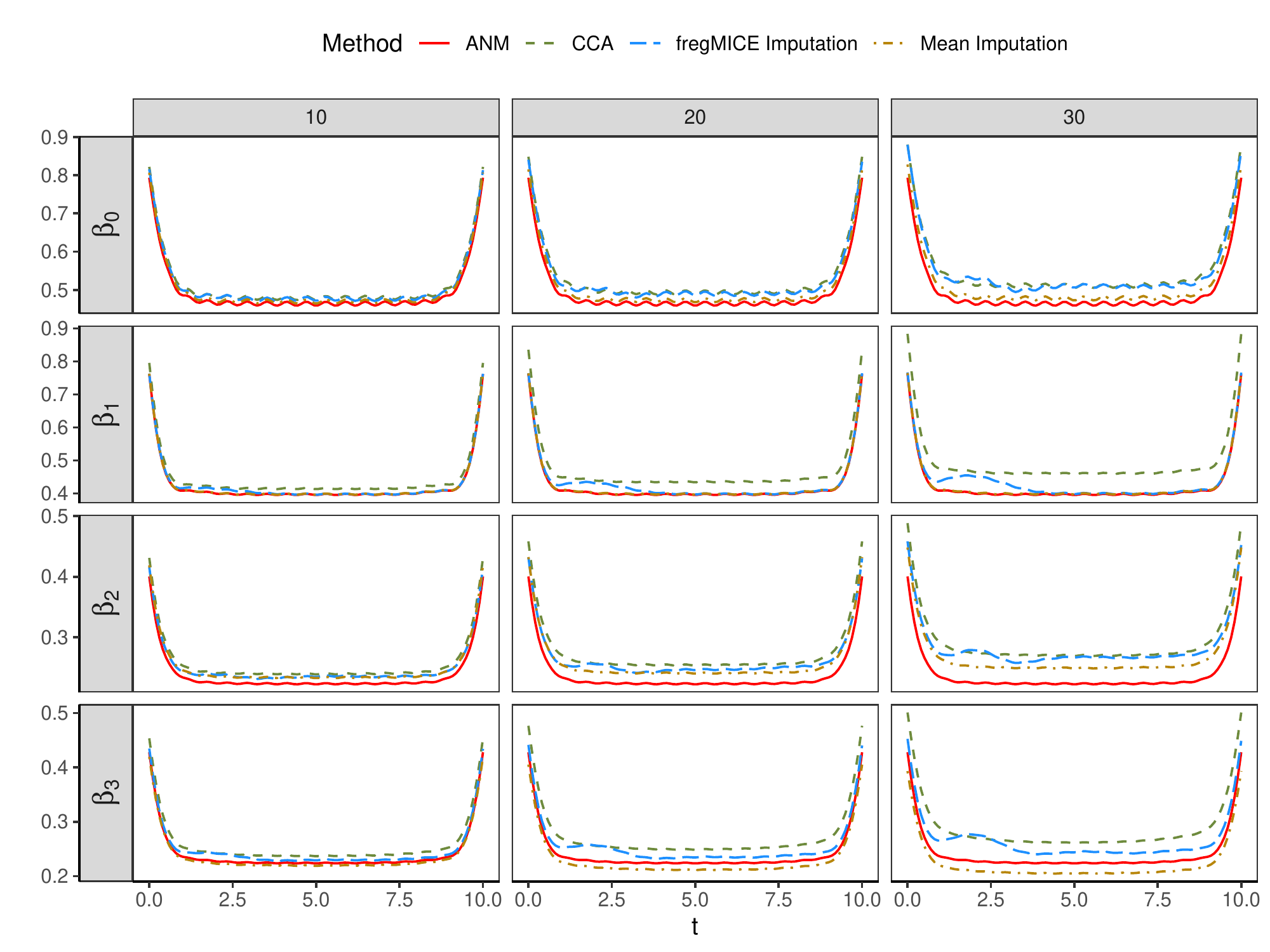}
  \includegraphics[width=0.85\linewidth]{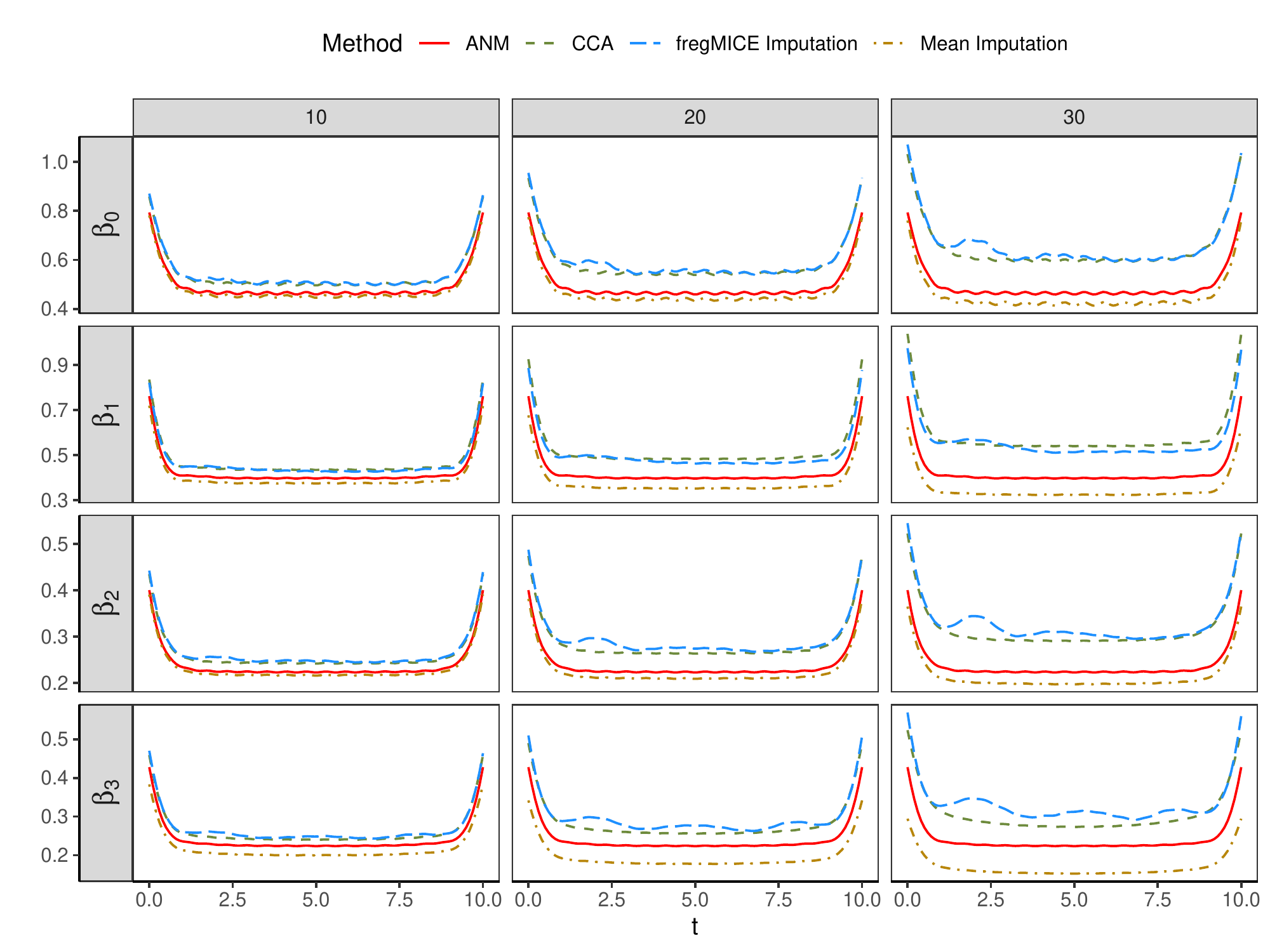}
  \caption{Point-wise 95\% confidence band width for Setting 1 Scenarios (a) (\textbf{Top}) and (b) (\textbf{Bottom}) with $n = 100$.  Columns (left to right) correspond to 10, 20, and 30\% missing data.  Rows (top to bottom) correspond to functional parameters $\beta_0, \beta_1, \beta_2,$ and $\beta_3$.}
  \label{fig:pww_1ab100}
\end{figure}

Figures \ref{fig:pwcov_1ab350} and \ref{fig:pwcov_1ab100} show the coverage probability for the estimated 95\% confidence bands at each point on $[0,10]$ for Setting 2 Scenarios (a) and (b) with $n = 350$ and $n = 100$ respectively, over the 500 simulation runs for each method of handling the missing data.  Figures \ref{fig:pww_1ab350} and \ref{fig:pww_1ab100} show the corresponding mean point-wise widths for the 95\% confidence bands.

\newpage 

\subsection{Simulation Study with a Scalar Response} \label{sim_sec_03}

Here we consider a simulation study where the analysis model has a scalar response with scalar and functional predictors.  As in the simulation study presented in the main article, we compare the complete case, mean imputation, and fregMICE approaches.   

\subsubsection{Data Generation}
\begin{figure}[!htbp]
\centering
  \includegraphics[width=0.50\linewidth]{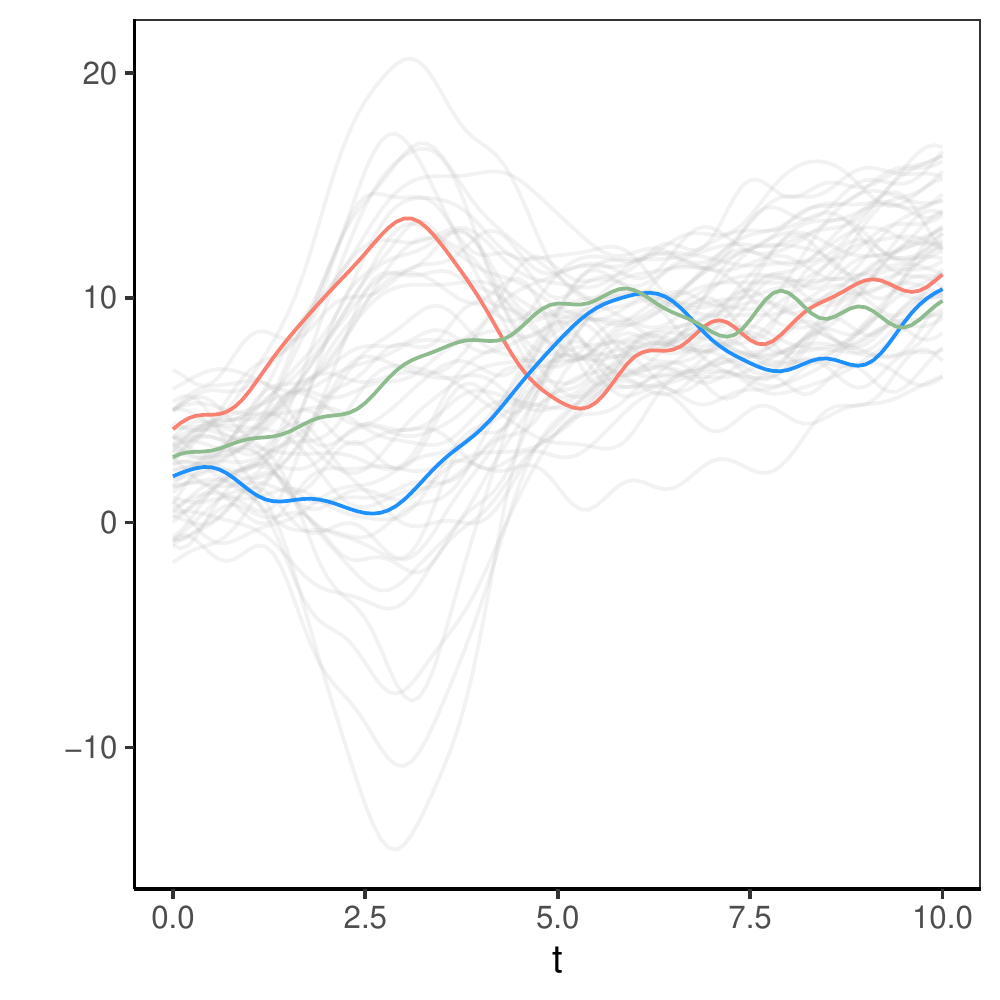}
  \caption{50 simulated functional predictors ($X$) with three highlighted observations.}
  \label{fig:simX_sc_resp}
\end{figure}

For this simulation, each complete observation consists of a scalar response, $y$, a scalar predictor, $z$, and a functional predictor, $X$. For each observation ($i = 1, \ldots, n$), we generated a scalar predictor $z_i \sim N(0,1)$ and a functional predictor, $X_i$, observed on a grid $ \{t_g = \frac{g}{10}: g = 0, 1, \ldots, 100 \}$ such that $X_i(t_g) = u_{i1} + u_{i2}t_{g} + u_{i3}\frac{20}{\sqrt{2\pi}}e^{-(t_g - 3)^{2}/2}  + \sum_{k = 1}^{10} \left \{ v_{ik1} sin \left (\frac{2 \pi k}{10} t_{g} \right ) + v_{ik2} cos \left (\frac{2 \pi k}{10} t_{g} \right ) \right \}$ where $u_{i1} \sim U(0,5)$, $u_{i2} \sim N(1,0.04)$, $u_{i3} \sim N(0,1)$, $v_{ik1}, v_{ik2} \sim N \left(0, \frac{1}{k^2} \right)$ for $k = 1, \ldots, 10$.  Figure \ref{fig:simX_sc_resp} displays a set of 50 simulated predictor functions.  We generated the scalar response according to the model,  
\begin{equation} \label{scal_resp_mod}
y_i = \theta_0 + \theta_1z_{i} + \int X_i(t)\beta_1(t) + \varepsilon_i,
\end{equation}
where $\theta_0 = 0$, $\theta_1 = 1$, $\beta_1(t) = sin\left(\frac{\pi t}{5} \right)$, and $\varepsilon_i \sim N(0, 0.5)$.  

We considered two settings: one in which the functional predictor is MCAR and a second in which it is MAR.  In both settings, $y_i$ and $z_i$ are always observed and the probability of $X_i$ being observed is given by $logit\{P(R_{X_i}=1)\} = \psi_0 + \psi_1y_i$.  In the MCAR setting,  we have $\psi_1 = 0$ and $\psi_0 = log \left(\frac{0.9}{0.1} \right), log \left(\frac{0.8}{0.2} \right)$, and $log \left(\frac{0.7}{0.3} \right)$ to achieve about $10\%, 20\%$, and $30\%$ missingness respectively.  In the MAR setting, we have $\psi_1 = -6$ and $\psi_0 = 6.2, 1$, and $-3$ to achieve about $10\%, 20\%$, and $30\%$ missingness respectively.  For each setting we considered two sample sizes: $n = 350$ and $n = 100$ and generated 500 data sets for each combination of setting, missingness, and sample size.  The average proportion of missing data for each combination is given in Table \ref{p_mis_tab}.

\begin{table}[h!]
\caption{\label{p_mis_tab} Mean (SD) proportion of missing $X$ functions.} 
\centering
\scriptsize
\begin{tabular}{llccc}
\hline
 & & 10\% Intended & 20\% Intended & 30\% Intended \\
\hline
\textbf{MCAR} & $n = 100$ & 0.10 (0.03) & 0.20 (0.04) & 0.27 (0.04)  \\
      & $n = 350$ & 0.10 (0.02) & 0.20 (0.02) & 0.27 (0.02)  \\
\textbf{MAR} & $n = 100$ & 0.11 (0.03) & 0.20 (0.04) & 0.30 (0.04)  \\
      & $n = 350$ & 0.10 (0.02) & 0.20 (0.02) & 0.30 (0.03)  \\
\hline
\end{tabular}
\end{table}

\subsubsection{Procedures Compared and Performance Measures}
For each simulation setting, we fit the correctly specified model for the scalar response, $y$, using PFR, implemented by the \verb+pfr+ function from the \verb+R+ package \verb+refund+ \citep{refund}.  For the PFR fitting procedure, the functional predictor was represented using functional principle components (FPCs) by smoothed covariance \citep{yao_2003} where the number of FPCs was selected to be the minimum number of components explaining at least 99\% of the variance in the functional observations.  The coefficient function, $\beta_1$, was represented using a basis set of 30 thin-plate regression splines and the fitting procedure penalized the magnitude of the second derivative. 

As a benchmark, we used all of the data, prior to imposing missingess on any of the variables to fit model (\ref{scal_resp_mod}).  In the results, we refer to this as ``all no missing'' (ANM).  For mean imputation, we filled in any missing $X$ functions with the point-wise mean function of the observed $X$ functions.  For complete case analysis (CCA), the analysis model was fit on observations with complete data only.  For our fregMICE procedure, the imputation model for the missing $X$ functions was given by $X_i(t) = \gamma_{0}(t) + \gamma_{1}(t)z_i + \gamma_{2}(t)y_i + \varepsilon(t)$ where the error term at each point is assumed to be from the $N(0, \sigma_{\varepsilon}^2)$ distribution with constant error variance, $\sigma_{\varepsilon}^2$.  We used the \verb+pffr+ function from the \verb+refund+ package \citep{refund} to fit this model in each iteration of the fregMICE procedure using 20 cubic B-splines to represent each coefficient function and we penalized the magnitude of the second derivative.  We ran the fregMICE procedure for 20 iterations and constructed 5 imputed data sets.  We fit model (\ref{scal_resp_mod}) on each imputed data set and used the extension of Rubin's rules described in Section 4.2 of the article to pool estimates from the 5 data sets.

We computed the mean squared error (MSE) and standardized bias, given by $(\bar{\hat{\theta}}_1 - \theta_1) / sd(\hat{\theta}_1)$, for estimates of the scalar coefficient $\theta_1$.  For the standardized bias computation, $\bar{\hat{\theta}}_1$ is the average of the 500 estimates of $\theta_1$ and $sd(\hat{\theta}_1)$ is the Monte-Carlo standard deviation of these estimates.  These values are given in Table \ref{sc_bias_mse}.  Mean coverage and width of the 95\% confidence intervals for $\theta_1$ are given in Table \ref{sc_ci_cov_w}.     

As in the main article, we provide point-wise mean and point-wise standardized bias curves for the estimates of the $\beta_1$ coefficient function from each method.  These plots are displayed in Figures \ref{fig:pw_meanSB_mcar} and \ref{fig:pw_meanSB_mar}.  Point-wise coverage and width of the 95\% confidence bands for $\beta_1$ are displayed in Figures \ref{fig:pw_coverwidth_mcar} and \ref{fig:pw_coverwidth_mar}. Across-the-function mean coverage and width of the 95\% confidence bands for $\beta_1$ are given in Table \ref{sc_pwcovW_tab}.

\subsubsection{Results}

\textbf{Point and Interval Estimates for $\bf{\theta}_1$} \\

Table \ref{sc_bias_mse} shows that the MSE and standardized bias for estimating $\theta_1$ for the fregMICE and CCA estimates are comparable across the different amounts of missingness and sample size in the MCAR setting.  Mean imputation yields higher MSE, but similar standardized biases.  In the MAR setting, the fregMICE estimates for $\theta_1$ have smaller magnitude MSE and standardized bias under all combinations of missingness and sample size in comparison to both mean imputation and CCA.  

\begin{table}[h!]
\caption{\label{sc_bias_mse} MSE (SD) and standardized bias for $\theta_1$ estimates from scalar response simulations. MSE (SD) values are $\times 10^{-4}$. } 
\centering
\scriptsize
\begin{tabular}{llcccccccc}
\hline
                        & & \multicolumn{2}{c}{$10\%$} & & \multicolumn{2}{c}{$20\%$} & & \multicolumn{2}{c}{$30\%$} \\
                        \cline{3-4} \cline{6-7} \cline{9-10}
\textbf{MCAR} & & $MSE \ (SD)$ & $Std. Bias$ & & $MSE \ (SD)$ & $Std. Bias$ & & $MSE \ (SD)$ & $Std. Bias$ \\
\hline
 $n = 100$ & ANM      & 5.42 (7.82) & -0.02 & & 5.42 (7.82) & -0.02 &  & 5.42 (7.82) & -0.02 \\  
 &           Mean     & 8.91 (12.35) & 0.05 & & 10.87 (15.14) & 0.03 & & 13.49 (20.07) & -0.01 \\
 &           CCA      & 6.25 (9.43) & -0.02 & & 7.14 (10.63) & -0.03 & & 8.07 (11.76) & -0.03 \\ 
 &           fregMICE & 6.29 (9.40) & -0.01 & & 6.98 (10.55) & -0.03 & & 7.85 (10.75) & -0.06 \\ 
 \hline                                                                                         
 $n = 350$ & ANM      & 1.43 (2.05) & 0.03 &  & 1.43 (2.05) & 0.03 &   & 1.43 (2.05) & 0.03 \\   
 &           Mean     & 2.33 (3.30) & 0.00 &  & 3.27 (4.43) & -0.01 &  & 3.72 (5.20) & -0.02 \\  
 &           CCA      & 1.64 (2.36) & 0.02 &  & 1.88 (2.73) & 0.01 &   & 2.05 (3.05) & 0.01 \\   
 &           fregMICE & 1.63 (2.35) & 0.01 &  & 1.92 (2.82) & 0.00 &   & 2.06 (2.98) & 0.02 \\   

\textbf{MAR} & & & & & & & & & \\
\hline
 $n = 100$ & ANM      & 5.42 (7.82) & -0.02 &  & 5.42 (7.82) & -0.02 &   & 5.42 (7.82) & -0.02 \\  
 &           Mean     & 45.13 (50.99) & 1.39 & & 61.48 (64.80) & 1.59 &  & 67.65 (71.32) & 1.51 \\ 
 &           CCA      & 8.94 (12.38) & -0.63 & & 13.89 (18.29) & -0.93 & & 21.17 (25.27) & -1.17 \\
 &           fregMICE & 6.64 (9.25) & -0.05 &  & 7.28 (9.90) & -0.09 &   & 9.08 (13.24) & -0.13 \\ 
 \hline                                                                                           
 $n = 350$ & ANM      & 1.43 (2.05) & 0.03 &   & 1.43 (2.05) & 0.03 &    & 1.43 (2.05) & 0.03 \\   
 &           Mean     & 31.97 (21.95) & 2.64 & & 46.95 (28.93) & 2.97 &  & 51.44 (32.71) & 2.81 \\ 
 &           CCA      & 3.43 (3.97) & -1.10 &  & 6.97 (6.39) & -1.70 &   & 12.54 (10.21) & -2.16 \\
 &           fregMICE & 1.68 (2.30) & 0.04 &   & 2.01 (2.73) & 0.02 &    & 2.47 (3.33) & -0.02 \\  
 \hline
\end{tabular}
\end{table}

\begin{table}[h!]
\caption{\label{sc_ci_cov_w} Coverages (SD) and widths (SD) of 95\% confidence intervals for $\theta_1$.} 
\centering
\scriptsize
\begin{tabular}{llcccccccc}
\hline
                        & & \multicolumn{2}{c}{$10\%$} & & \multicolumn{2}{c}{$20\%$} & & \multicolumn{2}{c}{$30\%$} \\
                        \cline{3-4} \cline{6-7} \cline{9-10}
\textbf{MCAR} & & Cov. $(SD)$ & Width $(SD)$ & & Cov. $(SD)$ & Width $(SD)$ & & Cov. $(SD)$ & Width $(SD)$ \\
\hline
 $n = 100$ & ANM      & 0.95 (0.21) & 0.29 (0.03) &  & 0.95 (0.21) & 0.29 (0.03) &  & 0.95 (0.21) & 0.29 (0.03) \\
 &           Mean     & 0.94 (0.23) & 0.36 (0.05) &  & 0.95 (0.23) & 0.42 (0.06) &  & 0.95 (0.23) & 0.46 (0.06) \\
 &           CCA      & 0.94 (0.24) & 0.30 (0.04) &  & 0.94 (0.24) & 0.32 (0.04) &  & 0.94 (0.24) & 0.34 (0.04) \\
 &           fregMICE & 0.94 (0.24) & 0.30 (0.04) &  & 0.94 (0.24) & 0.31 (0.04) &  & 0.93 (0.26) & 0.33 (0.05) \\
 \hline                                                                                                          
 $n = 350$ & ANM      & 0.96 (0.20) & 0.15 (0.01) &  & 0.96 (0.20) & 0.15 (0.01) &  & 0.96 (0.20) & 0.15 (0.01) \\
 &           Mean     & 0.96 (0.20) & 0.19 (0.01) &  & 0.94 (0.23) & 0.22 (0.02) &  & 0.95 (0.21) & 0.24 (0.02) \\
 &           CCA      & 0.95 (0.22) & 0.16 (0.01) &  & 0.94 (0.24) & 0.17 (0.01) &  & 0.94 (0.23) & 0.18 (0.01) \\
 &           fregMICE & 0.95 (0.21) & 0.16 (0.01) &  & 0.94 (0.23) & 0.17 (0.02) &  & 0.93 (0.25) & 0.17 (0.02) \\
 \textbf{MAR} & & & & & & & & & \\
\hline
 $n = 100$ & ANM      & 0.95 (0.21) & 0.29 (0.03) &  & 0.95 (0.21) & 0.29 (0.03) &  & 0.95 (0.21) & 0.29 (0.03) \\
 &           Mean     & 0.71 (0.45) & 0.49 (0.07) &  & 0.69 (0.46) & 0.56 (0.07) &  & 0.71 (0.45) & 0.61 (0.07) \\
 &           CCA      & 0.90 (0.30) & 0.31 (0.04) &  & 0.84 (0.37) & 0.33 (0.04) &  & 0.78 (0.41) & 0.36 (0.05) \\
 &           fregMICE & 0.95 (0.23) & 0.32 (0.04) &  & 0.95 (0.22) & 0.34 (0.06) &  & 0.93 (0.26) & 0.37 (0.07) \\
 \hline                                                                                                          
 $n = 350$ & ANM      & 0.96 (0.20) & 0.15 (0.01) &  & 0.96 (0.20) & 0.15 (0.01) &  & 0.96 (0.20) & 0.15 (0.01) \\
 &           Mean     & 0.25 (0.43) & 0.25 (0.02) &  & 0.19 (0.39) & 0.29 (0.02) &  & 0.21 (0.41) & 0.32 (0.02) \\
 &           CCA      & 0.84 (0.37) & 0.16 (0.01) &  & 0.62 (0.49) & 0.17 (0.01) &  & 0.45 (0.50) & 0.19 (0.01) \\
 &           fregMICE & 0.95 (0.21) & 0.16 (0.01) &  & 0.95 (0.21) & 0.18 (0.02) &  & 0.95 (0.23) & 0.20 (0.03) \\
 \hline
\end{tabular}
\end{table}

Table \ref{sc_ci_cov_w} shows that the 95\% confidence intervals from the fregMICE imputed data achieve or nearly achieve the nominal coverage in both the MCAR and MAR settings for both sample sizes.  Intervals based on mean imputed data and CCA perform similarly in the MCAR setting, but show considerably worse coverage in the MAR setting for each combination of sample size and missingness.  The widths of the 95\% confidence intervals from the fregMICE imputed data tend to be the same as or slightly larger than those for CCA with both being close to the ANM benchmark widths or slightly wider in both the MCAR and MAR settings. \\  

\noindent \textbf{Point and Interval Estimates for $\bf{\beta}_1$} \\

\begin{figure}[!htbp]
\centering
  \includegraphics[width=1\linewidth]{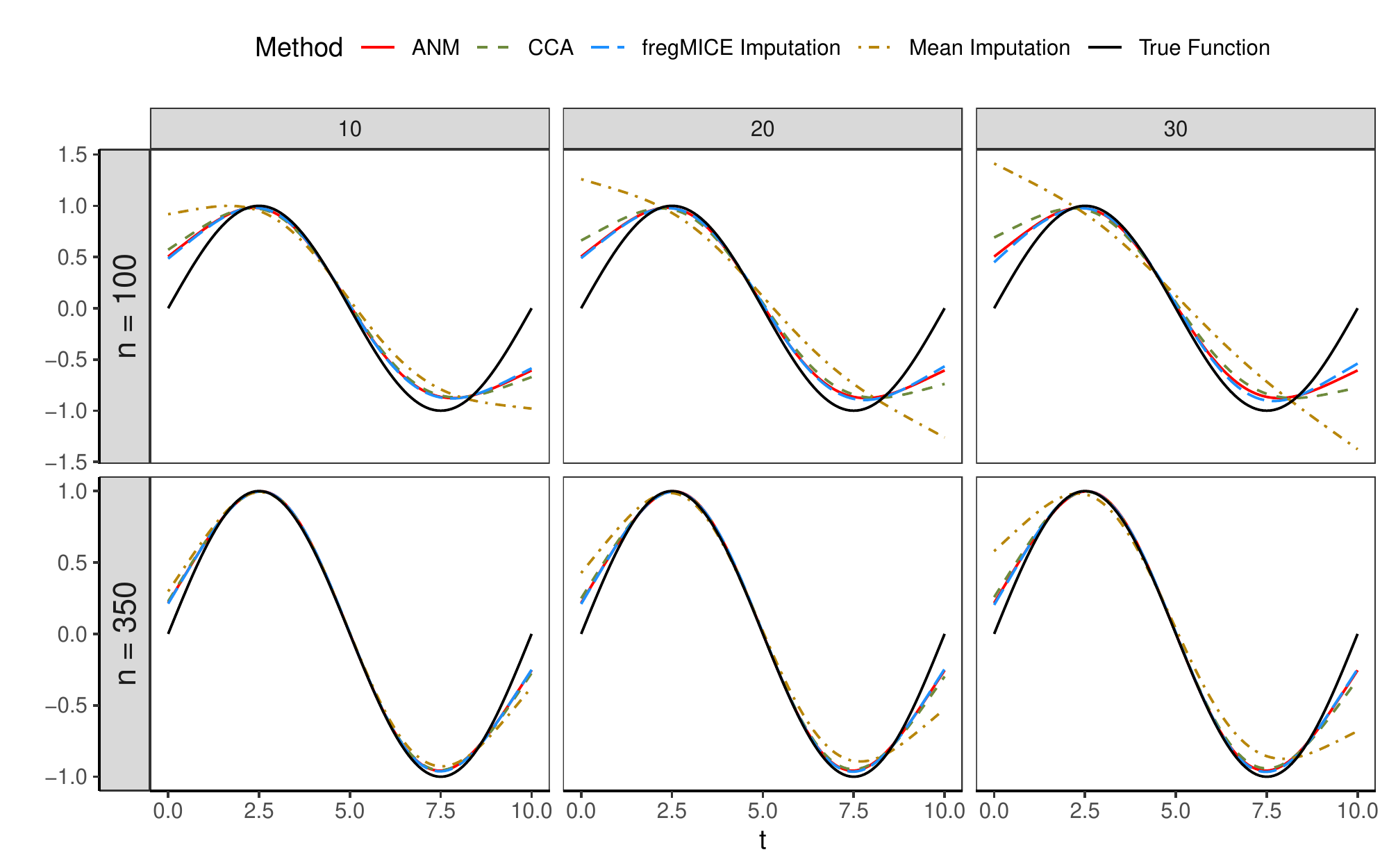}
  \includegraphics[width=1\linewidth]{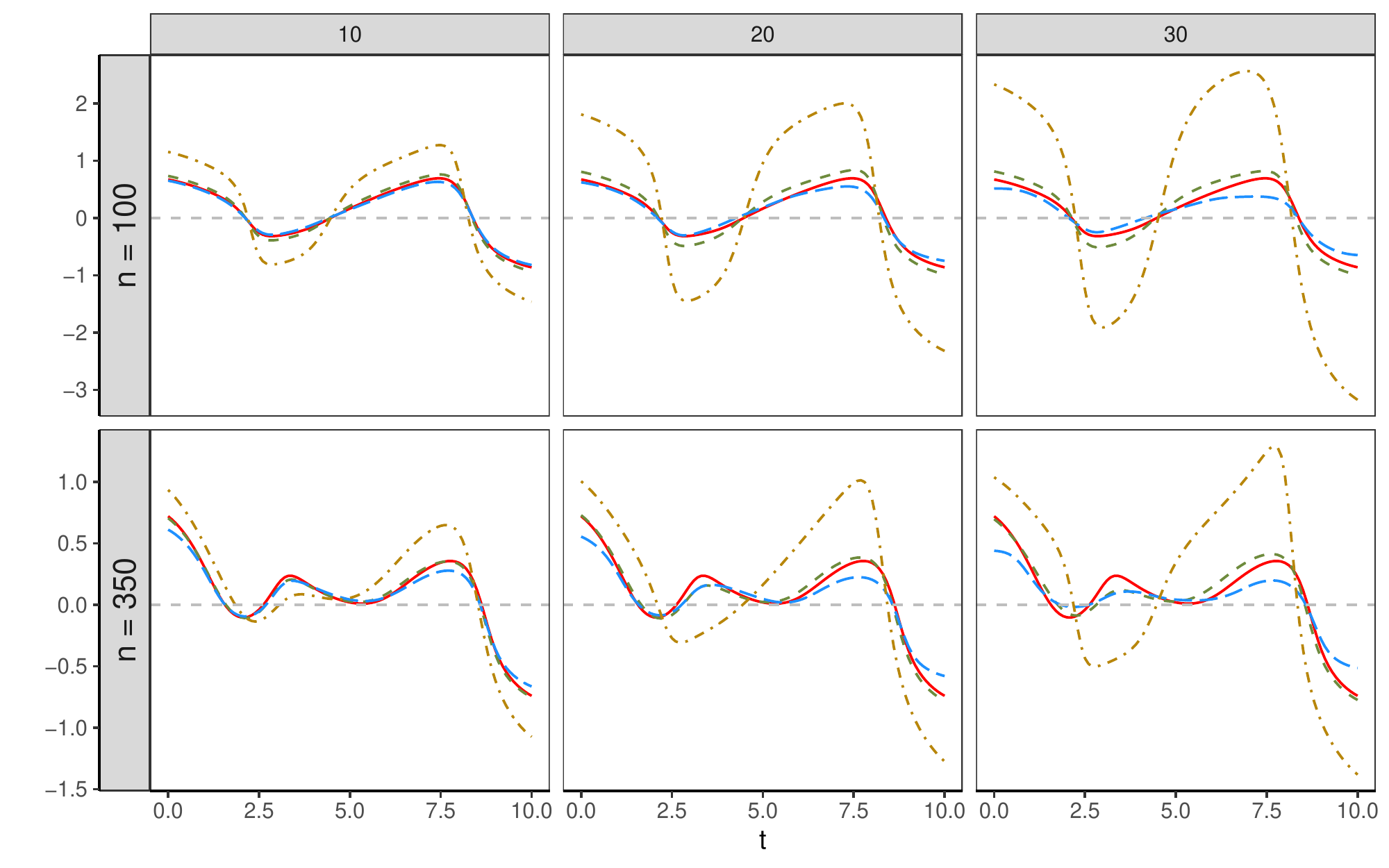}
  \caption{(\textbf{Top}) Point-wise mean curves and (\textbf{Bottom}) point-wise standardized bias for $\beta_1$ in MCAR setting. Columns (left to right) correspond to 10, 20, and 30\% missing data.  Rows correspond to sample size $n = 100$ or $n = 350$.}
  \label{fig:pw_meanSB_mcar}
\end{figure}

\begin{figure}[!htbp]
\centering
  \includegraphics[width=1\linewidth]{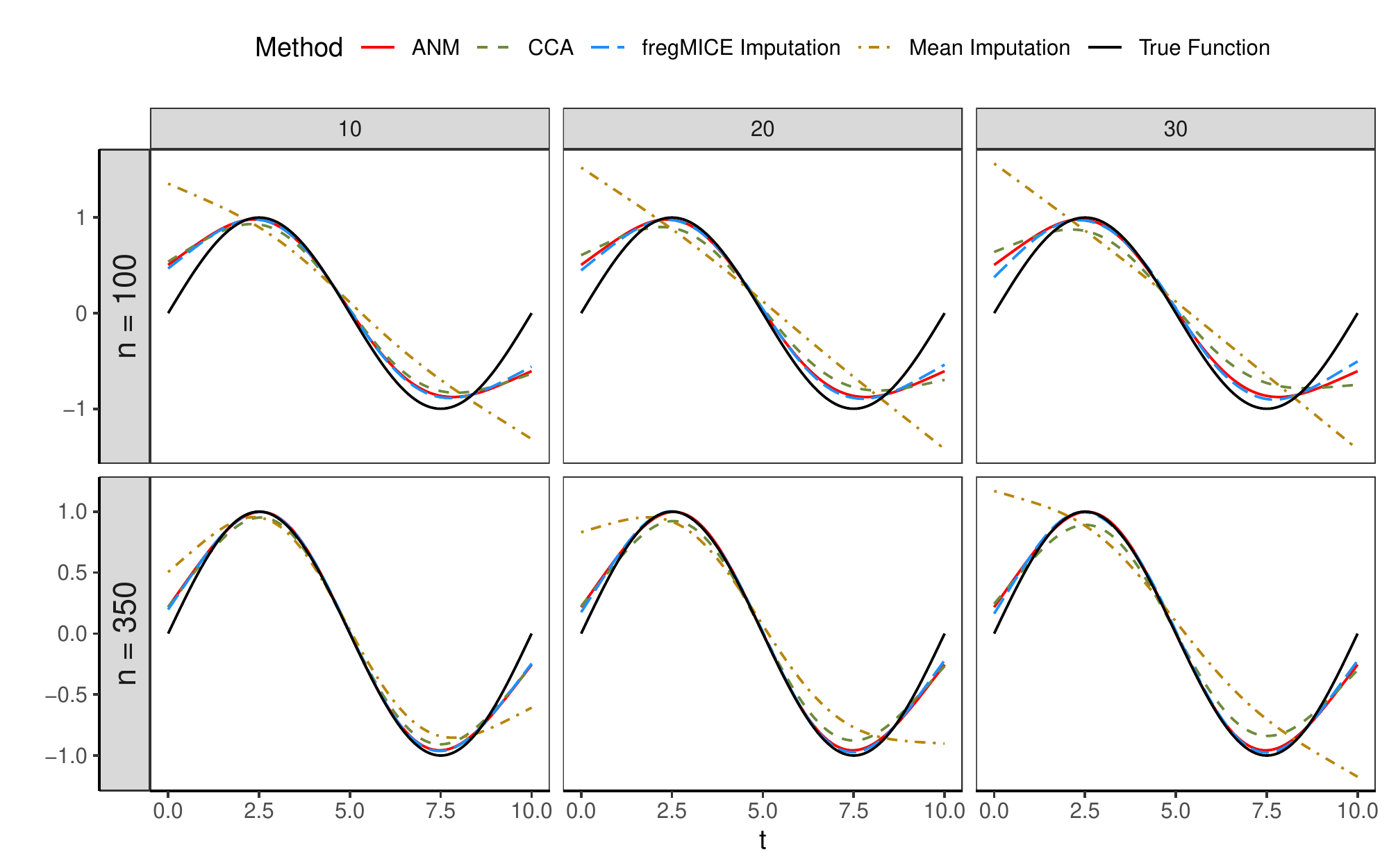}
  \includegraphics[width=1\linewidth]{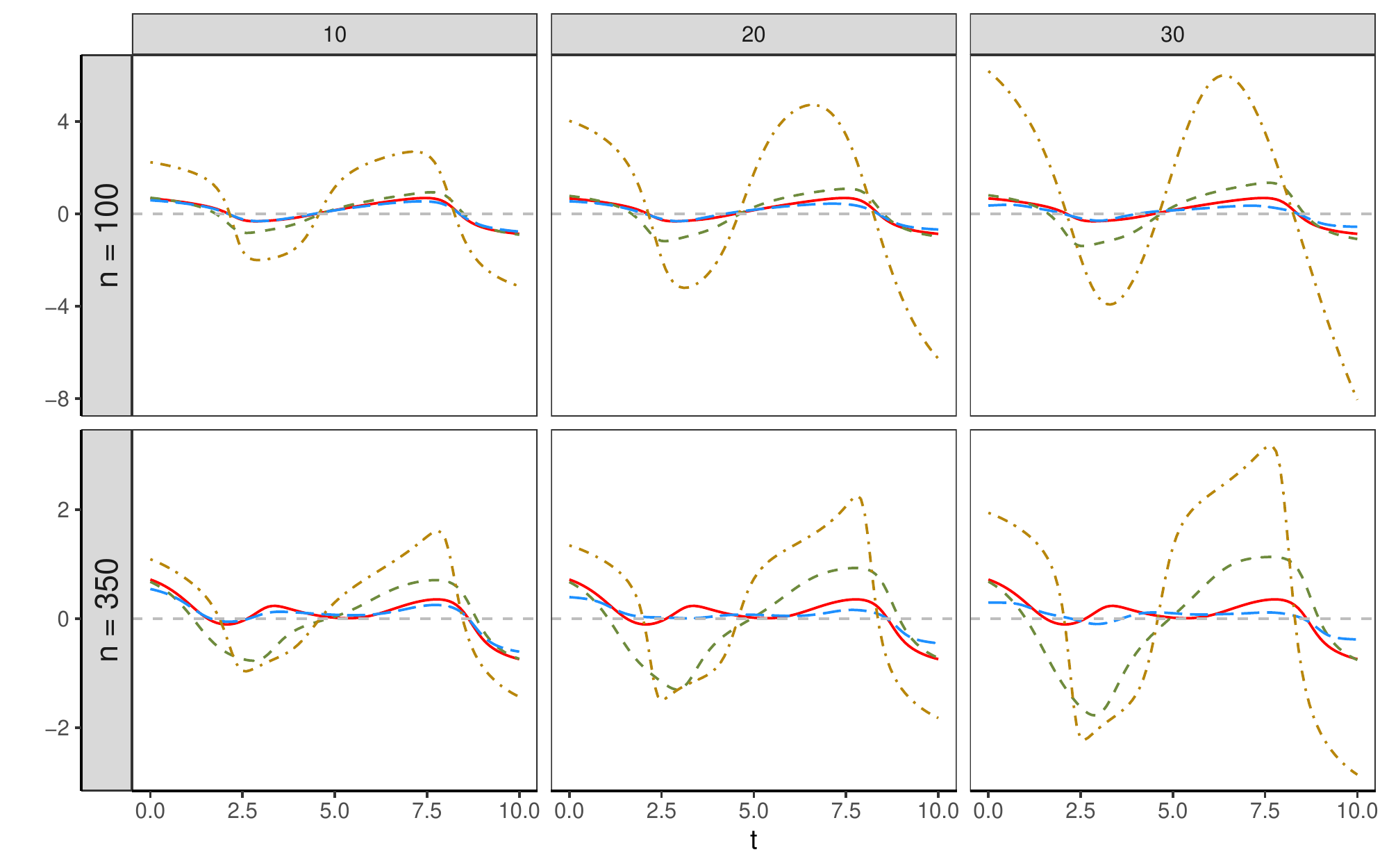}
  \caption{(\textbf{Top}) Point-wise mean curves and (\textbf{Bottom}) point-wise standardized bias for $\beta_1$ in MAR setting. Columns (left to right) correspond to 10, 20, and 30\% missing data.  Rows correspond to sample size $n = 100$ or $n = 350$.}
  \label{fig:pw_meanSB_mar}
\end{figure}

Figure \ref{fig:pw_meanSB_mcar} shows that, in the MCAR setting the point-wise means and point-wise standardized biases for the $\beta_1$ function are similar for the fregMICE imputation and CCA approaches with fregMICE showing slightly less bias for greater amounts of missing data.  Estimates derived from mean imputed data perform especially poorly for the smaller sample size of $n = 100$ and performance deteriorates with greater amounts of missing data.  Figure \ref{fig:pw_meanSB_mar} shows that, in the MAR setting, the fregMICE imputation approach yields estimates for $\beta_1$ that perform nearly identically to the ANM benchmark estimates while the estimates derived from mean imputed data and CCA show considerable amounts of bias.  The bias increases with smaller sample size and with greater amounts of missing data.  

\begin{figure}[!htbp]
\centering
  \includegraphics[width=1\linewidth]{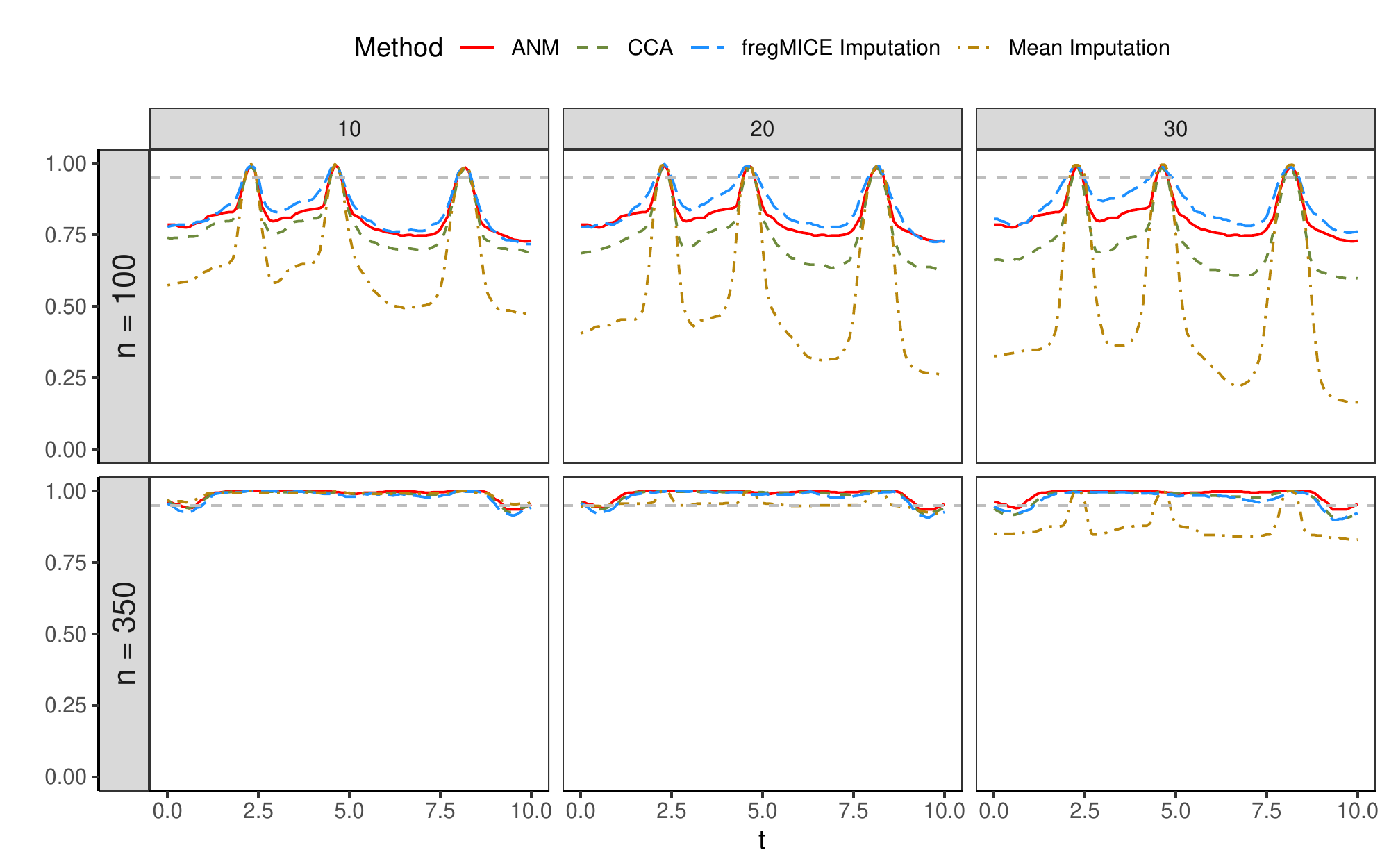}
  \includegraphics[width=1\linewidth]{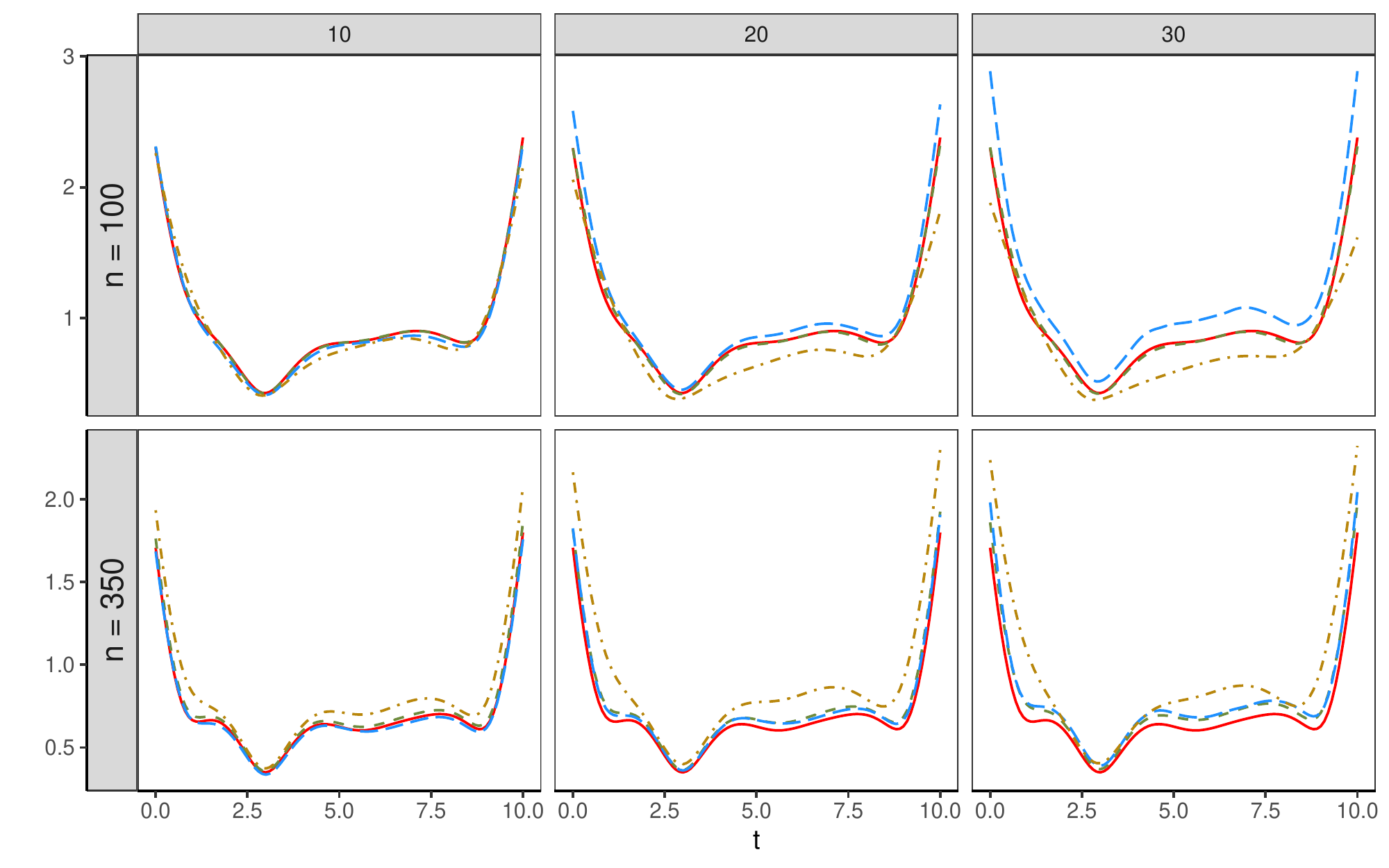}
  \caption{(\textbf{Top}) Point-wise 95\% confidence band coverage of $\beta_1$ in MCAR setting and (\textbf{Bottom}) point-wise 95\% confidence band width. Columns (left to right) correspond to 10, 20, and 30\% missing data. Rows correspond to sample size $n = 100$ or $n = 350$.}
  \label{fig:pw_coverwidth_mcar}
\end{figure}

For the smaller sample size of $n = 100$ in the MCAR setting, Figure \ref{fig:pw_coverwidth_mcar} shows that the the point-wise coverage for the 95\% confidence intervals derived from the fregMICE imputed data is similar to or better than that of the ANM benchmark and typically better than the coverage provided by intervals based on mean imputed data or CCA.  The point-wise width for the intervals derived from the fregMICE imputed data sets increases with greater amounts of missing data.  The coverage is poor for the intervals, regardless of the method used, since the small sample size tended to result in estimates of $\beta_1$ that were linear (i.e., the PFR fitting procedure placed a large penalty on the second derivative of the coefficient estimate).  The larger widths of the fregMICE-based intervals for larger amounts of missing data explain why the coverage of these intervals is better than than the coverage of intervals from the ANM benchmark as the amount of missing data increases.  For the larger sample size of $n = 350$ in the MCAR setting, Figure \ref{fig:pw_coverwidth_mcar} shows that point-wise coverage is typically greater than 95\% for fregMICE-based intervals and those based on CCA.  Intervals based on mean imputed data perform poorly with respect to coverage as the amount of missing data increases.  As in the cases with the smaller sample size, the widths of the fregMICE-based intervals increase for larger amounts of missing data. 

\begin{figure}[!htbp]
\centering
  \includegraphics[width=1\linewidth]{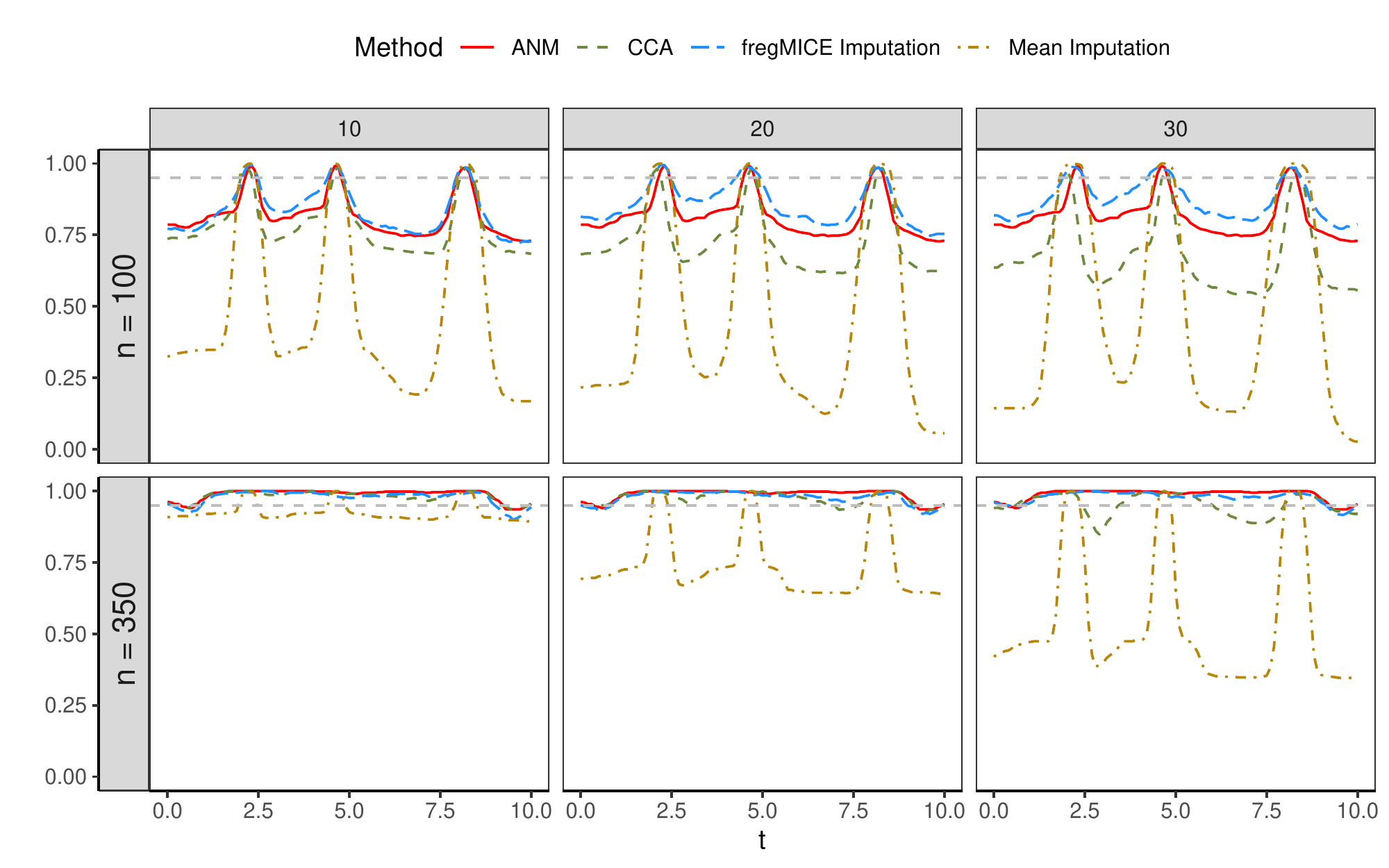}
  \includegraphics[width=1\linewidth]{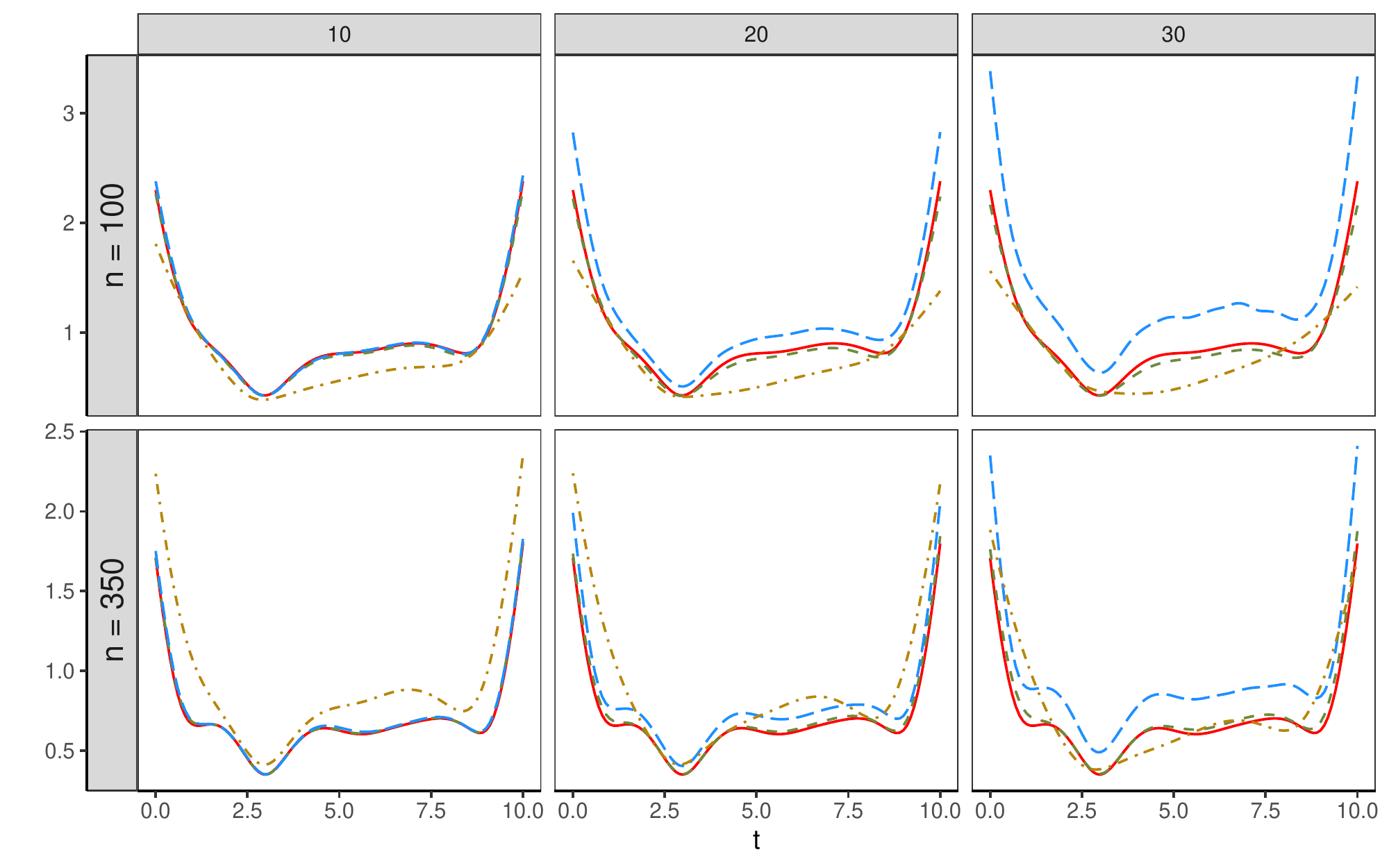}
  \caption{(\textbf{Top}) Point-wise 95\% confidence band coverage of $\beta_1$ in MAR setting and (\textbf{Bottom}) point-wise 95\% confidence band width. Columns (left to right) correspond to 10, 20, and 30\% missing data. Rows correspond to sample size $n = 100$ or $n = 350$.}
  \label{fig:pw_coverwidth_mar}
\end{figure}

In the MAR setting, Figure \ref{fig:pw_coverwidth_mar} shows similar results, with respect to point-wise 95\% confidence interval coverage and width, compared to those in the MCAR setting discussed above.  The main difference is that we see a more pronounced widening of the fregMICE-based intervals as the amount of missing data increases in both the small and large sample settings.  Table \ref{sc_pwcovW_tab} provides additional summary measures that support the findings in Figures \ref{fig:pw_coverwidth_mcar} and \ref{fig:pw_coverwidth_mar}.

\begin{table}[!htbp]
\caption{\label{sc_pwcovW_tab} Across-the-function mean (SD) point-wise 95\% confidence band coverage (pwCov) of $\beta_1$ and width (pwWidth) for the MCAR and MAR settings.} 
\centering
\scriptsize
\begin{tabular}{llcccccccc}
\hline
                       & & \multicolumn{2}{c}{$10\%$} & & \multicolumn{2}{c}{$20\%$} & & \multicolumn{2}{c}{$30\%$} \\
                           \cline{3-4} \cline{6-7} \cline{9-10}
\textbf{MCAR} & & pwCov $(SD)$ & pwWidth $(SD)$ & & pwCov $(SD)$ & pwWidth $(SD)$ & & pwCov $(SD)$ & pwWidth $(SD)$ \\
\hline
  $n = 100$ & ANM      & 0.82 (0.32) & 0.95 (0.35) &  & 0.82 (0.32) & 0.95 (0.35) &  & 0.82 (0.32) & 0.95 (0.35) \\
 &           Mean     & 0.65 (0.36) & 0.91 (0.49) &  & 0.53 (0.34) & 0.84 (0.54) &  & 0.49 (0.30) & 0.80 (0.53) \\
 &           CCA      & 0.78 (0.34) & 0.95 (0.38) &  & 0.75 (0.35) & 0.94 (0.41) &  & 0.72 (0.36) & 0.95 (0.43) \\
 &           fregMICE & 0.83 (0.29) & 0.92 (0.30) &  & 0.85 (0.27) & 1.02 (0.36) &  & 0.86 (0.25) & 1.13 (0.52) \\
 \hline                                                                                                          
 $n = 350$ & ANM      & 0.99 (0.04) & 0.70 (0.07) &  & 0.99 (0.04) & 0.70 (0.07) &  & 0.99 (0.04) & 0.70 (0.07) \\
 &           Mean     & 0.99 (0.08) & 0.80 (0.08) &  & 0.96 (0.18) & 0.89 (0.16) &  & 0.88 (0.29) & 0.91 (0.27) \\
 &           CCA      & 0.98 (0.06) & 0.72 (0.08) &  & 0.98 (0.06) & 0.75 (0.09) &  & 0.97 (0.09) & 0.77 (0.10) \\
 &           fregMICE & 0.98 (0.06) & 0.69 (0.07) &  & 0.98 (0.06) & 0.74 (0.11) &  & 0.97 (0.07) & 0.79 (0.19) \\
 \textbf{MAR} & & & & & & & & & \\
\hline
 $n = 100$ & ANM      & 0.82 (0.32) & 0.95 (0.35) &  & 0.82 (0.32) & 0.95 (0.35) &  & 0.82 (0.32) & 0.95 (0.35) \\
 &           Mean     & 0.48 (0.30) & 0.77 (0.48) &  & 0.45 (0.23) & 0.75 (0.42) &  & 0.48 (0.19) & 0.79 (0.37) \\
 &           CCA      & 0.78 (0.34) & 0.93 (0.36) &  & 0.73 (0.36) & 0.91 (0.40) &  & 0.68 (0.37) & 0.90 (0.43) \\
 &           fregMICE & 0.83 (0.28) & 0.96 (0.32) &  & 0.86 (0.25) & 1.11 (0.41) &  & 0.87 (0.23) & 1.32 (0.64) \\
 \hline                                                                                                          
 $n = 350$ & ANM      & 0.99 (0.04) & 0.70 (0.07) &  & 0.99 (0.04) & 0.70 (0.07) &  & 0.99 (0.04) & 0.70 (0.07) \\
 &           Mean     & 0.92 (0.23) & 0.92 (0.21) &  & 0.74 (0.36) & 0.89 (0.38) &  & 0.54 (0.36) & 0.76 (0.45) \\
 &           CCA      & 0.98 (0.07) & 0.70 (0.09) &  & 0.97 (0.08) & 0.72 (0.09) &  & 0.95 (0.11) & 0.73 (0.10) \\
 &           fregMICE & 0.98 (0.06) & 0.71 (0.10) &  & 0.98 (0.06) & 0.80 (0.18) &  & 0.98 (0.06) & 0.93 (0.28) \\
 \hline
\end{tabular}
\end{table}

 \newpage

\section{Additional Results from EMBARC EEG Application} 

\subsection{fregMICE Computational Considerations}

We conducted the EMBARC data analysis using \verb+R+ version 4.0.0 on an iMAC with a 4.2GHz Intel Core i7 processor.  To run one iteration of the fregMICE procedure on the subset of healthy control data ($n = 40$) it took 9.14 seconds, on average (sd = 2.52 seconds) based on 10 runs.  To run one iteration of the fregMICE procedure on the subset of MDD data ($n = 295$) it took 47.69 seconds, on average (sd = 6.78 seconds) based on 10 runs.   

\subsection{Functional Convergence Plots for MDD Subjects}
\begin{figure}[th]
 \center
  \includegraphics[width=\linewidth]{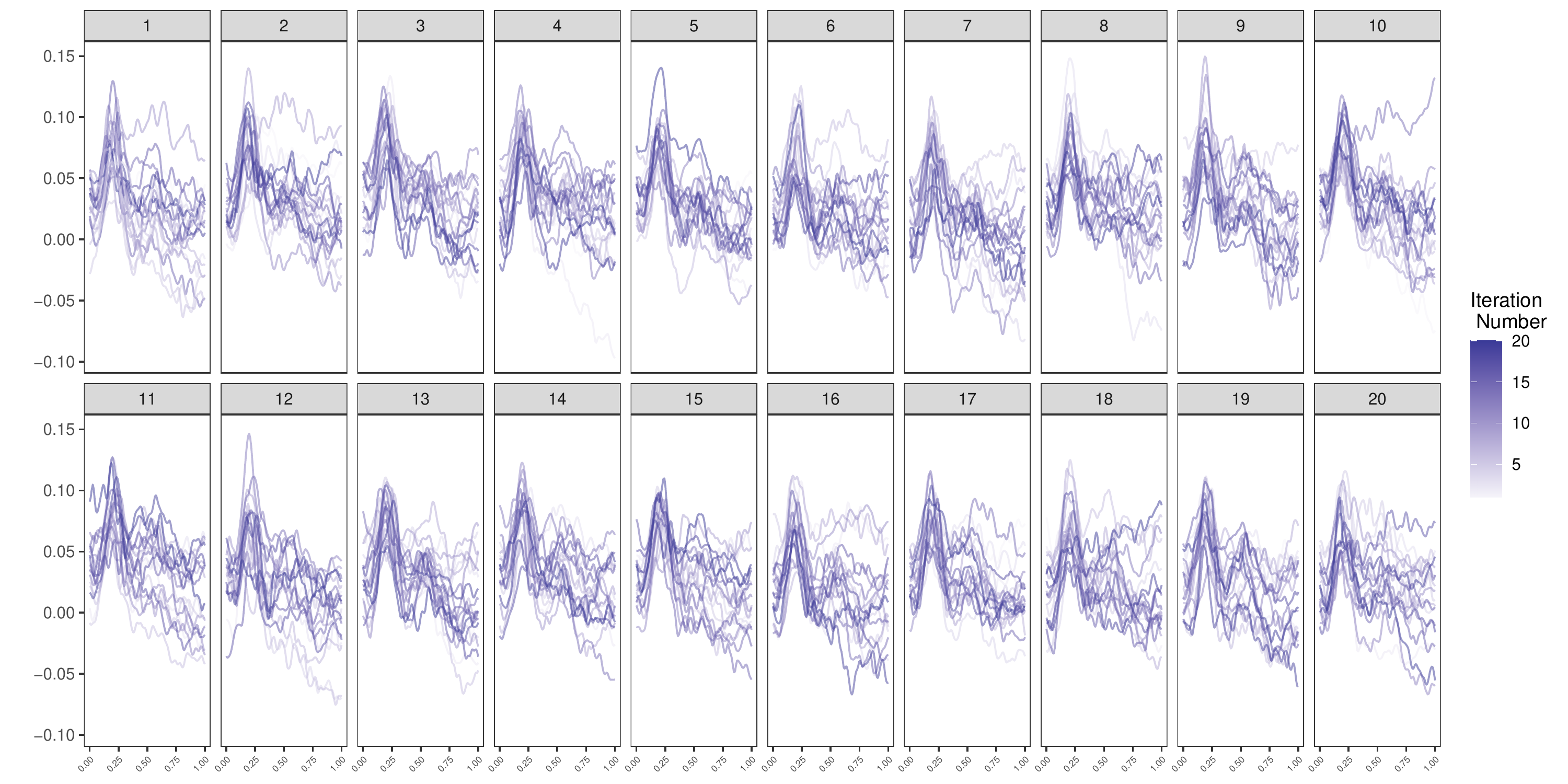}
  \caption{Functional convergence plots. Point-wise mean of the imputed function values for HC subjects.  Each panel corresponds to one imputation stream.  Dark colors correspond to later iterations.}
  \label{fig:func_conv_MDD}
\end{figure}

Figures \ref{fig:func_conv_MDD} shows the functional convergence plot for the MDD subset.  The figure for the HC subset is provided in the main article.  
 
\subsection{Functional Strip Plots}
\begin{figure}[!htbp]
\centering
  \includegraphics[width=0.99\linewidth]{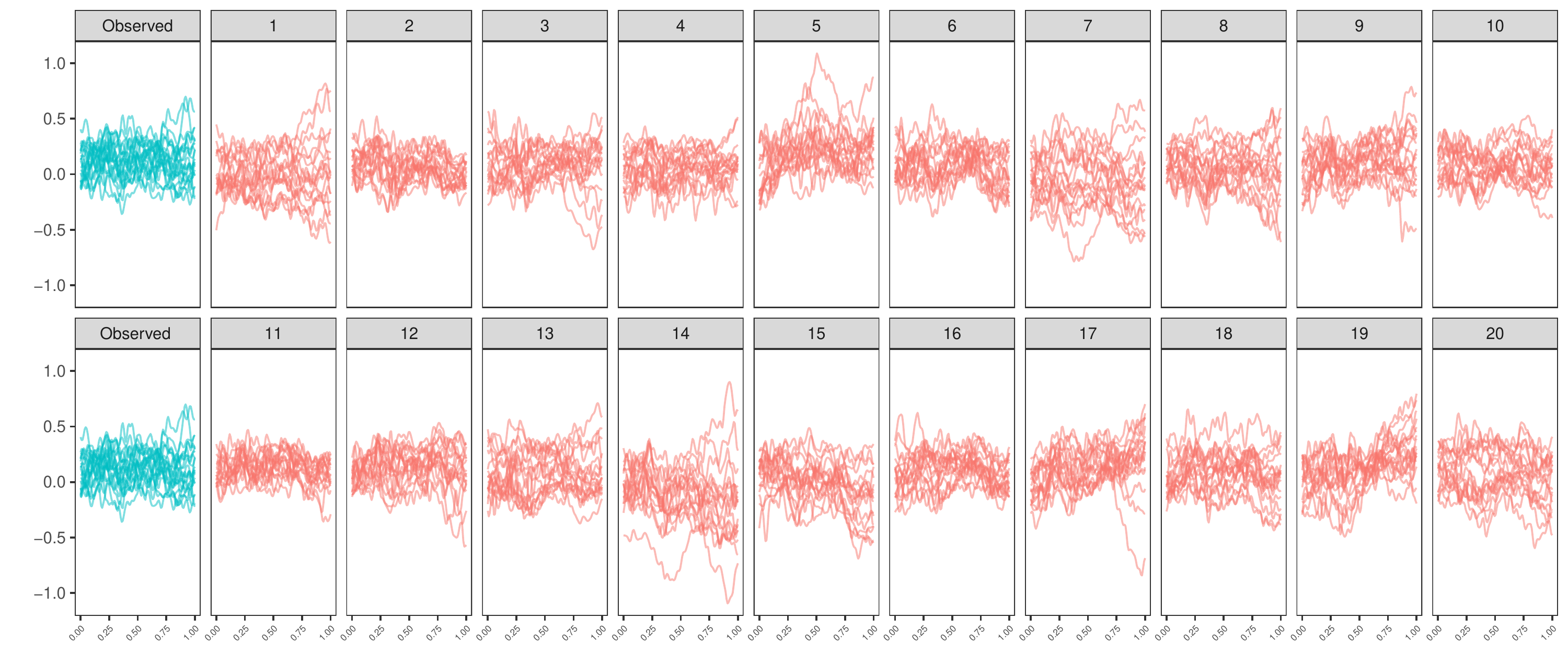}
  \caption{Functional strip plot for HC subset.}
  \label{fig:HC_func_strip_plot}
\end{figure}

\begin{figure}[!htbp]
\centering
  \includegraphics[width=0.99\linewidth]{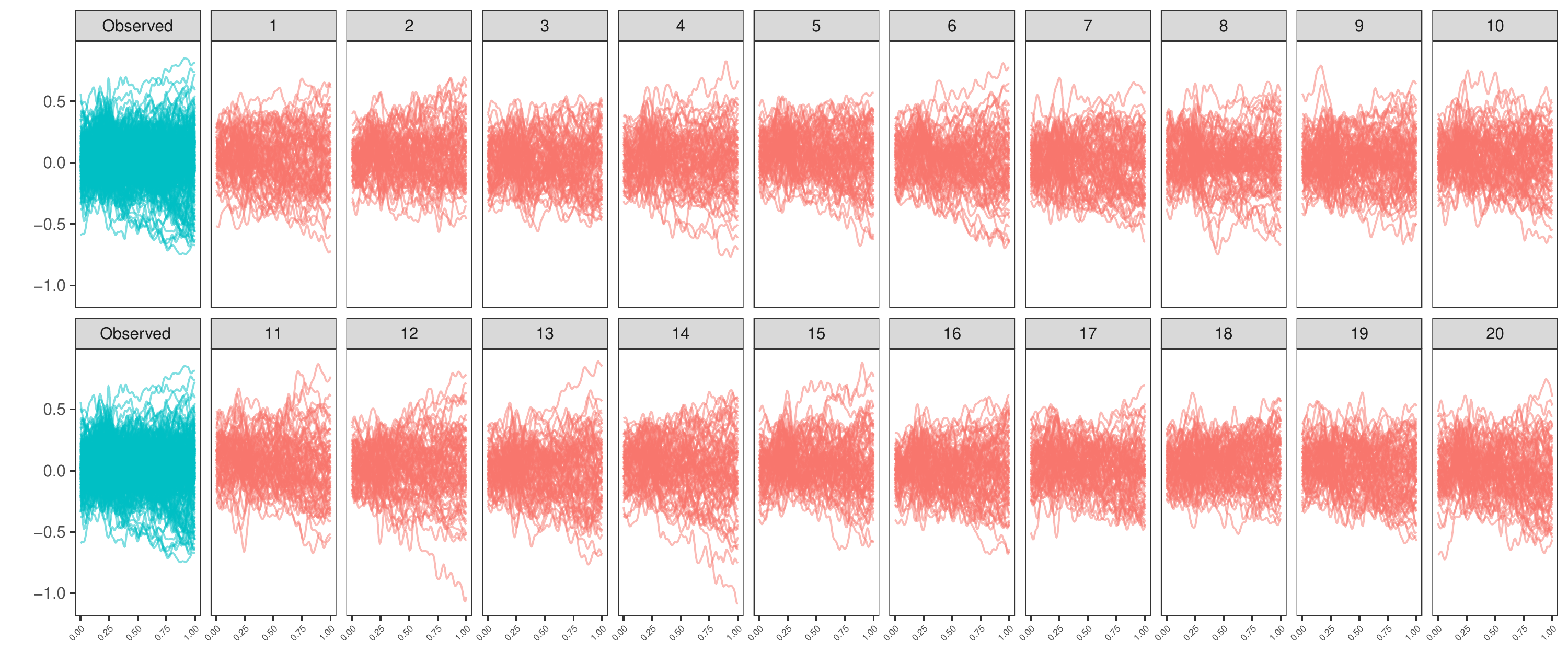}
  \caption{Functional strip plot for MDD subset.}
  \label{fig:MDD_func_strip_plot}
\end{figure}

Figures \ref{fig:HC_func_strip_plot} and \ref{fig:MDD_func_strip_plot} show strip plots for the frontal asymmetry (FA) curves from the Healthy Control (HC) and Major Depressive Disorder (MDD) subsets respectively.  In each figure, there is one panel for each of the 20 imputed data sets.  In each of the numbered panels, we show the imputed FA values.  The panels titled ``Observed'' show the observed FA curves for the sample.

 \newpage
 
\bibliographystyle{mybst}
\bibliography{fmd_refs}

\end{document}